\documentclass[10pt,journal,compsoc]{IEEEtran}

\ifCLASSOPTIONcompsoc
\usepackage[nocompress]{cite}
\else
\usepackage{cite}
\fi

\ifCLASSINFOpdf
\else
\fi

\usepackage{cite}
\usepackage{graphicx}
\usepackage{textcomp}
\usepackage{xcolor}
\usepackage{tcolorbox}
\usepackage{booktabs}
\usepackage[ruled,linesnumbered]{algorithm2e}
\usepackage{booktabs}
\usepackage[british]{babel}
\usepackage{paralist}
\usepackage{multirow}
\usepackage{amsmath}
\usepackage{amssymb}
\usepackage[marginal]{footmisc}
\usepackage{subfigure}
\usepackage{balance}
\usepackage{threeparttable}
\usepackage{hyperref}
\usepackage{color}
\usepackage{colortbl}
\usepackage{array}
\usepackage[framemethod=default]{mdframed}
\usepackage{showexpl}
\usepackage{comment}
\usepackage{float}
\usepackage{url}
\makeatletter
\g@addto@macro{\UrlBreaks}{\UrlOrds}
\makeatother
\usepackage{framed}
\usepackage{alltt}
\usepackage{enumitem}

\SetCommentSty{mycommfont}
\usepackage{algpseudocode}
\usepackage{tikz}
\usetikzlibrary{trees}
\usepackage{breqn}
\usepackage{pgfplots}
\usepackage{bchart}
\usepackage{makecell}
\usepackage{caption} 
\usepackage{booktabs}
\usepackage{array}
\usepackage{subfigure} 
\usepackage{graphicx}
\usepackage{footnote}
\usepackage{booktabs}
\usepackage{threeparttable}
\usepackage{multirow}
\usepackage{framed}
\usepackage{xcolor}
\usepackage{verbatim}
\usepackage[utf8x]{inputenc}
\usepackage{amsmath}
\usepackage{tikz}
\usepackage{amssymb}
\usepackage{bbding}
\usepackage{bm}
\usepackage{threeparttable}
\usepackage{scalerel}
\usepackage{amssymb}
\def\mcirc{\mathbin{\scalerel*{\bigcirc}{t}}}
\usepackage{diagbox}
\usepackage{balance}
\usepackage{url}
\usepackage[linesnumbered, ruled]{algorithm2e}

\usepackage{setspace}
\usepackage{textcomp}
\usepackage{tcolorbox}
\usepackage[normalem]{ulem}

\newcommand{\new}[1]{{\color{black} #1}}
\newcommand{\baidu}{Baidu}
\newcommand{\add}[1]{{\color{black} #1}}
\newcommand{\del}[1]{}
\newcommand{\ins}[1]{{\color{black} #1}}
\newcommand{\rme}[1]{}
\newcommand{\inss}[1]{{\color{black} #1}}
\newcommand{\rmee}[1]{}

\IEEEoverridecommandlockouts
\IEEEpubid{\makebox[\columnwidth]{978-1-6654-4407-1/21/\$31.00~\copyright2021 IEEE \hfill}
\hspace{\columnsep}\makebox[\columnwidth]{ }}

\begin{document}

\title{AGA: An Accelerated Greedy Additional Algorithm for Test Case Prioritization}

\author{Feng Li, Jianyi Zhou, Yinzhu Li, Dan Hao, Lu Zhang
\IEEEcompsocitemizethanks{\IEEEcompsocthanksitem Feng Li, Jianyi Zhou, Dan Hao, and Lu Zhang are with the Institute of Software, School of Computer Science, Peking University, Beijing, China and Key Laboratory of High Confidence Software Technologies (Peking University), MoE. Dan Hao is the corresponding author.\protect\\
 E-mail: \{lifeng2014, zhoujianyi, haodan, zhanglucs\}@pku.edu.cn  }
\IEEEcompsocitemizethanks{\IEEEcompsocthanksitem Yinzhu Li is with the Baidu Online Network Technology (Beijing) Co., Ltd.\protect\\
 E-mail: liyinzhu@baidu.com  }

}

\markboth{TRANSACTIONS ON SOFTWARE ENGINEERING, VOL. X, NO. X, MONTH YEAR}%
{Li \MakeLowercase{\textit{et al.}}: Bare Demo of IEEEtran.cls for Computer Society Journals}

\IEEEtitleabstractindextext{%
\begin{abstract}
In recent years, many test case prioritization (TCP) techniques have been proposed to speed up the process of fault detection. However, little work has taken the efficiency problem of these techniques into account. In this paper, we target the Greedy Additional (GA) algorithm, which has been widely recognized to be effective but less efficient, and try to improve its efficiency while preserving effectiveness. In our Accelerated GA (AGA) algorithm, we use some extra data structures to reduce redundant data accesses in the GA algorithm and thus the time complexity is reduced from $\mathcal{O}(m^2n)$ to $\mathcal{O}(kmn)$ \add{when $n > m$}, where $m$ is the number of test cases, $n$ is the number of program elements, and $k$ is the iteration number. Moreover, we observe the impact of iteration numbers on prioritization efficiency \inss{on our dataset} and propose to use a specific iteration number in the AGA algorithm to further improve the efficiency. We conducted \rmee{an experiment}\inss{experiments} on 55 open-source subjects. \new{In particular, we implemented each TCP algorithm with two kinds of widely-used input formats, adjacency matrix and adjacency list. Since a TCP algorithm with adjacency matrix is less efficient than the algorithm with adjacency list, the result analysis is mainly conducted based on TCP algorithms with adjacency list. The results show that AGA achieves 5.95X} speedup ratio over GA on average, while it achieves the same average effectiveness as GA in terms of Average Percentage of Fault Detected (APFD). Moreover, we conducted an industrial case study on 22 subjects, collected from \baidu, \del{which is a famous Internet service provider with over 600M monthly active users, }and find that the average speedup ratio of AGA over GA is \new{44.27X}, which indicates the practical usage of AGA in real-world scenarios.

\textbf{Note: This is a preprint of the accepted paper ``Feng Li, Jianyi Zhou, Yinzhu Li, Dan Hao, and Lu Zhang. AGA: An Accelerated Greedy Additional Algorithm for Test Case Prioritization. IEEE Transactions on Software Engineering, 2021'', which can be accessed at \url{https://ieeexplore.ieee.org/document/9662236}.}
\end{abstract}

}

\maketitle

\IEEEdisplaynontitleabstractindextext

\IEEEpeerreviewmaketitle

\section{Introduction}
\label{introduction}

Test case prioritization (abbreviated as TCP)~\cite{rothermel2001prioritizing,elbaum2001incorporating,elbaum2002test,qu2008configuration,rothermel1999test,wong1997study}, is proposed to schedule the execution order of test cases so as to detect faults as early as possible. To address this problem, a large number of TCP techniques have been proposed in the literature.

Among these TCP techniques, the Greedy Additional (GA) algorithm has received much attention since it was proposed in 1999~\cite{rothermel1999test} due to its widely recognized effectiveness~\cite{zhang2013bridging,li2007search,lin1965computer,jiang2009adaptive}. In particular, the GA algorithm iteratively selects the next test case which covers the largest number of elements (e.g., methods, branches, statements) that have not been covered by previously selected test cases. When the selected test cases cover all elements, this GA algorithm deals with the remaining unselected test cases with any prioritization technique (e.g., Greedy Total algorithm~\cite{rothermel1999test}, \add{which schedules these test cases based on the descendent order of the number of total covered program elements}). Later in 2002, Elbaum et al.~\cite{elbaum2002test} slightly modified this algorithm by reordering the remaining test cases with the GA strategy again after resetting all the elements to be ``uncovered''. This GA algorithm repeats the GA strategy until all the test cases are selected and thus its effectiveness is no worse than that of the original GA algorithm~\cite{rothermel1999test}. Therefore, the GA algorithm proposed by Elbaum et al.~\cite{elbaum2002test} is taken as the default GA algorithm by most researchers in TCP and in this paper\footnote{Without further clarification, the GA algorithm used in this paper refers to the one proposed by Elbaum et al.~\cite{elbaum2002test}.}. Moreover, the original GA algorithm is called the GA-first algorithm for distinction. \add{Note that we target GA rather than GA-first in this paper because the former is more widely used in the literature.} Although researchers have put dedicated efforts in TCP and have proposed a large number of TCP techniques since then, the GA approach~\cite{elbaum2002test} remains one of the most effective strategies in terms of fault-detection rate~\cite{zhang2013bridging,li2007search,jiang2009adaptive}, which is usually measured by the average percentage of faults detected (abbreviated as APFD). In other words, none of the existing TCP techniques can always outperform GA~\cite{elbaum2002test} in terms of effectiveness.

Besides effectiveness, time cost is widely recognized as another important issue influencing the application of an approach~\cite{cormen2009introduction}\add{\mbox{\cite{henard2016comparing,elbaum2014techniques,mantyla2015rapid}}, especially considering the limited available time}. In particular, the time cost of TCP, called TCP efficiency in this paper, refers to how much time a TCP approach consumes. As reported, Google~\cite{memon2017taming} runs 800K builds and 150M tests every day \add{(the same tests are run many times)}. If a TCP approach consumes much more time on prioritization, the time left for test running will be reduced to a large extent. Furthermore, software modification occurs dramatically frequently so that regression testing consumes about 80\% testing cost~\cite{Ashish:ISDC10}. For example, Google developers modify source code \del{more than 20 times per minute }\add{one time per second} on average~\del{\mbox{\cite{Chittimalli:TSE09}}}\add{\cite{memon2017taming}}. To improve the efficiency of regression testing, it is necessary to apply TCP more than once because frequent code modification may hamper the effectiveness of TCP~\cite{lu2016does}. That is, considering the practical application of TCP, including the GA algorithm, both effectiveness and efficiency are important.

However, existing TCP approaches, including the GA algorithm, suffer from the efficiency problem, e.g., the previous work shows that most existing TCP approaches cannot deal with large-scale application scenarios~\cite{miranda2018fast,elbaum2014techniques,memon2017taming}. Furthermore, some work~\cite{miranda2018fast,elbaum2014techniques,memon2017taming} points out that the GA algorithm spends dramatically long time on prioritization. Note that in the 20-year history of GA, there is no approach proposed to improve its efficiency while preserving the high effectiveness. 

In this paper, we make the first attempt to accelerate the GA algorithm and maintain the effectiveness. In particular, we analyze the efficiency problem of the GA algorithm and propose to accelerate the GA algorithm through two enhancements. The proposed algorithm is called the Accelerated Greedy Additional (abbreviated as AGA) algorithm. First, many redundant data accesses occur during prioritization in GA. Whenever a test case is selected, the GA algorithm scans the coverage information of all test cases to mark elements covered by this selected test case and calculates the number of unmarked elements covered by each unselected test case. Such scanning is less efficient and may contain many redundant data accesses. Therefore, we design some extra data structures (e.g., indices) to summarize the coverage information of each test case in the AGA algorithm. Supposed that $m$, $n$, $k$ are the number of test cases, the number of elements and the number of iterations to repeat GA strategy (which is called iteration number in this paper), \add{and given $n > m$ (which is true in most cases),} the time complexity of our AGA algorithm is $\mathcal{O}(kmn)$, while the time complexity of the GA algorithm is $\mathcal{O}(m^2n)$. \inss{The value of $k$ determines to what extent the former is superior to the latter. In practice, $k$ is usually much smaller than $m$, and in our approach, $k$ is fixed as a constant (by the second part below), so, our $\mathcal{O}(kmn)$ is superior to 
$\mathcal{O}(m^2n)$.} Second, the GA algorithm proposed by Elbaum et al.~\cite{elbaum2002test} repeats the GA strategy multiple times in TCP and thus the iteration number is usually larger than 1. Intuitively, when an element is covered for enough times, the probability that it still contains faults is low, so the remaining iterations may not contribute to the effectiveness but only decrease TCP efficiency. Therefore, we investigated their relation \inss{empirically} and applied it to modify the GA algorithm to improve efficiency but preserving effectiveness. To sum up, our AGA algorithm consists of two parts, time complexity reduction and iteration number reduction. Note that theoretical improvement is rather important and gives clear assurance for high-efficiency under any situations \inss{(especially in the first part of AGA)}. Also, our simple technique with theoretical improvement is meaningful in practice and can illustrate the simple nature of the problem.

We conducted controlled \rme{experimental study}\ins{experiments} by using 55 open-source projects from GitHub (whose total lines of code are from 1,621 to 177,546). \rme{In this paper we implement a TCP algorithm with two kinds of input formats, adjacency matrix and adjacency list, both of which are widely used to record the coverage information in existing TCP. Moreover, as the latter is more efficient, we conducted detailed analysis on the results of TCP algorithms with adjacency list and summarized the conclusions for TCP algorithms with adjacency matrix.}\ins{Because the algorithm input (program coverage) has two kinds of format, adjacency matrix and adjacency list, we conduct our experiments on both of them, which is discussed in Section~\ref{sec:example}.} In the experiments, we studied the contributions of the two parts of AGA separately, and found that both of them improve the efficiency to a large extent. Furthermore, we investigated the effectiveness and efficiency of AGA by comparing it with GA. The results showed that on average the speedup ratio of AGA over GA is \new{5.95X and} 27.72X \new{on two input formats}, which is a very large improvement. We also find that the average APFD of AGA and GA is the same, and \rme{the Kruskal-Wallis rank sum test~\mbox{\cite{kruskal1952use}}}\ins{Analysis of Covariance (ANCOVA)~\mbox{\cite{fisher1992statistical}}} shows no significant difference between them. Moreover, the effect size \rme{($\eta^2$)}\ins{(Cohen's $d$)} also indicates small effect.

We also empirically compared AGA with \emph{FAST}~\cite{miranda2018fast}, which focuses on the TCP efficiency problem. As \emph{FAST}~\cite{miranda2018fast} targets a different problem, improving the time efficiency by sacrificing effectiveness, such a comparison in terms of efficiency may be a bit unfair to our AGA approach. Surprisingly, the results showed that the average speedup ratio of AGA over \emph{FAST} is 4.29X (with significant difference and \rme{moderate}\ins{medium} effect), which means AGA even outperforms the technique that sacrifices effectiveness to achieve high efficiency. Also, the average APFD difference that AGA exceeds \emph{FAST} is 0.1702, and \rme{the Kruskal-Wallis rank sum test}\ins{ANCOVA} shows that the difference is statistically significant. Moreover, the effect size \rme{($\eta^2$)}\ins{(Cohen's $d$)} also indicates \rme{large}\ins{huge} effect.

We further performed an industrial case study in \baidu, a famous Internet service provider with over 600M monthly active users. In particular, we compared the performance of AGA and GA in 22 subjects of \baidu. In this industrial case study, the average speedup ratio of AGA over GA is \new{44.27X and} 61.43X \new{on two input formats}, which indicates the usefulness of AGA in real-world large-scale scenarios. Also, AGA is faster than \emph{FAST} on all 22 subjects and achieves \new{4.58X} speedup ratio on average, and the difference is statistically significant with \rme{large}\ins{very large} effect. Due to the commercial constraints, we cannot access the source code of these projects, and the developers in \baidu~also do not record the fault positions in the history, which are necessary to calculate the APFD results. So, we did not compare the effectiveness of these approaches in this study.

The contributions of this work are summarized as below.
\begin{itemize}
    \item The first attempt to improve the efficiency of GA while preserving its effectiveness, since GA is believed to have high effectiveness. \add{In particular, we resolve the efficiency issue of GA through theoretical improvement, which gives clear assurance for high-efficiency under any situations.}
    \item An approach to accelerating the widely-known GA algorithm through two parts, including time complexity reduction and iteration number reduction. With the former, the complexity is reduced from $\mathcal{O}(m^2n)$ to $\mathcal{O}(kmn)$ \add{given $n > m$}\inss{, which is theoretically proved}; with the latter, the corresponding AGA algorithm is more efficient and can be as competitive as GA regarding to effectiveness\inss{, which is empirically shown}. \ins{In fact, although it seems like an easy-to-implement algorithm, in the broad literature, nobody realizes this optimization and the subsequent reduction of complexity. Therefore, this paper is the first to systematically analyze this problem and propose and evaluate the optimization approach, which is helpful for the community.}
    \item Large scale experiments on 55 open-source projects demonstrating the effectiveness and efficiency of our AGA approach, compared with the GA algorithm.
    \item An empirical comparison of AGA with \emph{FAST}, which improves time efficiency but decreases effectiveness.
    \item An industrial case study on 22 subjects from \baidu, which indicates the practical usage of AGA in real-world scenarios.

\end{itemize}

\section{Time Complexity Reduction}
\label{sec:app1}

In this section, we review the Greedy Additional (GA) algorithm by an example (in Section~\ref{sec:example}). By analyzing its time complexity (in Section~\ref{sec:analysis}), we propose to accelerate GA through extra-defined data structures (in Section~\ref{sec:tech1}). Such modification improves the efficiency of GA so that the time complexity becomes $\mathcal{O}(kmn)$ \add{(given $n > m$)}, whereas the complexity of GA is $\mathcal{O}(m^2n)$, where $n$ is the number of program elements (e.g., statements, branches, methods) covered by the test suite, $m$ is the number of test cases in the test suite, and $k$ is the iteration number.

\subsection{Example}
\label{sec:example}

Table \ref{example} presents an example showing the coverage information of a test suite. This test suite consists of five test cases (i.e., T1, T2, $\dots$, and T5) and the test suite covers five program elements (i.e., E1, E2, $\dots$, and E5). \new{A common representation form of coverage information is adjacency matrix, which is shown in Table~\ref{matrix}.} $\mcirc$ represents that the test case covers the corresponding program element, while $\times$ represents the opposite. \new{Another representation form of coverage information is adjacency list, which is shown in Table~\ref{list}. In our example, the two forms represent totally the same information.}

\begin{table}[htbp]
	\centering
	\caption{An Example}
	\label{example}
	\subtable[Adjacency Matrix]{
	\label{matrix}
	\resizebox{0.32\textwidth}{!}{
	\begin{threeparttable}
	\begin{tabular}{cc|ccccc}
	\toprule
	\multicolumn{2}{c|}{\multirow{2}{*}{\textbf{Cover or Not}}} & \multicolumn{5}{c}{\textbf{Elements}}  \\
		\multicolumn{2}{c|}{} & E1 & E2 & E3 & E4 & E5 \\
	\midrule
		\multirow{5}{*}{\textbf{Test Cases}} & T1 & $\mcirc$ & $\mcirc$ & $\mcirc$ & $\times$ & $\times$ \\
		& T2 & $\times$ & $\times$ & $\mcirc$ & $\mcirc$ & $\mcirc$ \\
		& T3 & $\mcirc$ & $\mcirc$ & $\times$ & $\times$ & $\times$ \\
		& T4 & $\times$ & $\times$ & $\mcirc$ & $\mcirc$ & $\times$  \\
		& T5 & $\times$ & $\times$ & $\times$ & $\times$ & $\mcirc$ \\
	\bottomrule
	\end{tabular}
	\end{threeparttable}
	}
	}
	
	\subtable[\new{Adjacency List}]{
	\label{list}
	\resizebox{0.25\textwidth}{!}{
	\begin{threeparttable}
	\begin{tabular}{c|l}
	\toprule
	\textbf{Test Cases} & \textbf{Covered Elements} \\
	\midrule
		T1 & E1 \quad E2 \quad E3 \\
		T2 & E3 \quad E4 \quad E5 \\
		T3 & E1 \quad E2 \\
		T4 & E3 \quad E4 \\
		T5 & E5 \\
	\bottomrule
	\end{tabular}
	\end{threeparttable}
	}
	}
\end{table}

\new{If we take the adjacency matrix as input,} the GA algorithm runs as follow. First, no element has been covered before and this algorithm scans the whole table to calculate the number of elements covered by each test case. Then it chooses T1 or T2 since both of them cover the most elements. Supposed that this algorithm chooses T1, then T2, T3, T4, and T5 remain unselected. As the selected test case T1 covers elements E1, E2, and E3, the rest elements E4 and E5 remain uncovered. The algorithm scans the whole table again to find that T2, T3, T4, and T5 covers 2, 0, 1, and 1 of the 2 uncovered elements, respectively. So, the GA algorithm chooses T2 as the next test case. Now, all elements have been covered and the GA algorithm~\cite{elbaum2002test} starts another iteration by resetting all elements to ``uncovered''. Finally, the test execution sequence produced by the GA algorithm is ``T1, T2, T3, T4, T5''. \new{On the other hand, provided the adjacency list as input, GA runs similarly and produces the same output.}

\subsection{Analysis of the GA Algorithm}
\label{sec:analysis}

In this section, we analyze the time complexity of the GA algorithm through its general implementation. Suppose the coverage information is recorded in a table like Table~\ref{matrix}, the GA algorithm first scans the whole table to find the line with the most ``$\mcirc$'' entries and selects the corresponding test case into the prioritized sequence. When a test case is selected and added to the sequence, the GA algorithm scans the whole table to find the ``$\mcirc$''s whose corresponding element is covered by the latest selected test case. These ``$\mcirc$''s are replaced by ``$\times$''s. The GA algorithm repeats the proceeding process until all the entries in the table are ``$\times$''s or all the test cases have been selected. In the latter case the termination condition is satisfied and the GA algorithm ends by producing a prioritized test suite; otherwise, GA reuses the initial table by replacing ``$\mcirc$''s with ``$\times$''s for each selected test case and repeats the proceeding process again.

Supposed that there are $m$ test cases in the given test suite to be prioritized and $n$ program elements are covered by the test suite, the GA algorithm needs to scan the whole table for $m$ times and thus the time complexity is $\mathcal{O}(m^2n)$, as shown by previous work~\cite{li2007search,elbaum2002test,zhang2013bridging}. However, lots of accesses of the table are redundant. First and the most importantly, every time the coverage table is updated, the GA algorithm recalculates the total ``$\mcirc$'' entries of each unselected test case, without reusing previous calculation. Second, none of the accesses to ``$\times$''s in the table is necessary because the GA algorithm does not want to update them in the process. Third, in order to find the elements covered by the latest selected test case, the GA algorithm scans all elements in the table, which is also unnecessary. Let us illustrate the preceding redundant accesses by the example. When T1 is selected first, the GA algorithm scans Row T1 and finds three ``$\mcirc$''s. Among the five accesses (i.e., E1, E2, $\ldots$, E5), the accesses of E4 and E5 are redundant. Then, the GA algorithm changes the state of E1, E2, and E3 in other four test cases from ``$\mcirc$'' to ``$\times$''. During this process, it is also not necessary to access the state ``$\times$''. Then, the GA algorithm scans the whole table to select the next test case, but this process can be optimized by analyzing updated columns and the previous calculation on total number of ``$\mcirc$'' covered by each test case. To sum up, due to such a large number of redundant accesses in the GA algorithm, it is possible to reduce its time cost and improve its efficiency.

\new{If we take the adjacency list as input, similar analysis can be done. First, the accesses of ``$\times$''s to find covered elements in one row is reduced, while more time is spent on finding all test cases that cover a specific element (through scanning of the whole list). As a result, the overall time complexity remains $\mathcal{O}(m^2n)$. Second, lots of accesses of the list are redundant, too. Following our previous analysis, the time efficiency can be improved through reducing the unnecessary operations.}

\subsection{Improvement of Time Complexity}
\label{sec:tech1}

To reduce such redundant accesses, we propose the \textbf{AGA\_C} approach that defines extra data structures, which reuse previous information collected during its execution. In particular, we use a list to record the total number of elements covered by each test case and dynamically update it during prioritization, in order to alleviate the scanning of the coverage table. We also use forward and inverted indices to save the data accesses of ``$\times$'' entries in the table.

Our AGA\_C algorithm is shown in Algorithm~\ref{aga}. Line~\ref{ini} initializes several data structures. $TC$ is a list of length $m$ recording the number of elements covered by each test case. In our example, $TC$ is $[3, 3, 2, 2, 1]$ \inss{from Table~\ref{example} by definition}. $HS$ is a list of length $n$ recording whether each test case has been selected. $HC$ is a list of length $n$ recording whether each element has been covered by previous test cases. $FI$ are forward indices that index all elements covered by each test case, while $II$ are inverted indices that index all test cases that cover each element. \inss{From Table~\ref{example},} \ins{in our example, $FI$ records that T1 covers [E1, E2, E3], T2 covers [E3, E4, E5], etc. $II$ records that E1 is covered by [T1, T3], E2 is covered by [T1, T3], etc.} Line~\ref{pini} initializes $\mathbf{P}$ as the empty list. Then, in Line~\ref{loopstart} to Line~\ref{loopend}, the algorithm selects $m$ test cases in turn. First, it chooses the largest value in $TC$ whose test case $t$ is marked unselected in $HS$. The algorithm adds $t$ to the prioritized list $\mathbf{P}$ and marks it in $HS$. In our example, in the first loop, T1 is selected \inss{(since it covers the most program elements)}, marked in $HS$, and added to $\mathbf{P}$. Then, for every element $j$ in $FI[t]$ that is marked uncovered in $HC$, the algorithm marks it as covered and for every test case $i$ in $II[j]$, the algorithm substracts $TC[i]$ by 1. \ins{In our example, in the first loop, E1 and E2 are marked covered and the updated $TC$ is $[0, 2, 0, 1, 1]$} Finally, the algorithm continues to select the next test case by repeating the process. As shown from Line~\ref{ifstart} to Line~\ref{ifend}, if all elements have been covered by selected test cases, the algorithm completes current iteration and restores the original $TC$ to start the next iteration. \ins{In our example, after T1 and T2 are selected, the original $TC$ is restored.} The total number of iterations is called \textbf{iteration number}.

\begin{algorithm}[t]
\caption{AGA\_C algorithm}
\label{aga}
\KwIn{\new{Coverage information $\mathbf{M}$;}}
\KwOut{Prioritized test cases $\mathbf{P}$;}
Initialize $TC$, $HS$, $HC$, $FI$, and $II$ from $\mathbf{M}$, \ins{$t = 0$}\;
\label{ini}
Set $\mathbf{P}$ as empty list\;
\label{pini}
\While{\rme{$\textrm{length}(\mathbf{P})$ $\ne m$}\ins{$t < m$}}{
\label{loopstart}
Find the largest value in $TC$ that the corresponding test case $t$ has not been selected (take the use of $HS$)\;
\If{No test case can be selected}{
\label{ifstart}
Change $TC$ to the original value\;
Change $HC$ to the original value\;
Continue\;
}
\label{ifend}
Add $t$ into $\mathbf{P}$\;
Mark $t$ as selected in $HS$\;
\ForAll{$j$ in $FI[t]$}{
\If{$HC[j]$ is ``uncovered''}
{
Mark $j$ as ``covered'' in $HC$\;
\ForAll{$i$ in $II[j]$}{
Decrease $TC[i]$ by 1\;
}
}
}
\ins{$t = t + 1$}\;
}
\label{loopend}
Return $\mathbf{P}$\;
\end{algorithm}

Furthermore, we analyze the time complexity of our AGA\_C algorithm. All initialization operations consume $\mathcal{O}(mn)$ time. Each calculation of maximum value in $TC$ consumes $\mathcal{O}(m)$ time, which leads to $\mathcal{O}(m^2)$ time in total. The number of times to update $TC$ is equal to the elements in $FI$ (also equal to the test cases in $II$) in an iteration, which is the number of ``$\mcirc$'' entries in the coverage matrix. So, in each iteration, the algorithm updates $TC$ for up to $\mathcal{O}(mn)$ times, and the total time for updating $TC$ is $\mathcal{O}(kmn)$, where $k$ is the iteration number.\rme{In this paper, we focus on accelerating test case prioritization based on statement coverage, since statement coverage is the mostly studied coverage criterion and it low-efficiency problem is more severe than other granularities. Therefore, the complexity analysis is conducted mainly based on statement coverage as well.} \rme{In this case}\ins{Generally speaking}, the number of elements is often \del{much }larger than the number of test cases, which means \del{$n \gg m$}\add{$n > m$}. So, \add{according to the definition of Big O notation,} the total time complexity $\mathcal{O}(kmn + m^2)$ can be simplified \del{to $\mathcal{O}(kmn)$ }\add{as $\mathcal{O}(kmn + m^2) = \mathcal{O}(kmn + mn) = \mathcal{O}((k+1)mn) = \mathcal{O}(kmn)$}, where $k$ is the iteration number. \del{Note that for statement coverage, $n \gg m$ obviously holds, and can also be verified by subject statistics (see the subject information in Table~\ref{opensourceinfo}). For other coverage, this may not be true but we still improve the whole complexity a lot. }\add{Note that in most cases, $n > m$ obviously holds, and can also be verified by the subject statistics in this paper (given by Table~\ref{opensourceinfo}). For other special cases, the original time complexity $\mathcal{O}(kmn + m^2)$ is still a large improvement.}

In addition, in our algorithm, we use more storage space to maintain the extra data structures in order to improve time complexity. So, it is necessary to analyze the space complexity, too. In GA, \new{the coverage information (adjacency matrix/list)} takes $\mathcal{O}(mn)$ space, and additional $\mathcal{O}(1)$ space is used to store temporary variables in the algorithm, which means the overall space complexity of GA is $\mathcal{O}(mn)$. In AGA, the same $\mathcal{O}(mn)$ space is used to store the coverage table, while $TC$, $HS$, $HC$, $FI$, and $II$ need $\mathcal{O}(m)$, $\mathcal{O}(m)$, $\mathcal{O}(n)$, $\mathcal{O}(mn)$, and $\mathcal{O}(mn)$ spaces, respectively. So, the overall space complexity of AGA is $\mathcal{O}(mn)$, which is the same as GA, with the only difference lying in the constant factor.

\section{Iteration Number Reduction}
\label{sec:app2}

From Section~\ref{sec:app1}, we obtain a new approach with time complexity $\mathcal{O}(kmn)$, where $k$ is the iteration number. In practice, $k$ is often much smaller than $m$ in most projects because usually many test cases are needed to cover all elements in an iteration. However, in the worst case, $k$ may be equal to $m$, indicating the worst time complexity of our AGA algorithm is still the same as that of the GA algorithm.

To further improve the efficiency of the GA algorithm, especially in the worst case, in this section, we discuss the impact of the iteration number and introduce another modification adopted in our AGA algorithm. \ins{Finally, we present an experiment to evaluate the impact of iteration number on the GA algorithm.}

\ins{

\subsection{Modification with Iteration Number Reduction}
}
Let us re-examine the definition of ``an iteration''. In this paper, the process of selecting some test cases from covering $0$ element to covering all possible elements and then resetting them to be ``uncovered'' is called ``an iteration''. Intuitively, the iteration number may have large impact on time cost of the GA algorithm. The difference between the GA-first algorithm~\cite{rothermel1999test} and the GA algorithm~\cite{elbaum2002test} also indicates the influence of such an iteration number. Moreover, between these two algorithms, there exist many other potential algorithms, depending how many times the GA strategy is used (i.e., iteration number of the GA strategy) and what strategy is used to deal with the remaining unselected test cases (e.g., Greedy Total strategy, \add{which schedules test cases based on the descendent order of the number of total covered program elements}).

\begin{equation}
    \label{eq:nkl}
    n = k \times l
\end{equation}

Here, we define the average number of test cases selected in one iteration as $l$, so we can deduce Formula~(\ref{eq:nkl}). According to Formula~(\ref{eq:nkl}), if a large number of test cases are selected in one iteration, the total iteration number of this project is small; if few test cases are selected in one iteration, the total iteration number of this project is large. As our goal is to improve efficiency while preserving effectiveness, projects with small iteration number have already been efficient enough, and the time complexity $\mathcal{O}(kmn)$ can be reduced to $\mathcal{O}(mn)$. For those projects with large iteration number, it is necessary to optimize the iteration number to some extent.

In fact, everytime a program element is covered, the probability that it still contains faults decreases. After many iterations, all elements have been covered for enough times. On one hand, if all faults have been revealed after these iterations, the remaining iterations are useless for detecting faults but only increase the time cost. On the other hand, if there are still several faults existed after many iterations, they are supposed to be hard to reveal and the remaining iterations may only reveal them by chance, intuitively. So, we conjecture that after some iterations, the effectiveness of GA just fluctuates along with the remaining iterations.

Based on the above reasoning, we introduce another component of the proposed AGA algorithm, \textbf{AGA\_I}. AGA\_I reduces the time cost by reducing the iteration number. Different from the GA algorithm, AGA\_I does not repeat applying the GA strategy until all the test cases are prioritized, but stops when the specified iteration number is achieved. Regarding to the remaining unselected test cases, AGA\_I applies other less costly techniques (e.g., the Greedy Total technique (GT)~\cite{rothermel1999test}, which is usually used in previous work and also in this paper). \ins{Take Table~\ref{example} as an example, the original iteration number is 2. If we reduce it to be 1, AGA\_I does not repeat the additional strategy after selecting T1 and T2 and prioritizes the remaining test cases using GT.}

\ins{

\subsection{Experiment}
\label{sec:app2:exp}
We conjecture that AGA\_I does not influence the effectiveness (e.g., APFD) much but can improve efficiency (i.e., time cost) a lot. \rme{In our experiment in Section~\ref{setup}, we will verify this conjecture and determine a proper iteration number to be used in practice.}\ins{To verify our conjecture, we design an experiment to investigate how the iteration number impacts TCP in terms of both effectiveness and efficiency.}

Specifically, we use the same setup as the comprehensive experiments in Section~\ref{setup}. More details about the subjects, faults, implementation and supporting tools, and measurement are given in Section~\ref{setup}.
}

We applied the GA algorithm to all subjects, and recorded the total number of iterations the GA strategy is applied during the process for each project, which is denoted as $k$. Then we applied to each project $k$ modified GA algorithms, each of which is denoted as algorithm $algo_i$ ($1 \leq i \leq k$), recording their APFD values and time spent during prioritization. In particular, algorithm $algo_i$ repeats the GA strategy $i$ times and prioritizes the remaining unselected test cases by the Greedy Total strategy~\cite{rothermel1999test}. Note that algorithm $algo_1$ is actually the GA-first algorithm, whereas algorithm $algo_k$ is actually the GA algorithm.

\begin{table*}[h]
	\centering
	\caption{Statistics of the Impact of Iteration Number}
	\label{rq1_summary}
	\resizebox{0.99\textwidth}{!}{
	\begin{threeparttable}
		\begin{tabular}{c|c|cccccc|cccccc|cccccc}
			\toprule
			\multirow{2}{*}{\textbf{Subjects}} & \multirow{2}{*}{\textbf{\#Projects}} & \multicolumn{6}{c|}{\textbf{Iteration Number}} & \multicolumn{6}{c|}{\textbf{Time\_GA / Time\_GA-first} \tnote{*}} & \multicolumn{6}{c}{\textbf{APFD\_range} \tnote{**}} \\ 
			&                             & \textbf{min.}        & \add{\textbf{Q1}}& \add{\textbf{Q2}}& \add{\textbf{Q3}}& \textbf{max.}       & \textbf{ave.}          & \textbf{min.}& \add{\textbf{Q1}}& \add{\textbf{Q2}}& \add{\textbf{Q3}}& \textbf{max.}           & \textbf{ave.}          & \textbf{min.} & \add{\textbf{Q1}} & \add{\textbf{Q2}} &
			 \add{\textbf{Q3}} & \textbf{max.}        & \textbf{ave.}      \\ 
			\midrule
			\textbf{Open-Source}                & 55                          & 1 & \add{6} & \add{10} & \add{16.5} & 679         & 29.20         & 1.00   & \add{1.21} & \add{1.57} & \add{1.84} & 17.56             & 2.14         & 0.0000   & \add{0.0004} & \add{0.0013} & \add{0.0039}  & 0.1328         & 0.0085     \\ 

			\bottomrule
		\end{tabular}
		\begin{tablenotes}
			\footnotesize
			\item[*] The time ratio between the GA algorithm and the GA-first algorithm.
			\item[**] The highest APFD subtracts the lowest APFD among all iteration numbers.
		\end{tablenotes}
	\end{threeparttable}
	}
	
\end{table*}

Due to space limit, we only present some statistics of the experimental results in Table~\ref{rq1_summary}, \add{that is minimum, maximum, average, quartiles (Q1, Q2, Q3),} and the detailed results are given on the website of this project. From the \del{fifth }\add{eighth} column, the average iteration number among all open-source subjects is $29.20$. \del{Such a large iteration number implies the possibility of iteration number reduction, i.e., the second part of our AGA approach. }The \del{sixth }\add{ninth} to the \del{eighth }\add{fourteenth} columns present the ratio between the time cost of the GA approach and that of the GA-first approach~\cite{rothermel1999test}. The big gap between the maximal and minimal time ratio indicates the influence of the iteration number. \add{To better analyze the relationship between iteration number and time cost, we put detailed results in Appendix~\ref{appendb}. We draw a line chart of iteration number and time cost for each project. Note that in order to see the trend, we only present the projects whose iteration number is no less than \rme{$5$}\ins{$20$} (\rme{$k \geq 5$}\ins{$k \geq 20$}). The plots also support our claim that the iteration number contributes much to the time cost.} \add{As $k$ is the coefficient of time complexity, it largely determines the actual efficiency in practice, so, we think there is a large space to reduce time complexity.}

The last \del{three }\add{six} columns in Table~\ref{rq1_summary} present the APFD ranges of each project with different iteration numbers, that is, the highest APFD value minus the lowest APFD value. \add{From the quartiles, we conclude that although some outliers exist, most of the APFD ranges are very small.} \add{And} the average APFD range is only 0.0085 among all open-source subjects, indicating that little fluctuation of APFD occurs as the iteration number varies. 

To sum up, we have two main observations. First, along with the increase of the iteration number, the time cost also increases, indicating that the iteration number contributes much to the time cost. Second, the APFD value varies a little when the iteration number varies, which means a too large iteration number contributes little to the APFD value. These two observations also verify our conjectures in Section~\ref{sec:app2}. 

As we discuss in Section~\ref{sec:app2}, projects with small iteration numbers are efficient enough by using AGA\_C, so, we need to decide a proper reduced iteration number for projects with a large iteration number. In fact, this reduced iteration number is not fixed, which means it can be adjusted for specific usage. In this papar, we determine this value from some heuristics. On one hand, although we conjecture that there is no need to conduct too many iterations to detect faults, we still prefer to choose a relatively high value to ensure the effectiveness. On the other hand, if we assume that every time an element is covered, the probability that it still contains faults decreases to half of the original probability, given that the initial probability is $1$, we need to cover an element $10$ times to reduced the probability to be less than $1$\textperthousand~($(1/2)^{10} = 1/1024$). As a result, in the remaining of this paper, we implement our AGA approach by using $10$ as the reduced iteration number\rme{, and we discuss the influence of this choice in Section~\ref{sec:discussion}}.

\begin{tcolorbox}
	\textbf{Finding:} The iteration number has large influence on the efficiency of the GA algorithm, while it impacts little on effectiveness. In this paper we set the iteration number to be $10$ in implementing the AGA approach.
\end{tcolorbox}

\inss{Note that this finding is confirmed on our dataset empirically and may have bias considering the diversity of different datasets. However, the constraint on $k$ does reduce the overall time complexity from $\mathcal{O}(kmn + m^2)$ to $\mathcal{O}(mn + m^2)$. When $n > m$, which is general in most cases, the reduction is from $\mathcal{O}(kmn)$ to $\mathcal{O}(mn)$.}

\subsection{\ins{Discussion on the Chosen Iteration Number}}
In this paper, we set the iteration number to be $10$ in implementing AGA through some heuristics\rme{ and found that it can achieve our goal well}. Here we discuss the influence of this choice. First, we analyzed the APFD results of the GA algorithm with various iteration number (i.e., algorithm $algo_1$ ($1 \leq i \leq k$) in Section~\ref{sec:rq2}). In particular, for each project we recorded the highest APFD value (denoted as $\mathrm{APFD}_{\mathrm{max}}$) among these algorithms, and found the smallest iteration number $r$ whose corresponding APFD value is no smaller than $\mathrm{AFPD}_{\mathrm{max}}*99\%$. Surprisingly, the smallest iteration number $r$ for all projects are no larger than 10, which indicates that only several iterations is enough for maintaining original effectiveness, even in projects with the iteration number up to $679$. Second, although we set the iteration number to be $10$ in this paper, it may not be the best choice. We respectively applied $algo_8$, $algo_9$, $algo_{10}$, $algo_{11}$, and $algo_{12}$ to all projects with $k > 8$ as Section~\ref{sec:rq2}, and found that the gap between the maximum and minimum APFD value of these $algo$s is 0.0006 on average, which means that there might be many possible choices of the reduced iteration number in practice. \new{In other words, the value of $k$ in our evaluation is decided by reasoning, but it can have various values, depending on the choices of developers.} \ins{For example, they can use historical faults or seeded faults to empirically decide the value of $k$.}

\ins{

\section{Research Method}

To investigate the performance of our proposed AGA approach, we design comprehensive experiments. In this section, we briefly introduce each component of our experiments \inss{and their intentions}.

1) The main experiment of this paper is designed to confirm the contributions of our approach, and thus we investigate the improvement of AGA and its component (i.e., AGA\_I and AGA\_C) over the GA algorithm. In particular, this experiment is conducted on 55 open-source subjects. Details of this experiment are referred to Sections~\ref{setup} and \ref{results}. \inss{Note that the experiment in Section~\ref{sec:app2:exp} also shares the same setup and RQ1 complements the experiment in Section~\ref{sec:app2:exp}.} \inss{This part of experiment can show the superiority of AGA in widely-used open-source subjects.}

2) Although we aim to improve the efficiency of GA, we are also curious about how AGA performs compared with other TCP techniques. Specifically, \emph{FAST} targets the TCP efficiency problem and its goal is close to ours. Therefore, we first compare AGA with \emph{FAST}, and then with other representative TCP techniques, including ART-D, GA-S, and GE. This experiment is in Section~\ref{sec:comp}. \inss{This part of experiment can show that AGA even outperforms techniques that aim to reduce TCP time cost while sacrificing effectiveness.}

3) To show the practical usage of our approach, we conduct an industrial case study on Baidu, which is a famous Internet service providers with over 600M monthly active users. Specifically, we compare AGA with GA, \emph{FAST}, ART-D, GA-S, and GE, respectively and the experiment is in Section~\ref{sec:case}. \inss{This part of experiment can show that AGA works also well in real-world industrial applications and we receive positive feedback from Baidu.}

}
\section{Evaluation Design}
\label{setup}

We conducted experiments to evaluate our AGA approach. The experiments was performed on a server whose CPU is Intel(R) Xeon(R) E5-2683 2.10GHz with 132GB memory and whose operating system is Ubuntu 16.04.5 LTS. To make a fair comparison of time cost, we conducted all experiments on a single thread without parallel execution. 

\add{In order to make our results more reliable and let readers reuse the artefacts, we share our data, analysis scripts, and detailed data tables online.} \del{Our tool and experimental data}\add{They} are publicly available on our website: 
\new{\textbf{\url{https://github.com/Spiridempt/AGA}}}\add{, and also on figshare: \textbf{\url{https://figshare.com/s/cf8cc6ba9259c0e0754d}}.}

\subsection{Research Questions}

As our AGA approach consists of two parts, time complexity reduction (AGA\_C) and iteration number reduction (AGA\_I), the first two research questions are to investigate their impacts, separately. \inss{Note that the first research question also complements the experiment in Section~\ref{sec:app2:exp}.} The third research question is designed to investigate the performance of the whole AGA approach by comparing it with the GA algorithm. \ins{To investigate the influence of coverage type, the fourth research question is designed to investigate whether AGA  can also improve the efficiency of GA with method coverage.}

To sum up, this experiment is to answer the following three research questions.

\noindent{\bf RQ1:} \rme{How does the iteration number impact TCP in terms of both effectiveness and efficiency?}\ins{How does our reduction of iteration number perform compared with the GA algorithm in terms of efficiency?}

\noindent{\bf RQ2:} How does our reduction of the time complexity perform compared with the GA algorithm in terms of efficiency?

\noindent{\bf RQ3:} How does our AGA approach perform compared with the GA algorithm in terms of effectiveness and efficiency?

\ins{\noindent{\bf RQ4:} Can our AGA approach also improve the efficiency when method coverage is used?}

\subsection{Subjects and Faults}
\noindent{\bf Subjects.} \del{In this work, we randomly select 55 open-source projects from Github~\mbox{\cite{github}}. }\add{In this work, we use 55 open-source projects in total. Among these projects, 33 are widely used in prior work~\cite{luo2018static,lu2016does,zhou2021parallel}, the others are the most popular subjects selected from GitHub according to the number of stars.} \ins{Specifically, we target Github subjects whose primary programming language is Java and order them according to the number of stars in Jan 2019. Then, we check the first 100 subjects and keep only the ones that are code repository and the required tools (e.g., Maven, Clover, PIT, which is explained in Section~\ref{sec:implementation}) could work.} All the open-source projects used in this work are written in \emph{Java}, whose number of lines of code is from 1,621 to 254,284. Each of these projects has a test suite written in JUnit Testing Framework. The detailed information is given in \rme{Table~\ref{opensourceinfo}}\ins{Appendix~\ref{appenda} (Table~\ref{opensourceinfo})}. It is worth noting that compared with the experimental dataset used in recent TCP work~\cite{miranda2018fast,chen2018optimizing,wang2017qtep}, our dataset is larger and contains more large-scale projects, which can make our experimental results more reliable and convincing.

\noindent{\bf Faults.}
As existing work~\cite{just2014mutants,andrews2005mutation,do2006use} have demonstrated mutation faults to be suitable for software testing experimentation and mutation faults are widely used in prior work~\cite{lou2015mutation,lu2016does,zhang2013bridging,mei2012static,arafeen2013test,do2010effects,chen2018optimizing,luo2016large} to evaluate test case prioritization, we use a widely-used mutation testing tool PIT~\cite{pitest} to generate mutants for all open-source subjects. \rme{In particular, for each of these subjects we first generate all mutants and select those that were detected by at least one test case. Then, following prior work~\mbox{\cite{henard2016comparing}}, for each subject we construct one mutation group by containing all the selected mutation faults.}\ins{In particular, for each subject, first, we generate all mutants. Second, we keep the mutants that are killed by at least one failing test case\footnote{That is, the subject and the mutant produce different outputs on at least one test case}. Third, we construct one mutation group for each subject by containing all the remaining mutation faults, which is also consistent with previous work~\mbox{\cite{henard2016comparing}}.} 

\subsection{Implementation and Supporting Tools}
\label{sec:implementation}
We used Clover~\cite{clover} to collect code coverage information \ins{including both statement coverage and method coverage} for each open-source subject. In this work, \rme{we collected \textbf{statement coverage} and all}\ins{most} \rme{following }experiments are conducted on statement coverage \new{because it is the mostly studied test case prioritization granularity and its low-efficiency problem is severe}. \ins{In other word, the number of statements is larger than the number of methods and branches.} \ins{Additionally, we also design a research question to investigate whether AGA still improves efficiency in the scenario of method coverage.} The implementation code and all scripts used in this work are written with \emph{Python}.

\new{In prior work on coverage based test case prioritization, some takes the adjacency matrix as input~\cite{henard2016comparing}, while some uses the adjacency list~\cite{miranda2018fast}. In this work, in order to make a more general comparison, on one hand, we utilize the GA implementation in~\cite{miranda2018fast}, which is a relatively efficient implementation and uses adjacency list as input, and we implement AGA based on adjacency list. On the other hand, we implement GA and AGA based on adjacency matrix, too. Due to the space limit, in the experimental results, we only report the results based on adjacency list~\cite{miranda2018fast}, which can be more reliable, and the detailed results based on adjacency matrix are put on the website.}

\new{It is worth mentioning that in our experiments, when ties happen (i.e., more than one test case has the same number of covered elements), AGA/GA selects the topmost test case in the test-list (given by developers).}

\subsection{Compared Prioritization Approaches} 
Besides the proposed AGA approach and the GA approach~\cite{elbaum2002test}, in this study we also implemented the GA-first approach proposed by Rothermel et al.~\cite{rothermel1999test}. The GA-first approach~\cite{rothermel1999test} applies the greedy additional strategy only in the first iteration, and deals with the remaining test cases by other prioritization approach, e.g., the Greedy Total approach in this paper, which schedules these test cases based on the descendent order of the number of covered program elements.

\subsection{Measurement}
In this study, similar to existing work~\cite{rothermel1999test,elbaum2002test}, we used the Average Percentage of Fault Detected (APFD) to measure the effectiveness of TCP approaches. Formula~(\ref{eq:apfd}) presents how to calculate APFD values for a subject with $n$ tests and $m$ faults. Typically, $TF_i$ represents the first test case's position in the test suite that detects the $i$th fault.

\begin{equation}
    \label{eq:apfd}
    \textrm{APFD} = 1 - \frac{TF_1 + TF_2 + ... + TF_m}{nm} + \frac{1}{2n}
\end{equation}

Besides, we used the total time spent during the TCP process to measure the efficiency of a TCP approach. For fair comparison, we included the preparation time for a TCP approach, i.e., the time spent in constructing extra data structures in the AGA approach.

\subsection{Threats}
The internal threats to validity mainly lie in the implementation of studied approaches and scripts used in the experiments. To reduce this threat, the first two authors reviewed all the implementation and scripts used in this work. \add{Also, to improve the reliability of our work, we reuse some implementation code in previous work~\cite{miranda2018fast} to reduce the threats.}

The external threats to validity mainly lie in the subjects and faults. To reduce the former threat, we used 55 widely used open-source subjects in our study\add{, which consist of 33 previously used subjects~\cite{luo2018static,lu2016does} and 22 popular subjects selected from GitHub}. \ins{At the same time, as AGA is a general approach, it is not biased towards the chosen projects.} \inss{Note that because the second part of our approach (iteration number reduction) is empirically verified on our dataset, the large dataset itself also addresses the threat that our approach may be biased.} Also, some prior work~\cite{gopinath2014mutations,luo2018assessing} shows that the relative performance of different test case prioritization techniques on mutation faults may not strongly correlate with the performance on real faults, depending upon the attributes of the studied subjects, but we follow the common practice to use mutation faults for open-source projects following the preceding TCP work~\cite{lou2015mutation,lu2016does,zhang2013bridging,mei2012static,arafeen2013test,do2010effects,chen2018optimizing,luo2016large}. \ins{Additionally, to complement this experiment, in Section~\ref{sec:comp}, we also evaluate our approach on real faults.} In the future, we plan to conduct an extensive study by using more projects with more real faults. In addition, in this paper, we only target the GA algorithm and compare AGA with it. On one hand, it is widely accepted that GA remains one of the most effective strategies in terms of fault-detection rate~\cite{zhang2013bridging,li2007search,jiang2009adaptive}. On the other hand, the results of a recent work~\cite{henard2016comparing} shows that other black-box techniques that do not use coverage information (e.g.,~\cite{papadakis2014sampling,petke2013efficiency}) are often less effective than GA. \ins{At the same time, we also design another experiment in Section~\ref{sec:comp} to compare AGA with other representative prioritization techniques.} \ins{Additionally, most of our experiments are conducted on statement coverage because its wide usage and severe low-efficiency problem. In fact, our analysis of AGA is regardless of the scale of coverage matrix, and our theoretical improvement is general for all types of coverage. We also include RQ4 to empirically verify our improvement on method coverage.} \inss{Another minor threat is induced by the diversity of used subjects, which may lead to misleading statistics of our results. To address this threat, besides reporting the mean and median values, we also draw violin plots to learn the data distribution, which are shown on our website.}

\section{Results and Analysis}
\label{results}

In this section, we analyze the experimental results on open-source projects and answer the \rme{three}\ins{four} research questions.

\subsection{RQ1: \rme{Influence of the Iteration Number}\ins{Efficiency of Iteration Number Reduction}}
\label{sec:rq2}

\rme{In this section, we investigate the influence of the iteration number and determine its proper value in implementing our AGA approach.}

\ins{
In this section, we further investigate the efficiency improvement of the iteration number reduction. According to Section~\ref{sec:app2:exp}, we implement our approach with iteration number reduction alone by setting $k = 10$ and call this implementation AGA\_I. In other words, in this subsection, we assess the contribution of iteration number reduction alone (without the time complexity reduction). 
}

\ins{

The results on the 55 open-source projects are given in Table~\ref{opensource} \ins{(Appendix~\ref{appendc})}\footnote{Due to the space limit, we put the results of several research questions into one table \ins{and put the table in Appendix~\ref{appendc}}.}, where the projects are sorted in ascending order of source lines of code (SLOC) and the first two columns present the results for RQ1. $\mathrm{Time}_{\mathrm{GA}}$ presents the time cost of the GA approach, whereas $\mathrm{Time}_{\mathrm{I}}$ represents that of AGA\_I.
The speedup ratio of AGA\_I over GA is 1.08X. It is apparent that most subjects have a small iteration number in GA (less than or slightly more than 10). Therefore, AGA\_I does not improve the efficiency much for them. However, for those subjects with a large iteration number, AGA\_I could reduce their time cost.

To statistically check the differences between AGA\_I and GA, we adopt hypothesis test. We first use Shapiro-Wilk test~\cite{shapiro1965analysis} to check the normality \inss{of residuals}, and the \textit{p-value} in AGA\_I and GA is $9.416 * 10^{-16}$ and $5.239 * 10^{-16}$, which reject the hypothesis that they are normally distributed. Therefore, we need to adopt a non-parametric test. As we need to include project size as a control variable, Wilcoxon rank sum test~\cite{mann1947test} cannot be used. We seek for the proportional odds regression~\cite{mccullagh1980regression}, which is a class of generalized linear models and is equivalent to Wilcoxon rank sum test when there is a single binary covariate. We introduce a variable ``group'' representing AGA\_I and GA and take project size as a control variable. The results show that the \textit{p-value} of ``group'' is $1.380 * 10^{-6}$, indicating significant difference between AGA\_I and GA, and the effect size (Cohen's $d$~\cite{cohen2013statistical}) is 0.274 (medium effect). \inss{Here, because statistical tests of normality (e.g., Shapiro-Wilk test) might be impacted by characteristics of the data, we draw the normal probability plots additionally and put them on our website. Note that this applies to all normality checks in the following of the paper.}

}

\subsection{RQ2: Efficiency of Time Complexity Reduction}

The extra data structures defined in our AGA\_C approach do not affect the prioritization results, but reduce the time complexity of prioritization. In this section, we compared AGA\_C with GA only in terms of time cost. Note that we did not implement AGA\_I in this research question.

\rme{The results on the 55 open-source projects are given in Table~\ref{opensource}, where the projects are sorted in ascending order of source lines of code (SLOC) and the first four columns present the results for RQ2. We first focus on the third and the forth columns. $\mathrm{Time}_{\mathrm{GA}}$ presents the time cost of the GA approach, whereas $\mathrm{Time}_{\mathrm{C}}$ represents that of AGA\_C.}\ins{The results are given by the first five columns (except the third column) of Table~\ref{opensource} (Appendix~\ref{appendc}), where $\mathrm{Time}_{\mathrm{GA}}$ presents the time cost of the GA approach and $\mathrm{Time}_{\mathrm{C}}$ represents that of AGA\_C.} Moreover, we mark the results of $\mathrm{Time}_{\mathrm{C}}$ with $\checkmark$ only if $\mathrm{Time}_{\mathrm{C}} < \mathrm{Time}_{\mathrm{GA}}$.

The last row summarizes the total number of subjects where AGA\_C outperforms the GA approach. The results show that in most projects (\new{48} out of 55 open-source subjects), the time cost of AGA\_C is lower than the GA approach~\cite{elbaum2002test}, which confirms our previous theoretical analysis in Section~\ref{sec:app1}. \new{As we can see, in smaller subjects, the differences between GA and AGA\_C are very small, which may be caused by precision errors resulting from calculation or the operating system. In larger subjects, their differences are very large, which indicates the efficiency of AGA\_C.} 

In order to make our experiments comprehensive, we compared AGA\_C with the GA-first approach, whose time cost is given by the \rme{second}\ins{fourth} column $\mathrm{Time}_{\mathrm{GAF}}$ of Table~\ref{opensource} \ins{(Appendix~\ref{appendc})}. In \new{36} open-source subjects, AGA\_C is even more efficient than the GA-first approach, which applies the time-consuming additional strategy for only one iteration. 

\begin{figure}[htbp]
    \centering
	\includegraphics[width=0.49\textwidth]{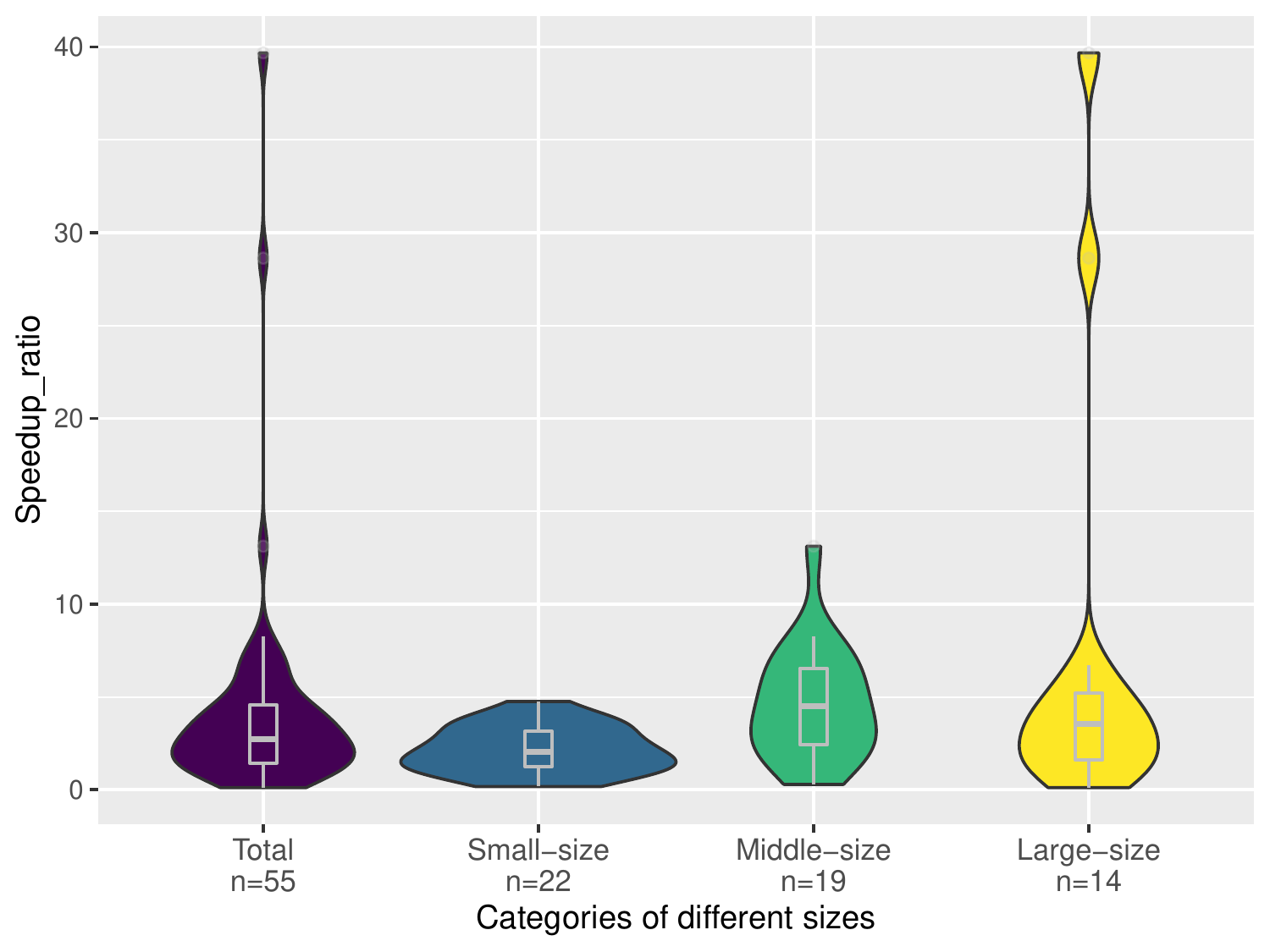}
	\caption{\add{Speedup Ratios Distribution of AGA\_C over GA on Open-Source Projects}}
	\label{violin1}
\end{figure}

In general, the efficiency improvement of AGA\_C is usually very large. In particular, if we define $\mathrm{Time}_{\mathrm{GA}}/\mathrm{Time}_{\mathrm{C}}$ as the speedup ratio of AGA\_C over GA for a project, the average speedup ratio is \new{4.37X}. As small time cost may yield biased speedup ratio, also in order to show the performance of AGA in projects with different sizes, we classify all 55 projects into small-size, middle-size, and large-size, according to the SLOC. The small-size projects ($\mathcal{S}_1$ to $\mathcal{S}_{22}$ in Table~\ref{opensource} \ins{(Appendix~\ref{appendc})}) all have less than 5,000 SLOC, the middle-size projects ($\mathcal{S}_{23}$ to $\mathcal{S}_{41}$) all have 5,000-20,000 SLOC, and the other large-size projects have more than 20,000 SLOC. The results show that the average speedup ratio in the three categories is \new{2.16X}, \new{4.65X}, and \new{7.44X}, respectively. So, the reduction of time complexity (AGA\_C) performs well, especially in projects with large sizes. \add{In order to give a more deep view into the distribution and variation of speedup ratios, we further present the violin plot with included box plot in Figure \ref{violin1}. The X-axis represents all projects and projects in three categories, respectively. We put the violin plots and box plots together to better present the distributions. From the plots, the speedup ratio of large-size projects tends to be slightly larger than that of small-size projects. Moreover, from the plot of large-size projects, several projects have very large speedup ratio because their scale is also large.}

\ins{To statistically check the differences between AGA\_C and GA, we perform hypothesis testing similar to the above. We first use Shapiro-Wilk test~\cite{shapiro1965analysis} to check the normality \inss{of residuals}, and the \textit{p-value} in AGA\_C and GA is $4.207 * 10^{-15}$ and $5.239 * 10^{-16}$, which reject the hypothesis that they are normally distributed. We also use the proportional odds regression~\cite{mccullagh1980regression} and include project size as a control variable. The results show that the \textit{p-value} of ``group'' is 0.038, indicating significant difference between AGA\_C and GA, and the effect size (Cohen's $d$~\cite{cohen2013statistical}) is 0.234 (medium effect).}

\add{Besides, we also calculate the speedup ratios of AGA\_C over GA-first for a more complete comparison. The average speedup ratio is 3.01X, and the average speedup ratio in the three categories is 1.26X, 3.31X, and 5.35X, respectively. This shows our AGA approach is also superior to GA-first.}

\ins{To statistically check the differences between AGA\_C and GA-first, we perform the similar procedure as above. We first use Shapiro-Wilk test to check the normality \inss{of residuals}, and the \textit{p-value} in AGA\_C and GA-first is $4.207 * 10^{-15}$ and $3.828 * 10^{-16}$, which reject the hypothesis that they are normally distributed. We also use the proportional odds regression~\cite{mccullagh1980regression} and include project size as a control variable. The results show that the \textit{p-value} of ``group'' is 0.399, indicating no significant difference between AGA\_C and GA-first, and the effect size (Cohen's $d$) is 0.208 (medium effect).}

\new{Provided adjacency matrix as input, we also implemented GA and AGA\_C, and the detailed results are on our website. Specifically, the average speedup ratio of AGA\_C over GA is 24.18X, and the average speedup ratio in the three categories is 5.47X, 28.16X, and 48.19X, respectively.}

Besides, the speedup ratios of the AGA\_C approach vary a lot in different projects. On \new{19} open-source subjects AGA\_C is less efficient than GA-first. On the one hand, the iteration numbers of these projects are high so that AGA\_C becomes a bit costly. On the other hand, in the only iteration of GA-first, few test cases are needed to cover all statements and they are selected fast so that GA-first is efficient on these projects.

\ins{
To sum up, AGA\_C addresses the high-complexity problem of GA well and successfully reduces its time complexity. For any project, any scale of coverage matrix, our approach could improve the efficiency a lot.
}

\begin{tcolorbox}
	\textbf{Conclusion to RQ2: } The time complexity reduction strategy used in our AGA approach demonstrates great efficiency improvement compared to GA. Specifically, the average speedup ratio of AGA\_C over GA is \new{4.37X/24.18X on two types of input}.
\end{tcolorbox}

\subsection{RQ3: Comparison with Greedy Additional Approaches}
\label{sec:rq3}

In this section, we compare the effectiveness and efficiency between the proposed AGA approach and two Greedy Additional approaches (including both GA and GA-first), whose results are given by the first \rme{eight}\ins{ninth} columns (except the \rme{fourth}\ins{third and fifth} column) of Table~\ref{opensource} \ins{(Appendix~\ref{appendc})}, where $\textrm{APFD}_{\textrm{AGA}}$ and $\textrm{Time}_{\textrm{AGA}}$ represent the APFD results and time cost of the AGA approach whose iteration number is set to be 10. Moreover, when the GA approach~\cite{elbaum2002test} does not outperform the corresponding AGA approach~\cite{elbaum2002test}, i.e., $\textrm{APFD}_{\textrm{AGA}} \geq \textrm{APFD}_{\textrm{GA}}$ or $\textrm{Time}_{\textrm{AGA}} < \textrm{Time}_{\textrm{GA}}$, the corresponding results of the AGA approach is marked with \checkmark.

\subsubsection{Effectiveness}

The proposed AGA approach has the same or better APFD performance as the GA approach in 51 out of 55 open-source subjects, and the average APFD value of AGA is 0.8870, which is the same as GA. On some subjects (e.g., the open-source project whose ID is $\mathcal{S}_{44}$), the AGA approach does not outperform the GA approach, but their APFD difference is usually very small (e.g., 0.0021 for this subject). We also make extra comparisons of AGA and GA-first \add{and find that AGA has the same or better APFD performance as GA-first in 45 out of 55 open-source subjects and their average APFD values are the same}. On 14 projects, neither the AGA approach nor the GA approach outperforms the GA-first approach, but their differences are small. Through our analysis, we suspect that after the first iteration, although all elements have been covered, the numbers of times that each element is covered still differ. This means test cases with a small number of times being covered should have higher priority, but in later iterations, this information is ignored.

Moreover, we statistically analyze whether the AGA approach and the Greedy Additional approaches have significant difference on their APFD values\rme{ at a significance level of 5\%}. \ins{First, we conduct the Shapiro-Wilk test to check the normality \inss{of residuals}. The \textit{p-value} of AGA, GA, and GAF is 0.328, 0.298, and 0.283, indicating we cannot reject the hypothesis that they are normally distributed. \inss{We additionally perform Shapiro-Wilk test to check the normality of residuals, and the \textit{p-value} of AGA, GA, and GAF is 0.328, 0.298, and 0.283, indicating we cannot reject the hypothesis that they are normally distributed.} Therefore, we can use parametric test in the following. We use Bartlett's test~\cite{bartlett1937properties} to check the homogeneity of variance, and the \textit{p-value} is 0.880, indicating we cannot reject the hypothesis that they have equal variance. Then, as we need to take project size as a control variable (covariate), we use Analysis of Covariance (ANCOVA)~\cite{fisher1992statistical}, a parametric test that works on two or more groups to check whether different groups have the same means. The \textit{p-value} is 0.641, indicating we cannot reject that they have the same means. Then, pairwise ANCOVA tests show that the \textit{p-values} of AGA vs. GA, AGA vs. GAF, and GA vs. GAF are 0.981, 0.427, and 0.414. In other words, the probability that AGA is as competitive as GA is more than 98\%. Then, we employ Cohen's $d$~\cite{cohen2013statistical} to compute the effect size (ES), and the results in AGA vs. GA, AGA vs. GAF, and GA vs. GAF are 0.005, 0.151, and 0.156, which are all small effects. Furthermore, we conduct Tukey's range test~\cite{tukey1949comparing} to check the 95\% confidence intervals for all pairwise differences, and the results are \mbox{[-0.022, 0.022]}, \mbox{[-0.030, 0.015]}, and \mbox{[-0.030, 0.014]}.} \rme{Similar to the previous work~\mbox{\cite{miranda2018fast}}, since we cannot assume the APFD data are normally distributed, we conducted a \textbf{non-parametric statistical hypothesis test}, the Kruskal-Wallis rank sum test~\mbox{\cite{kruskal1952use}} and the null hypothesis is the differences in the APFD values for different approaches are not statistically significant. The \textit{p-value} is \del{0.7189}0.7229, indicating no statistically significant difference among AGA, GA, and GA-first in terms of APFD. To measure the magnitude of their APFD difference, we compute the effect size~\mbox{\cite{kelley2012effect}} for Kruskal-Wallis test as the eta squared ($\eta^2$). $\eta^2 < 0.06$ indicates small effect, $0.06 <= \eta^2 < 0.14$ indicates moderate effect, and $\eta^2 >= 0.14$ indicates large effect. The effect size ($\eta^2$) here is $<0.01$, indicating small effect. So, AGA is as competitive as GA in terms of effectiveness.}

\subsubsection{Efficiency}

\begin{figure}[htbp]
    \centering
	\includegraphics[width=0.49\textwidth]{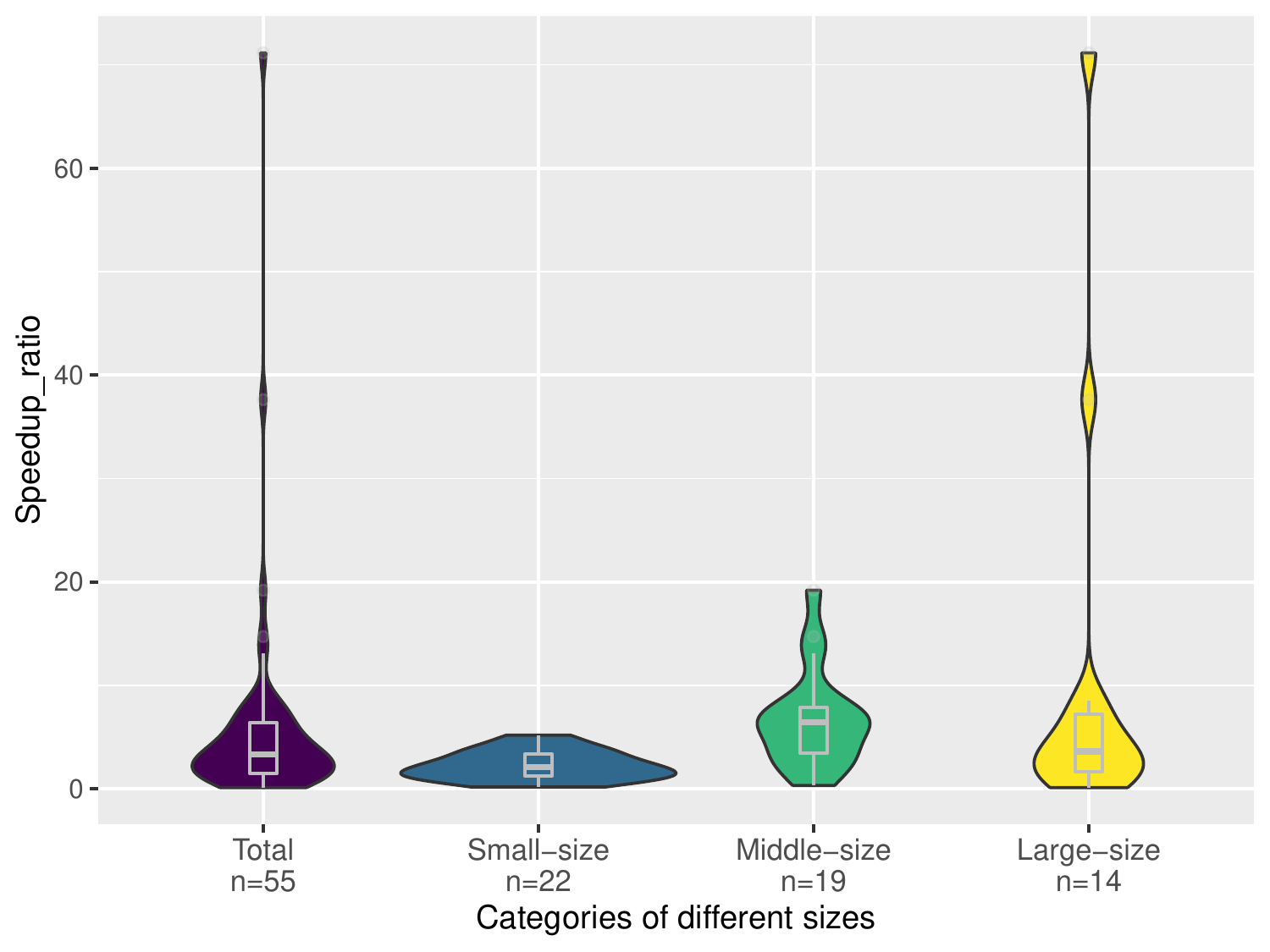}
	\caption{\add{Speedup Ratios Distribution of AGA over GA on Open-Source Projects}}
	\label{violin2}
\end{figure}

According to Table~\ref{opensource} \ins{(Appendix~\ref{appendc})}, in almost all subjects (i.e., 44 out of 55), the time cost of AGA is much lower than the GA approach. On average, the speedup ratio of AGA over GA is \new{5.95X}. Moreover, the speedup ratios in small-size, middle-size, large-size projects are \new{2.26X}, \new{6.69X}, and \new{10.76X}, respectively. \add{To learn the distribution of speedup ratios in small-size, middle-size, large-size projects, we also present the violin plot with included box plot in Figure~\ref{violin2}. From this figure, most medium-size and large-size projects achieve higher speedup ratios than small-size projects. Moreover, AGA achieves very large speedup ratios on some large-size projects.} So, AGA scales up well in large-size projects. Furthermore, we compared the time cost of the AGA approach with the GA-first approach, which requires less time than the GA approach, and find that the AGA approach even outperforms the GA-first approach in \new{37} open-source subjects. \add{The average speedup ratio is 3.95X, and the average speedup ratio in the three categories is 1.36X, 4.39X, and 7.44X, respectively.} \ins{Here, we notice that the speedup ratio of AGA over other approaches is sometimes less than 1 (e.g., $\mathcal{S}3$, $\mathcal{S}4$, $\mathcal{S}7$). In fact, the overall time complexity analysis is meaningful only when the parameters are large enough. In our dataset, some projects have a relatively small $m$ value. In this case, although $\mathcal{O}(mn)$ seems to be small, its coefficient is not negligible compared to $m$. In other words, the preliminary data structure setup consumes much time and it impacts the overall running time in some cases. This is also consistent with the empirical results that AGA performs better on large projects. On the other hand, the adjacency lists in some projects are very dense, which takes much time in the preparation of data structure, and further leads to a large coefficient. For example, $\mathcal{S}7$ and $\mathcal{S}42$ have relatively small $m$ values (45 and 34) and dense adjacency lists.}

\new{Provided adjacency matrix as input, we also implemented GA and AGA. The average speedup ratio of AGA over GA is 27.72X, and the average speedup ratio in the three categories is 5.84X, 35.47X, and 51.59X, respectively.}

To sum up, \ins{not surprisingly, the speedup ratio of AGA is higher than AGA\_C and AGA\_I. After combining AGA\_I and AGA\_C, our whole AGA approach obtains more efficient results while preserving high effectiveness.} At the same time, the proposed AGA approach is demonstrated to be efficient especially on large-scale projects. In fact, the surprisingly high efficiency of the AGA approach also indicates the existence of many redundant accesses of data and it is ubiquitous in most projects.

\begin{tcolorbox}
	\textbf{Conclusion to RQ3:} The AGA approach requires much less time in prioritization than the GA approach and the average speedup ratio is \new{5.95X and 27.72X on two types of input}. Also, AGA is as competitive as the latter in terms of APFD values (with no significant difference). This means that we achieve our goal in this paper and it has promising use in practice. 
\end{tcolorbox}

\ins{\subsection{RQ4: Performance on Method Coverage}}
\label{sec:rq4}

\ins{

In previous research questions, we focus on statement-level coverage because it is the mostly studied coverage criterion and its low-efficiency problem is more severe than other granularities. In this section, we collect the method-level coverage for each of our 55 subjects and compare the efficiency of AGA and GA. The results are shown in Table~\ref{opensource-method}. For each subject, we report the running time (in seconds) of GA and AGA.

According to Table~\ref{opensource-method}, in almost all subjects, the time cost of AGA is much lower than GA. On average, the speedup ratio of AGA over GA is 6.02X. Moreover, the speedup ratios in small-size, middle-size, large-size projects are 2.28X, 7.32X, and 10.13X, respectively. Compared to the results on statement coverage in Section~\ref{sec:rq3}, the speedup ratios are almost the same for all projects and projects in different sizes. This confirms that AGA also works well on method coverage.

In fact, the complexity analysis of GA and our AGA approach is based on a general (0,1) matrix, regardless of the meaning behind it. In other words, the type of program element (e.g., statement, method) does not affect any aspect of AGA, which means our approach works on any coverage and has a stable improvement.

\begin{tcolorbox}
	\textbf{Conclusion to RQ4:} \ins{The AGA approach also works on method-level coverage. Specifically, the average speedup ratio of AGA over GA is 6.02X.}
\end{tcolorbox}

}

\begin{table*}[htbp]
	\centering
	\caption{\ins{Results of Open-Source Subjects (Method-Level)}}
	\label{opensource-method}
	\begin{threeparttable}
\begin{tabular}{c|rr|c|rr|c|rr}
\toprule
\midrule

\ins{\textbf{Project}} & \ins{$\mathbf{Time}_{\mathbf{GA}}$} & \ins{$\mathbf{Time}_{\mathbf{AGA}}$} & \ins{\textbf{Project}} & \ins{$\mathbf{Time}_{\mathbf{GA}}$} & \ins{$\mathbf{Time}_{\mathbf{AGA}}$} & \ins{\textbf{Project}} & \ins{$\mathbf{Time}_{\mathbf{GA}}$} & \ins{$\mathbf{Time}_{\mathbf{AGA}}$} \\
\midrule

 \ins{$\mathcal{S}_{1}$} & \ins{0.0030} & \ins{0.0031} & \ins{$\mathcal{S}_{2}$} & \ins{0.0011} & \ins{0.0010} & \ins{$\mathcal{S}_{3}$} & \ins{0.0014} & \ins{0.0013} \\
 \ins{$\mathcal{S}_{4}$} & \ins{0.0021} & \ins{0.0116} & \ins{$\mathcal{S}_{5}$} & \ins{0.0051} & \ins{0.0014} & \ins{$\mathcal{S}_{6}$} & \ins{0.0019} & \ins{0.0007} \\
 \ins{$\mathcal{S}_{7}$} & \ins{0.0007} & \ins{0.0012} & \ins{$\mathcal{S}_{8}$} & \ins{0.0032} & \ins{0.0028} & \ins{$\mathcal{S}_{9}$} & \ins{0.2587} & \ins{0.0504} \\
 \ins{$\mathcal{S}_{10}$} & \ins{0.0145} & \ins{0.0081} & \ins{$\mathcal{S}_{11}$} & \ins{0.0158} & \ins{0.0080} & \ins{$\mathcal{S}_{12}$} & \ins{0.0027} & \ins{0.0020} \\
 \ins{$\mathcal{S}_{13}$} & \ins{0.0068} & \ins{0.0048} & \ins{$\mathcal{S}_{14}$} & \ins{0.0158} & \ins{0.0076} & \ins{$\mathcal{S}_{15}$} & \ins{0.0036} & \ins{0.0008} \\
 \ins{$\mathcal{S}_{16}$} & \ins{0.0106} & \ins{0.0024} & \ins{$\mathcal{S}_{17}$} & \ins{0.0741} & \ins{0.0146} & \ins{$\mathcal{S}_{18}$} & \ins{0.0047} & \ins{0.0013} \\
 \ins{$\mathcal{S}_{19}$} & \ins{0.2907} & \ins{0.1007} & \ins{$\mathcal{S}_{20}$} & \ins{0.0024} & \ins{0.0023} & \ins{$\mathcal{S}_{21}$} & \ins{0.0098} & \ins{0.0064} \\
 \ins{$\mathcal{S}_{22}$} & \ins{0.0111} & \ins{0.0055} & \ins{$\mathcal{S}_{23}$} & \ins{0.0502} & \ins{0.0147} & \ins{$\mathcal{S}_{24}$} & \ins{0.0035} & \ins{0.0031} \\
 \ins{$\mathcal{S}_{25}$} & \ins{1.1046} & \ins{0.1654} & \ins{$\mathcal{S}_{26}$} & \ins{0.0131} & \ins{0.0043} & \ins{$\mathcal{S}_{27}$} & \ins{0.0041} & \ins{0.0020} \\
 \ins{$\mathcal{S}_{28}$} & \ins{0.1945} & \ins{0.0347} & \ins{$\mathcal{S}_{29}$} & \ins{1.2958} & \ins{0.3499} & \ins{$\mathcal{S}_{30}$} & \ins{0.0624} & \ins{0.0177} \\
 \ins{$\mathcal{S}_{31}$} & \ins{0.4390} & \ins{0.0495} & \ins{$\mathcal{S}_{32}$} & \ins{0.3495} & \ins{0.0589} & \ins{$\mathcal{S}_{33}$} & \ins{0.1714} & \ins{0.0162} \\
 \ins{$\mathcal{S}_{34}$} & \ins{0.8556} & \ins{0.0824} & \ins{$\mathcal{S}_{35}$} & \ins{0.2642} & \ins{0.0330} & \ins{$\mathcal{S}_{36}$} & \ins{31.6469} & \ins{3.8975} \\
 \ins{$\mathcal{S}_{37}$} & \ins{0.7977} & \ins{0.0594} & \ins{$\mathcal{S}_{38}$} & \ins{0.5507} & \ins{0.0445} & \ins{$\mathcal{S}_{39}$} & \ins{6.6268} & \ins{0.3601} \\
 \ins{$\mathcal{S}_{40}$} & \ins{0.6570} & \ins{0.0963} & \ins{$\mathcal{S}_{41}$} & \ins{0.3203} & \ins{0.0463} & \ins{$\mathcal{S}_{42}$} & \ins{0.0011} & \ins{0.0006} \\
 \ins{$\mathcal{S}_{43}$} & \ins{0.2057} & \ins{0.0211} & \ins{$\mathcal{S}_{44}$} & \ins{1.3687} & \ins{0.1088} & \ins{$\mathcal{S}_{45}$} & \ins{0.0202} & \ins{0.0034} \\
 \ins{$\mathcal{S}_{46}$} & \ins{0.0182} & \ins{0.0044} & \ins{$\mathcal{S}_{47}$} & \ins{0.0183} & \ins{0.0045} & \ins{$\mathcal{S}_{48}$} & \ins{0.0010} & \ins{0.0003} \\
 \ins{$\mathcal{S}_{49}$} & \ins{0.0230} & \ins{0.0183} & \ins{$\mathcal{S}_{50}$} & \ins{0.0011} & \ins{0.0007} & \ins{$\mathcal{S}_{51}$} & \ins{1.0812} & \ins{0.0883} \\
 \ins{$\mathcal{S}_{52}$} & \ins{0.2423} & \ins{0.0652} & \ins{$\mathcal{S}_{53}$} & \ins{0.1631} & \ins{0.0369} & \ins{$\mathcal{S}_{54}$} & \ins{15.7745} & \ins{1.1856} \\
 \ins{$\mathcal{S}_{55}$} & \ins{190.9669} & \ins{2.9963} & \ins{} & \ins{} & \ins{} & \ins{} & \ins{} & \ins{} \\

\midrule
\bottomrule
\end{tabular}
\end{threeparttable}
\end{table*}

\section{Empirical Comparison with \rme{\emph{FAST}} \ins{Representative Prioritization Techniques}}
\label{sec:comp}

\ins{In this section, we present an experiment comparing AGA with some representative prioritization techniques. In particular, as \emph{FAST} targets the TCP efficiency problem and thus is closet to our goal, we first present the comparison study with \emph{FAST} in Section~\ref{sec:comp-fast}. Then we present the comparison study with other representative TCP techniques in Section~\ref{sec:compother}.

\subsection{Comparison with \emph{FAST}}
\label{sec:comp-fast}
}

In this section, we investigate the performance of AGA with its most related work \emph{FAST}~\cite{miranda2018fast}. In particular, \emph{FAST} is proposed as a TCP approach to address the general TCP efficiency problem by sacrificing the TCP effectiveness, and it is shown to be more efficient than other TCP techniques~\cite{miranda2018fast}. Note that there is no other work in the literature focusing on the same objective as ours, and thus we compare AGA against 
\emph{FAST}. However, AGA and \emph{FAST} target at slightly different goals: \emph{FAST} approach focuses on the efficiency problem of test prioritization, not specific to GA approaches. \del{Recently, Miranda et al.~\mbox{\cite{miranda2018fast}} proposed a TCP approach \emph{FAST} to address the general TCP efficiency problem by sacrificing the TCP effectiveness, and their experiments show that \emph{FAST} is more efficient than other techniques. }Although \emph{FAST} targets a different goal, it is still interesting to learn how AGA performs compared with \emph{FAST} in terms of time cost since both AGA and \emph{FAST} can be viewed as addressing the efficiency problem. \add{However, as \emph{FAST} improves efficiency while sacrifices effectiveness, the comparison 
in terms of time cost is a bit ``unfair'' for AGA.}
\del{Therefore, in this study, we compared the performance of AGA and \emph{FAST} on the 55 open-source projects in terms of effectiveness and efficiency.}

\add{In this study, we compare the performance of AGA and \emph{FAST} on both the 55 open-source projects used in Section~\ref{setup} and Defects4J~\cite{just2014defects4j}, which is the largest real-fault benchmark (i.e., a set of projects with reproducible real bugs) widely used in test case prioritization~\cite{luo2018assessing, paterson2018using, hasan2017test, noor2015similarity, haghighatkhah2018test} and fault localization~\cite{li2019deepfl, li2017transforming, pearson2017evaluating, sohn2017fluccs, zhang2017boosting}. For ease of understanding, we present the results of the former subjects with seeded faults and the results of the latter subjects with real faults separately.}

The \emph{FAST} approach borrows algorithms commonly used in the big data domain to find similar items and contains a family of similarity-based test case prioritization approaches. In general, the authors proposed two categories of \emph{FAST}, While-box (WB) and Black-box (BB). \ins{BB approaches take test code as input, while WB approaches take program coverage as input.} As WB approaches \ins{have the same input as us and} are much faster than BB approaches, we compare our work with WB approaches~\cite{miranda2018fast}. WB approaches include five algorithms \emph{FAST-pw}, \emph{FAST-all}, \emph{FAST-1}, \emph{FAST-log}, and \emph{FAST-sqrt}, whose difference lies in how many test cases are randomly selected for prioritization at a time. In this section, we implemented this family, and for each subject, we compared \textbf{the best results of this family} with AGA. \ins{Specifically, according to prior work~\cite{miranda2018fast}, none of the algorithms in \emph{FAST} family always performs the best. Therefore, to show the superiority of our approach, we run all \emph{FAST} algorithms and select the best one for each project. In other words, when comparing APFD, we keep the highest APFD, and when comparing time cost, we keep the lowest time cost.} Moreover, due to the randomness in \emph{FAST}, for each subject we applied each of these approaches 10 times and used their median effectiveness and efficiency results. Regarding the time cost, the same as Section~\ref{setup}, we measure the efficiency of a TCP approach by including its preparation time, i.e., the preparation time used in \emph{FAST}\footnote{\new{The previous work \emph{FAST}~\cite{miranda2018fast} separated their total running time into preparation time and prioritization time in their evaluation. However, preparation happens only once in BB approaches while not in WB approaches, because the input of BB approaches is test code. Given updated source code but out-of-date coverage information (from the previous version), we need not prioritize again and TCP results will not change. Otherwise, with updated coverage information, the whole process (including preparation) has to be repeated.}}.

\subsubsection{\add{FAST Results on Seeded Faults}} 
\label{sec:compfast1}

The results of \emph{FAST} are shown by the \rme{ninth}\ins{tenth} and \rme{eleventh}\ins{twelfth} columns in Table~\ref{opensource} \ins{(Appendix~\ref{appendc})}. Due to space limit, we do not present the results of all the five \emph{FAST} algorithms, but the largest APFD value and smallest time cost among them for each subject. Note that usually a \emph{FAST} algorithm cannot achieve both the largest APFD value and the smallest time cost. As the APFD results and time cost of AGA is already given by the \rme{seventh}\ins{eighth} and \rme{eighth}\ins{ninth} columns, we use column $\mathrm{Win}_{\mathrm{APFD}}$ and column $\mathrm{Win}_{\mathrm{Time}}$ to show whether $\mathrm{APFD}_{\mathrm{AGA}} \geq \mathrm{APFD}_{\mathrm{FAST}}$ and $\mathrm{Time}_{\mathrm{AGA}} < \mathrm{Time}_{\mathrm{FAST}}$, respectively.

Regarding to APFD values, the AGA approach is much better than \emph{FAST} in all subjects. More specifically, the differences between them are from 0.0456 to 0.3039, and 0.1702 on average. \ins{To statistically check their differences, we follow the similar procedure as above. We first use Shapiro-Wilk test to check the normality \inss{of residuals}, and the \textit{p-value} in AGA and \emph{FAST} is 0.328 and 0.137, which cannot reject the hypothesis that they are normally distributed. Then, taken project size as a control variable, the Analysis of Covariance (ANCOVA) shows that \textit{p-value} $< 2 * 10^{-16}$, indicating the statistically significant difference between AGA and \emph{FAST}. Moreover, the effect size (Cohen's $d$) is 2.96 (huge effect) and Tukey's range test shows that the 95\% confidence interval of their difference is [0.149, 0.192].}\rme{Through the Kruskal-Wallis rank sum test, the \textit{p-value} is $< 2.2 \times 10^{-16}$, indicating the statistically significant difference between AGA and \emph{FAST}. To measure the magnitude of their APFD difference, we compute the effect size $\eta^2 = 0.669$, indicating large effect.} To sum up, AGA significantly outperforms \emph{FAST} in terms of APFD because \emph{FAST} algorithms are designed to sacrifice prioritization accuracy to achieve high efficiency by using hash signatures.

\begin{figure}[htbp]
    \centering
	\includegraphics[width=0.49\textwidth]{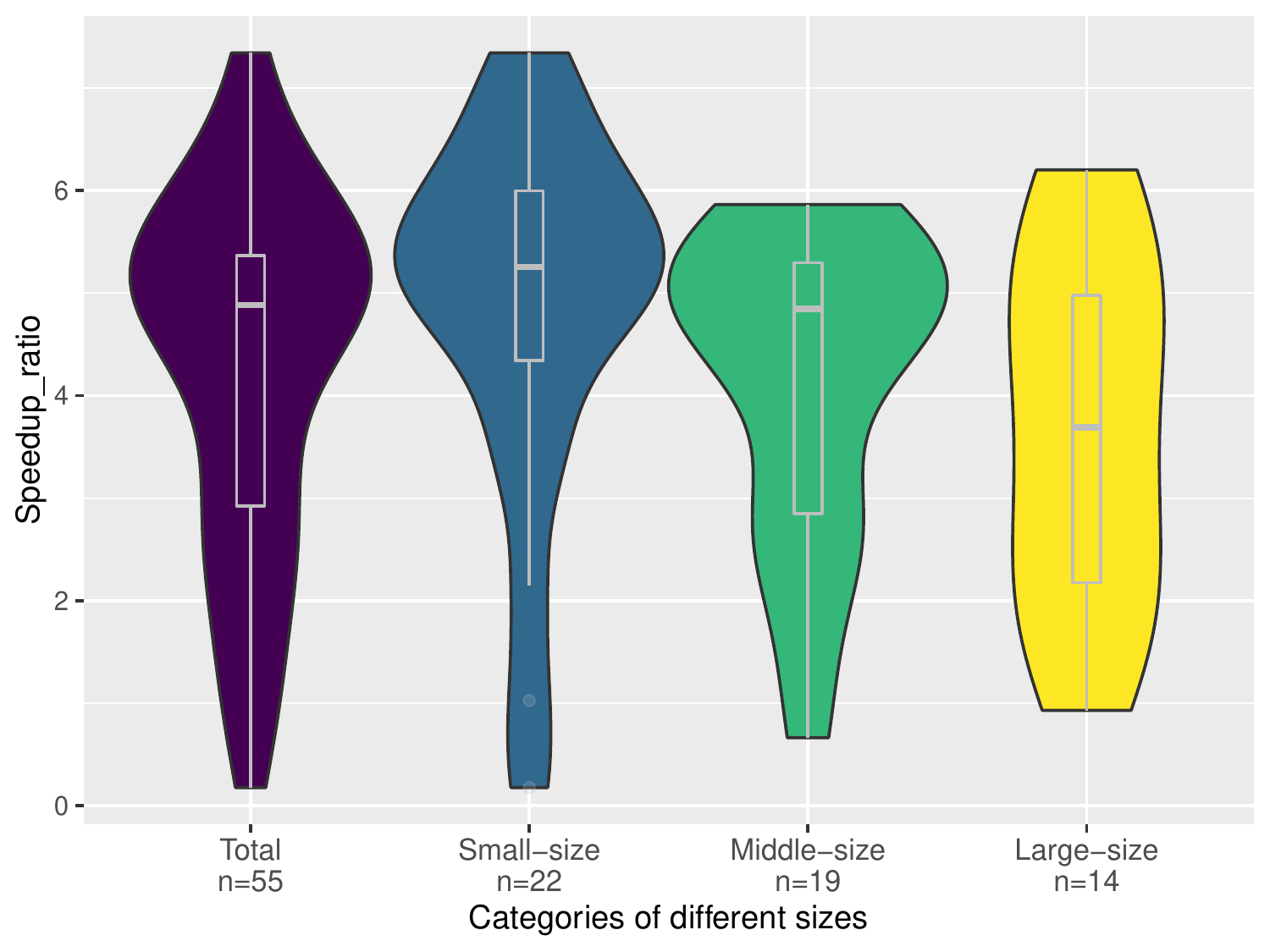}
	\caption{\add{Speedup Ratios Distribution of AGA over \emph{FAST} on Open-Source Projects}}
	\label{violin3}
\end{figure}

Regarding to the time cost, the time cost of AGA outperforms \emph{FAST} on \new{52} out of 55 open-source subjects, and the speedup ratio of AGA over \emph{FAST} is \new{4.29X}. \ins{To statistically check their differences, we follow the similar procedure as above. We first use Shapiro-Wilk test to check the normality \inss{of residuals}, and the \textit{p-value} in AGA and \emph{FAST} is $4.92 * 10^{-15}$ and $4.05 * 10^{-15}$, which reject the hypothesis that they are normally distributed. Therefore, we use the proportional odds regression~\cite{mccullagh1980regression} and include project size as a control variable. The results show that the \textit{p-value} of ``group'' is $4.250 * 10^{-4}$, indicating significant difference between AGA and \emph{FAST}, and the effect size (Cohen's $d$) is 0.286 (medium effect).} \rme{The Kruskal-Wallis rank sum test shows that the \textit{p-value} in all open-source subjects is $0.0023$, meaning statistically significant difference exists between them. To measure the magnitude of their time difference, we compute the effect size $\eta = 0.0770$, indicating moderate effect.} That is, the proposed AGA is more efficient to \emph{FAST} (with \new{4.29X} speedup ratio). This is a \textbf{surprising result} because AGA can even be faster than a technique that is designed to sacrifice effectiveness to reduce time cost. \add{We also present the violin plot with included box plot in Figure~\ref{violin3}. On larger projects, the speedup ratios are smaller, which means \emph{FAST} also scales up well on large-size projects, whereas it is less efficient than AGA.}

\begin{table*}[htbp]
    \centering
    \caption{Results of Some Defects4J Projects}
    \label{defects4j}
    \begin{threeparttable}
    \begin{tabular}{c|cr|ccrc|r|cr}
    \toprule
\multirow{2}{*}{\textbf{Projects}} & \multicolumn{2}{c|}{\textbf{\emph{FAST}}} & \multicolumn{4}{c|}{\textbf{AGA}}  & \multirow{2}{*}{\textbf{\ins{APFD\_range}}}  &  \multicolumn{2}{c}{\textbf{\ins{GAF}}}      \\
                                   & \textbf{APFD}   & \textbf{Time}   & \textbf{APFD} & $\mathbf{Win}_{\mathbf{APFD}}$ & \textbf{Time} & $\mathbf{Win}_{\mathbf{Time}}$ & & \textbf{\ins{APFD}} & \textbf{\ins{Time}} \\
    \midrule
Closure & 0.5219 & 11.7302 & 0.4347 &  & 2.0408 & \checkmark & \ins{0.0006} & \ins{0.4354} & \ins{9.4078} \\
    Math & 0.5471 & 5.2600 & 0.6992 & \checkmark & 0.9586 & \checkmark & \ins{0.0000} & \ins{0.6992} & \ins{7.4710} \\
    Lang & 0.5627 & 0.4280 & 0.6094 & \checkmark & 0.1513 & \checkmark & \ins{0.0000} & \ins{0.6094} & \ins{0.2150} \\
    Time & 0.5463 & 2.8633 & 0.5469 & \checkmark & 0.5042 & \checkmark & \ins{0.0034} & \ins{0.5436} & \ins{0.9870} \\
    Chart & 0.5264 & 5.1456 & 0.7128 & \checkmark & 0.9030 & \checkmark & \ins{NA} & \ins{0.7128} & \ins{5.0767} \\
    \ins{Mockito} & \ins{0.5197} & \ins{2.7311} & \ins{0.5975} & \ins{\mbox{\checkmark}} & \ins{0.4537} & \ins{\mbox{\checkmark}} & \ins{0.0014} & \ins{0.5961} & \ins{2.7539} \\
    \midrule
    \textbf{Total} &  &  &  & \rme{4}\ins{5} &  & \rme{5}\ins{6} \\
    \bottomrule
    \end{tabular}
    \end{threeparttable}
\end{table*}

\subsubsection{\add{FAST Results on Real Faults}}
\label{sec:compfast2}
Besides, as \emph{FAST} is evaluated by some subjects of Defects4J~\cite{just2014defects4j} in the previous work~\cite{miranda2018fast}, we apply the AGA approach to these subjects by reusing their artifact package (including subjects and code) for fair comparison. \ins{Moreover, we add the experiment on Mockito, which is also in Defects4J but does not appear in the experiment of \emph{FAST}.} \add{Defects4J is the largest real-fault benchmark, so this experiment complements the previous experiments on seeded faults and can evaluate AGA on real faults.} The comparison results are given by Table~\ref{defects4j}, where $\textrm{Win}_{\textrm{APFD}}$ and $\textrm{Win}_{\textrm{Time}}$ show whether the proposed AGA approach outperforms \emph{FAST} in terms of APFD and time cost, respectively. From this table, AGA is more effective than \emph{FAST} algorithms on \rme{4}\ins{5} out of \rme{5}\ins{6} projects and it achieves better time efficiency on all \rme{5}\ins{6} projects (with \rme{5.09X}\ins{5.24X} as average speedup ratio), which indicates the superiority of AGA. \add{Also, from this experiment, we show that AGA is superior on real faults, too.}

Actually, it is worth pointing out that as the authors of \emph{FAST}~\cite{miranda2018fast} stated, no single \emph{FAST} algorithm can be the best, which means the most effective algorithm in \emph{FAST} may lead to somewhat higher time cost and the most efficient algorithm in \emph{FAST} may lead to somewhat lower APFD value. That is, the results of \emph{FAST} in Table~\ref{opensource} \ins{(Appendix~\ref{appendc})} and Table~\ref{defects4j} are not results of one \emph{FAST} algorithm, but the best results of all \emph{FAST} algorithms. Moreover, even compared with these results, AGA is still promising considering both effectiveness and efficiency.

\add{Considering the advantageous of AGA over \emph{FAST}, it is interesting to analyze the secrets behind the observation. \emph{FAST} approach achieves the efficiency improvement by using the algorithms used in big data domain to summarize the key information in coverage, but suffers from the effectiveness loss to some extent because some information is missing in the summarization. AGA consists of two parts, time complexity reduction and iteration number reduction. In particular, the former part is to use some extra data structures (e.g., indices) to summarize the coverage information of each test case, i.e., the statements covered by each test. With these data structures, AGA does not need to scan the coverage table whenever a test case is selected, and thus the time cost of AGA reduces but its effectiveness maintains. To sum up, \emph{FAST} suffers from effectivness loss because it uses simplified information, while AGA does not because it uses the same information as before but in an easy-to-access way.}

\ins{Moreover, similar to Table~\ref{rq1_summary}, we compute the gaps between the highest and lowest APFD among all iteration numbers for Defects4J subjects, which are shown in Column ``APFD\_range'' of Table~\ref{defects4j}. The range of ``Chart'' is marked as ``NA'' because it has only one iteration. As we can see, the gaps are extremely small, which also confirms the conclusion in Section~\ref{sec:app2}.}

\ins{Additionally, Column ``GAF'' of Table~\ref{defects4j} shows the results of GA-first. AGA is much more efficient than GAF while achieves larger APFD, which is consistent with the conclusion in Section~\ref{sec:rq3}.
}

\begin{tcolorbox}
	\textbf{Conclusion:} Surprisingly, AGA can achieve \new{4.29X} speedup ratio compared to \emph{FAST}, which targets improving time efficiency while sacrificing effectiveness. At the same time, the experimental results show that AGA is significantly better than \emph{FAST} in terms of APFD values, and the average difference between them is 0.1702.
\end{tcolorbox}

\subsection{\ins{Comparison with other TCP Techniques}}
\label{sec:compother}

\ins{

Although only \emph{FAST} has a close goal to ours, to better evaluate AGA, we also compare it with more representative TCP techniques. In particular, in this study we use the following TCP techniques whose input is only coverage information and which have been widely used in the literature~\cite{miranda2018fast, luo2016large}.  \begin{itemize}
    \item \textbf{ART-D}~\cite{jiang2009adaptive} is a family of adaptive random-based TCP techniques guided by coverage information. At each iteration, a candidate set is dynamically created by randomly picking test cases from the set of not-yet-prioritized test cases as long as they can increase coverage. The test case in the candidate set that is the farthest away from the set of prioritized test cases is selected. 
    \item \textbf{GA-S (Additional Spanning)}~\cite{marre2003using} is a variant of GA that at each iteration picks the test case that covers the largest number of uncovered elements among those in the ``spanning set''. Here, an element subsumes another if covering the former guarantees covering the latter: The notion of a spanning set denotes the subset of non-subsumed elements. 
    \item \textbf{GE}~\cite{li2007search} is a genetic algorithm, which is a representative of search-based prioritization techniques and is evaluated to be effective. In each iteration, it uses a fitness function to select individuals and then applies crossover and mutation operators to generate new individuals. Specifically, an individual (a sequence) is encoded as an array where each value indicates the position of a test case; The fitness function is defined by Baker's linear ranking algorithm~\cite{baker1985adaptive}; The crossover operator selects two parents and each of the two offspring is formed by combining the first several values in one parent and the remaining values in the other parent; The mutation operator randomly selects two values in an individual and exchanges their positions. 
\end{itemize}

In this section, we reuse the implementation of ART-D, GA-S, and GE in \cite{miranda2018fast, chen2018optimizing} and compare them with AGA on the 55 open-source projects. Considering the randomness of these techniques, each of them is run 10 times. The remaining setting of this experiment is the same as Section~\ref{setup}. Due to the space limit of Table~\ref{opensource} \ins{(Appendix~\ref{appendc})}, we put the results in Table~\ref{opensource-othertcp}. In Table~\ref{opensource-othertcp}, each row represents one project, and the running time and APFD of AGA, ART-D, GA-S, and GE are shown separately.

The average speedup ratio of AGA over ART-D is 144.58X. Moreover, in all 55 projects, the APFD values of AGA are larger than ART-D, and the average APFD difference is 0.1384. That is, AGA always outperforms ART-D in terms of both effectiveness and efficiency. The average speedup ratio of AGA over GA-S is 182.27X. In 54 out of 55 projects, the APFD values of AGA is larger than GA-S, and the average APFD difference is 0.0708. The average speedup ratio of AGA over GE is 285.91X. In 50 out of 55 projects, the APFD values of AGA are larger than GE, and the average APFD difference is 0.0459. That is, compared with the three TCP techniques, our proposed AGA achieves both effectiveness and efficiency. Moreover, the time cost and APFD values of the compared TCP techniques distribute in a larger range than AGA, indicating that the latter can achieve stably promising performance.

}

\begin{tcolorbox}
	\textbf{Conclusion:} \ins{As AGA aims to largely improve the TCP efficiency while preserving the high-effectiveness of GA, it outperforms ART-D, GA-S, and GE in terms of both efficiency and effectiveness.}
\end{tcolorbox}

\begin{table*}[htbp]
	\centering
	\caption{\ins{Comparison with Other TCP Techniques on Open-Source Subjects}}
	\label{opensource-othertcp}
	\begin{threeparttable}
\begin{tabular}{c|rc|rc|rc|rc}
\toprule
\midrule
& \multicolumn{2}{c|}{\ins{\textbf{AGA}}} & \multicolumn{2}{c|}{\ins{\textbf{ART-D}}}  & \multicolumn{2}{c|}{\ins{\textbf{GA-S}}}  & \multicolumn{2}{c}{\ins{\textbf{GE}}}              \\
\multirow{-2}{*}{\ins{\textbf{Project}}} & \ins{\textbf{Time}} & \ins{\textbf{APFD}} & \ins{\textbf{Time}} & \ins{\textbf{APFD}} & \ins{\textbf{Time}} & \ins{\textbf{APFD}} & \ins{\textbf{Time}} & \ins{\textbf{APFD}}  \\
\midrule
 \ins{$\mathcal{S}_{1}$} & \ins{0.0157} & \ins{0.9070} & \ins{0.0832} & \ins{0.8440} & \ins{0.5878} & \ins{0.8812} & \ins{0.0650} & \ins{0.8747}  \\
 \ins{$\mathcal{S}_{2}$} & \ins{0.0058} & \ins{0.8380} & \ins{0.0224} & \ins{0.7508} & \ins{0.3312} & \ins{0.7870} & \ins{0.0660} & \ins{0.8373}  \\
 \ins{$\mathcal{S}_{3}$} & \ins{0.0222} & \ins{0.8848} & \ins{0.0196} & \ins{0.7681} & \ins{0.1593} & \ins{0.8108} & \ins{0.0900} & \ins{0.8705}  \\
 \ins{$\mathcal{S}_{4}$} & \ins{0.0789} & \ins{0.8509} & \ins{0.0478} & \ins{0.5933} & \ins{0.1079} & \ins{0.6555} & \ins{0.4440} & \ins{0.8061}  \\
 \ins{$\mathcal{S}_{5}$} & \ins{0.0089} & \ins{0.8527} & \ins{0.1361} & \ins{0.7405} & \ins{0.6358} & \ins{0.7662} & \ins{0.9430} & \ins{0.8310}  \\
 \ins{$\mathcal{S}_{6}$} & \ins{0.0046} & \ins{0.8101} & \ins{0.0957} & \ins{0.6909} & \ins{0.2560} & \ins{0.6800} & \ins{0.9140} & \ins{0.7935}  \\
 \ins{$\mathcal{S}_{7}$} & \ins{0.0076} & \ins{0.9059} & \ins{0.0150} & \ins{0.7711} & \ins{0.1116} & \ins{0.8255} & \ins{0.0970} & \ins{0.8855}  \\
 \ins{$\mathcal{S}_{8}$} & \ins{0.0180} & \ins{0.8898} & \ins{0.0752} & \ins{0.7459} & \ins{2.7672} & \ins{0.8153} & \ins{0.1160} & \ins{0.8985}  \\
 \ins{$\mathcal{S}_{9}$} & \ins{0.3363} & \ins{0.9144} & \ins{53.9033} & \ins{0.8821} & \ins{33.7447} & \ins{0.8995} & \ins{38.4580} & \ins{0.8949}  \\
 \ins{$\mathcal{S}_{10}$} & \ins{0.0247} & \ins{0.9518} & \ins{0.5895} & \ins{0.8009} & \ins{0.9681} & \ins{0.9017} & \ins{0.8060} & \ins{0.9172}  \\
 \ins{$\mathcal{S}_{11}$} & \ins{0.0553} & \ins{0.8766} & \ins{0.4818} & \ins{0.7640} & \ins{10.6250} & \ins{0.7950} & \ins{0.4070} & \ins{0.8501}  \\
 \ins{$\mathcal{S}_{12}$} & \ins{0.0106} & \ins{0.8864} & \ins{0.0708} & \ins{0.7589} & \ins{0.3043} & \ins{0.8375} & \ins{0.2970} & \ins{0.8667}  \\
 \ins{$\mathcal{S}_{13}$} & \ins{0.0385} & \ins{0.8615} & \ins{0.1763} & \ins{0.7816} & \ins{2.0406} & \ins{0.7667} & \ins{0.2000} & \ins{0.7783}  \\
 \ins{$\mathcal{S}_{14}$} & \ins{0.0385} & \ins{0.9188} & \ins{0.6426} & \ins{0.7800} & \ins{0.8556} & \ins{0.8779} & \ins{0.6880} & \ins{0.9011}  \\
 \ins{$\mathcal{S}_{15}$} & \ins{0.0055} & \ins{0.8582} & \ins{0.1225} & \ins{0.7285} & \ins{0.2513} & \ins{0.7117} & \ins{0.7360} & \ins{0.8470}  \\
 \ins{$\mathcal{S}_{16}$} & \ins{0.0152} & \ins{0.8031} & \ins{0.3687} & \ins{0.6578} & \ins{0.9745} & \ins{0.6509} & \ins{3.0660} & \ins{0.7760}  \\
 \ins{$\mathcal{S}_{17}$} & \ins{0.1317} & \ins{0.9183} & \ins{7.8831} & \ins{0.8316} & \ins{7.8195} & \ins{0.8893} & \ins{2.7440} & \ins{0.8785}  \\
 \ins{$\mathcal{S}_{18}$} & \ins{0.0226} & \ins{0.9028} & \ins{0.1431} & \ins{0.7783} & \ins{1.5687} & \ins{0.8039} & \ins{0.3120} & \ins{0.8950}  \\
 \ins{$\mathcal{S}_{19}$} & \ins{0.6995} & \ins{0.9033} & \ins{34.4428} & \ins{0.7553} & \ins{374.3919} & \ins{0.8381} & \ins{4.1710} & \ins{0.8528}  \\
 \ins{$\mathcal{S}_{20}$} & \ins{0.0469} & \ins{0.8013} & \ins{0.2180} & \ins{0.7021} & \ins{36.2190} & \ins{0.7293} & \ins{0.3410} & \ins{0.8185}  \\
 \ins{$\mathcal{S}_{21}$} & \ins{0.0248} & \ins{0.8328} & \ins{0.4520} & \ins{0.7156} & \ins{1.5459} & \ins{0.7676} & \ins{1.5660} & \ins{0.7899}  \\
 \ins{$\mathcal{S}_{22}$} & \ins{0.0215} & \ins{0.8642} & \ins{0.2999} & \ins{0.7660} & \ins{0.8994} & \ins{0.8151} & \ins{0.5080} & \ins{0.8570}  \\
 \ins{$\mathcal{S}_{23}$} & \ins{0.0962} & \ins{0.8198} & \ins{4.8919} & \ins{0.6979} & \ins{4.2931} & \ins{0.7782} & \ins{3.6190} & \ins{0.7866}  \\
 \ins{$\mathcal{S}_{24}$} & \ins{0.0187} & \ins{0.9858} & \ins{0.0420} & \ins{0.9213} & \ins{12.3507} & \ins{0.9857} & \ins{0.1230} & \ins{0.9860}  \\
 \ins{$\mathcal{S}_{25}$} & \ins{1.0579} & \ins{0.8401} & \ins{70.1140} & \ins{0.8099} & \ins{387.4319} & \ins{0.8426} & \ins{3.5190} & \ins{0.8283}  \\
 \ins{$\mathcal{S}_{26}$} & \ins{0.0294} & \ins{0.8339} & \ins{0.4066} & \ins{0.6026} & \ins{0.3047} & \ins{0.6927} & \ins{2.6380} & \ins{0.7833}  \\
 \ins{$\mathcal{S}_{27}$} & \ins{0.0303} & \ins{0.9614} & \ins{0.0254} & \ins{0.7695} & \ins{0.2441} & \ins{0.8579} & \ins{0.2200} & \ins{0.9501}  \\
 \ins{$\mathcal{S}_{28}$} & \ins{0.1132} & \ins{0.9164} & \ins{11.6980} & \ins{0.7642} & \ins{4.6810} & \ins{0.8734} & \ins{8.5760} & \ins{0.8534}  \\
 \ins{$\mathcal{S}_{29}$} & \ins{2.3900} & \ins{0.9490} & \ins{130.3359} & \ins{0.8254} & \ins{7955.7540} & \ins{0.9180} & \ins{3.7060} & \ins{0.9131}  \\
 \ins{$\mathcal{S}_{30}$} & \ins{0.1804} & \ins{0.9617} & \ins{6.9994} & \ins{0.8572} & \ins{28.4835} & \ins{0.9202} & \ins{1.6830} & \ins{0.9285}  \\
 \ins{$\mathcal{S}_{31}$} & \ins{0.3170} & \ins{0.9426} & \ins{46.5734} & \ins{0.8342} & \ins{15.9675} & \ins{0.9191} & \ins{11.7620} & \ins{0.9065}  \\
 \ins{$\mathcal{S}_{32}$} & \ins{0.1270} & \ins{0.8911} & \ins{22.8351} & \ins{0.7165} & \ins{3.1369} & \ins{0.8231} & \ins{53.4990} & \ins{0.7696}  \\
 \ins{$\mathcal{S}_{33}$} & \ins{0.0921} & \ins{0.8662} & \ins{11.0712} & \ins{0.6612} & \ins{5.5717} & \ins{0.7356} & \ins{23.6020} & \ins{0.7568}  \\
 \ins{$\mathcal{S}_{34}$} & \ins{0.7903} & \ins{0.9328} & \ins{120.8907} & \ins{0.8271} & \ins{128.1646} & \ins{0.8861} & \ins{21.5730} & \ins{0.8780}  \\
 \ins{$\mathcal{S}_{35}$} & \ins{0.3631} & \ins{0.9467} & \ins{32.4553} & \ins{0.7277} & \ins{20.0292} & \ins{0.8694} & \ins{5.1690} & \ins{0.9105}  \\
 \ins{$\mathcal{S}_{36}$} & \ins{23.7849} & \ins{0.9371} & \ins{2,397.2400} & \ins{0.8207} & \ins{1,976.4397} & \ins{0.8597} & \ins{11.6880} & \ins{0.8570} \\
 \ins{$\mathcal{S}_{37}$} & \ins{0.3582} & \ins{0.8507} & \ins{33.6262} & \ins{0.6876} & \ins{23.6568} & \ins{0.7621} & \ins{1,979.6000} & \ins{0.7165}  \\
 \ins{$\mathcal{S}_{38}$} & \ins{0.2794} & \ins{0.8657} & \ins{115.8419} & \ins{0.7753} & \ins{35.4853} & \ins{0.7747} & \ins{337.0780} & \ins{0.8072}  \\
 \ins{$\mathcal{S}_{39}$} & \ins{2.2078} & \ins{0.9545} & \ins{1,114.6899} & \ins{0.7931} & \ins{265.5048} & \ins{0.9289} & \ins{168.0300} & \ins{0.7733}  \\
 \ins{$\mathcal{S}_{40}$} & \ins{0.5942} & \ins{0.9244} & \ins{118.3193} & \ins{0.7621} & \ins{17.4111} & \ins{0.8672} & \ins{112.3900} & \ins{0.8156}  \\
 \ins{$\mathcal{S}_{41}$} & \ins{0.1562} & \ins{0.9106} & \ins{19.2647} & \ins{0.7195} & \ins{4.2101} & \ins{0.8454} & \ins{40.8410} & \ins{0.8116}  \\
 \ins{$\mathcal{S}_{42}$} & \ins{0.0287} & \ins{0.8569} & \ins{0.0123} & \ins{0.7409} & \ins{0.2401} & \ins{0.8176} & \ins{0.1300} & \ins{0.8649}  \\
 \ins{$\mathcal{S}_{43}$} & \ins{0.2558} & \ins{0.8924} & \ins{32.1075} & \ins{0.7331} & \ins{38.5073} & \ins{0.7915} & \ins{24.0450} & \ins{0.8208}  \\
 \ins{$\mathcal{S}_{44}$} & \ins{0.8465} & \ins{0.9240} & \ins{162.8027} & \ins{0.8437} & \ins{147.3191} & \ins{0.9090} & \ins{34.0820} & \ins{0.8864}  \\
 \ins{$\mathcal{S}_{45}$} & \ins{0.0597} & \ins{0.8464} & \ins{0.8793} & \ins{0.6822} & \ins{7.4841} & \ins{0.7811} & \ins{4.6150} & \ins{0.7918}  \\
 \ins{$\mathcal{S}_{46}$} & \ins{0.0858} & \ins{0.8656} & \ins{3.7066} & \ins{0.6834} & \ins{17.5147} & \ins{0.7494} & \ins{4.0400} & \ins{0.8003}  \\
 \ins{$\mathcal{S}_{47}$} & \ins{0.0629} & \ins{0.8750} & \ins{0.7825} & \ins{0.6884} & \ins{1.0181} & \ins{0.7588} & \ins{3.6900} & \ins{0.8225}  \\
 \ins{$\mathcal{S}_{48}$} & \ins{0.0025} & \ins{0.7939} & \ins{0.0095} & \ins{0.6455} & \ins{0.0386} & \ins{0.7089} & \ins{0.0790} & \ins{0.7873}  \\
 \ins{$\mathcal{S}_{49}$} & \ins{0.1120} & \ins{0.8009} & \ins{0.8459} & \ins{0.6463} & \ins{5.6348} & \ins{0.7600} & \ins{4.3070} & \ins{0.7884}  \\
 \ins{$\mathcal{S}_{50}$} & \ins{0.0047} & \ins{0.8517} & \ins{0.0164} & \ins{0.5108} & \ins{0.0585} & \ins{0.6855} & \ins{0.1830} & \ins{0.8559}  \\
 \ins{$\mathcal{S}_{51}$} & \ins{1.1036} & \ins{0.8671} & \ins{203.2114} & \ins{0.6595} & \ins{505.1426} & \ins{0.8080} & \ins{61.9450} & \ins{0.7261}  \\
 \ins{$\mathcal{S}_{52}$} & \ins{0.4052} & \ins{0.9542} & \ins{26.1336} & \ins{0.8454} & \ins{200.9616} & \ins{0.8889} & \ins{20.1360} & \ins{0.9233}  \\
 \ins{$\mathcal{S}_{53}$} & \ins{0.2298} & \ins{0.8710} & \ins{13.8199} & \ins{0.6956} & \ins{32.9077} & \ins{0.8101} & \ins{33.7620} & \ins{0.7768}  \\
 \ins{$\mathcal{S}_{54}$} & \ins{2.6648} & \ins{0.9089} & \ins{3,373.5142} & \ins{0.7578} & \ins{252.3093} & \ins{0.8478} & \ins{13,384.3880} & \ins{0.7902}  \\
 \ins{$\mathcal{S}_{55}$} & \ins{18.1040} & \ins{0.9544} & \ins{55,120.6523} & \ins{0.8613} & \ins{4536.0047} & \ins{0.9292} & \ins{13,516.8050} & \ins{0.8747}  \\

\midrule
\bottomrule
\end{tabular}
\end{threeparttable}
\end{table*}

\section{Industrial Case Study}
\label{sec:case}

To show the practical usage of our approach, we then conducted an industrial case study as follows.

\baidu~is a famous Internet service provider with over 600M monthly active users. In their regression testing infrastructure, test case prioritization is frequently needed and they have been adopting Greedy Additional (GA) strategy for a long time because of its simple idea and relatively high effectiveness. However, they often complain about the long running time of GA, which deviates from the original intention of test case prioritization, that is to accelerate the process of detecting faults.

To check the performance of our AGA approach in real-world scenarios, we collected 22 versions of five industrial projects from \baidu, each of which is taken as a subject in this study. More specifically, these subjects are collected from Dec.~2017 to Feb.~2018 and Oct.~2018 to Nov.~2018, and all of them are written in \emph{C}. As shown in the first three columns in Table~\ref{tab:case} \ins{(Appendix~\ref{appendd})}, we summarize the SLOC and number of test cases of each subject. The SLOCs range from 20K to 500K while the numbers of test cases range from 202 to 4,246. Besides, we used C-Cover~\cite{ccover} to collect statement coverage for each industrial subject.

\begin{figure}[htbp]
    \centering
	\includegraphics[width=0.23\textwidth]{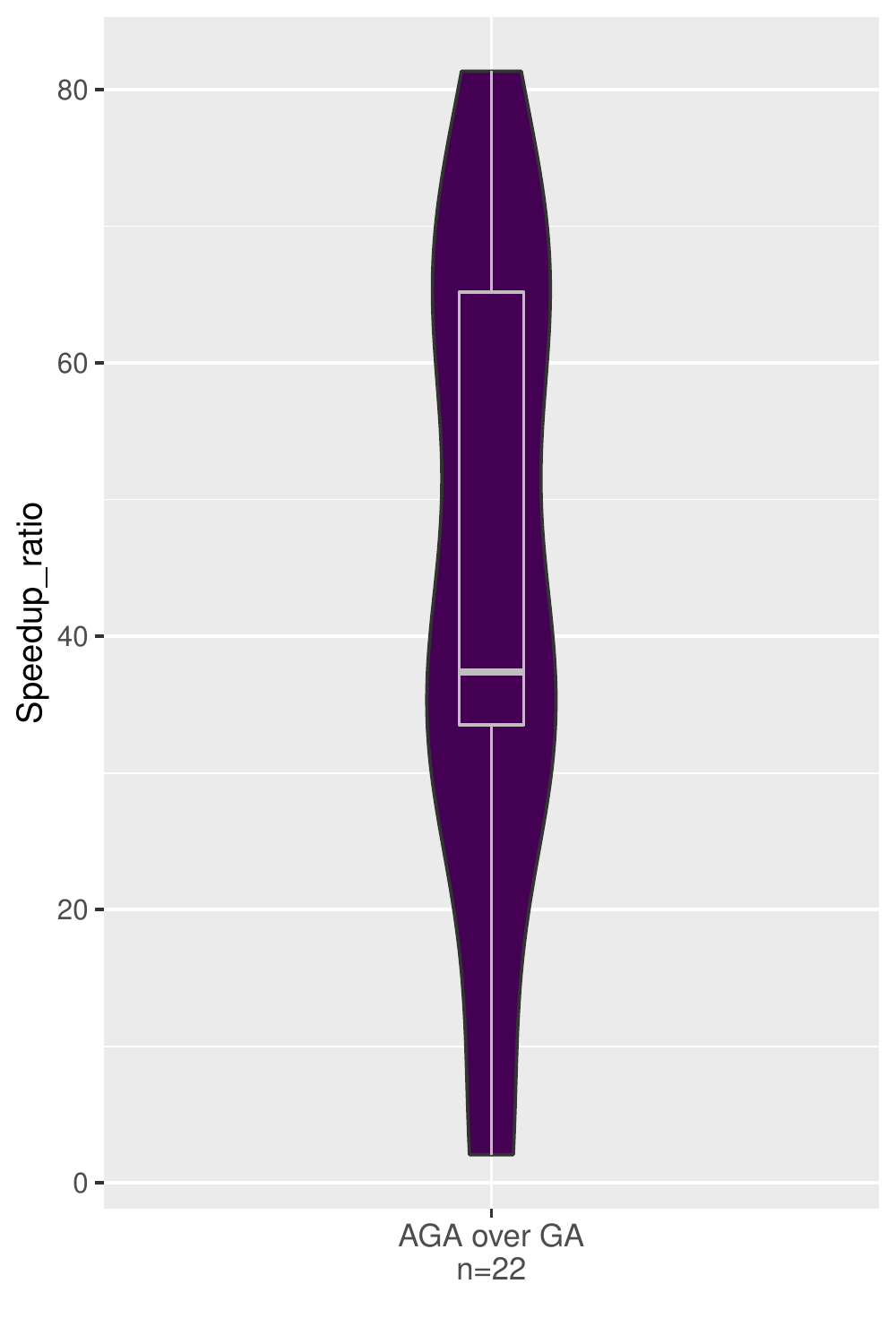}
	\includegraphics[width=0.23\textwidth]{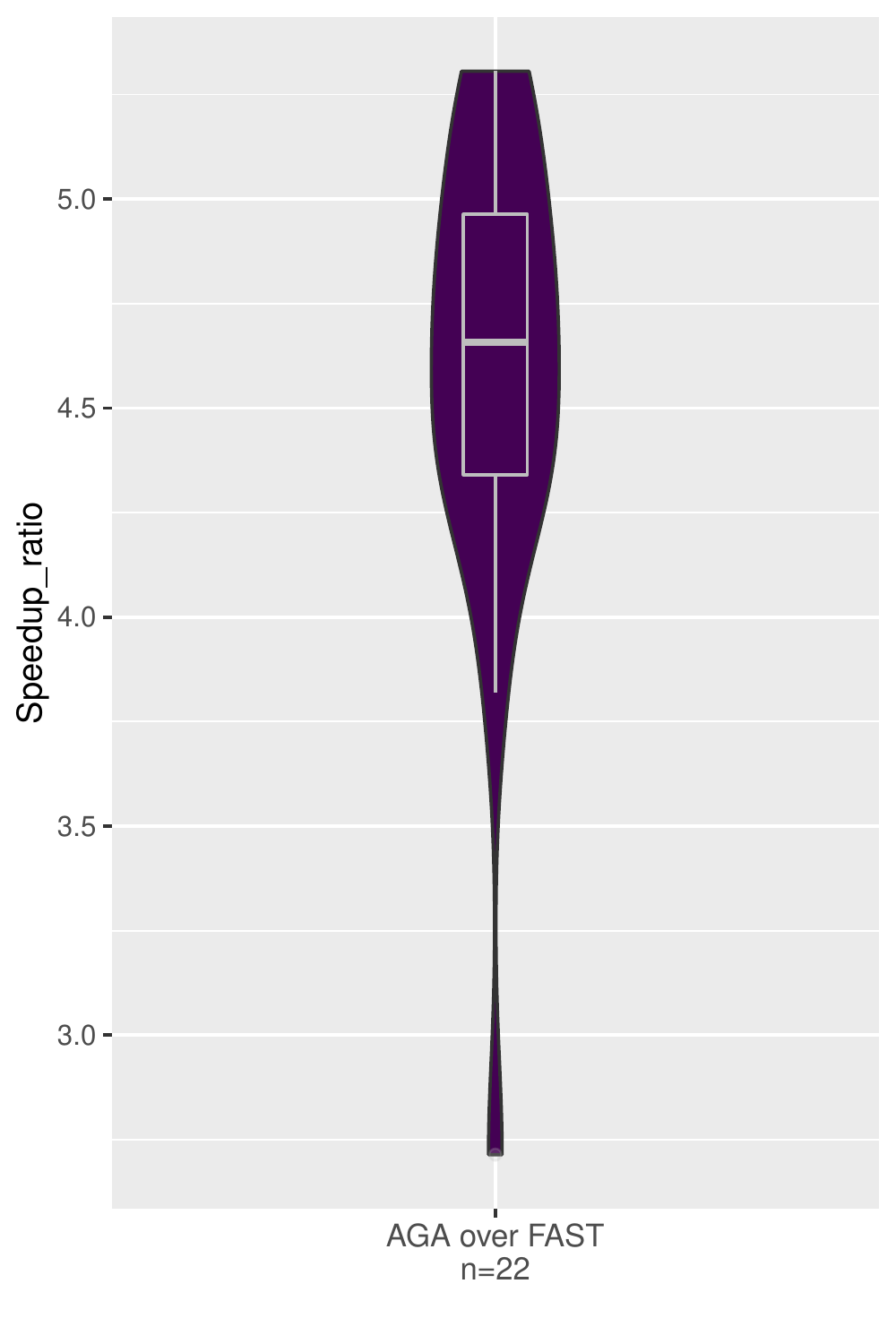}
	\caption{\add{Speedup Ratios Distribution of AGA over GA and \emph{FAST} on Industrial Projects}}
	\label{violin4}
\end{figure}

In Table~\ref{tab:case} \ins{(Appendix~\ref{appendd})}, we report the time cost of GA, AGA, and \emph{FAST}, respectively. When the time cost of AGA is less than GA, we mark it with $\checkmark$. As we can see, in all 22 subjects, the time cost of AGA is much lower than that of GA, and the speedup ratio is \new{44.27X} on average. In general, our AGA approach is demostrated to be efficient on industrial subjects from \baidu. For example, for the subject $\mathcal{I}_1$, its original prioritization time is larger than \new{29,000} seconds, which may be unbearable in practice. However, through AGA, the prioritization time is reduced to less than \new{360} seconds. On the other hand, the surprisingly high efficiency of AGA also indicates the ubiquitous existence of many redundant accesses of data in industrial projects. \add{We also present the violin plot with included box plot in Figure~\ref{violin4} to show the distribution and variation. As we can see, on most projects, AGA has a large improvement compared to GA.}

\new{Provided adjacency matrix as input, the average speedup ratio of AGA over GA is 61.43X.}

\new{After we report the results, developers in \baidu~verified (1) the time cost of our implementation of GA is close to their inner implementations, and (2) the speedup ratio is significant and our technique improves their prioritization efficiency, because their implementation only works on small projects, not large projects.}

In addition, we also compared our approach with \emph{FAST}, and the experimental setup is the same with Section~\ref{sec:comp}. Surprisingly, AGA outperforms \emph{FAST} again. Specifically, on all 22 subjects, AGA is faster than \emph{FAST} and the average speedup ratio is \new{4.58X}. \ins{To statistically check their differences, we follow the similar procedure as above. We first use Shapiro-Wilk test to check the normality \inss{of residuals}, and the \textit{p-value} in AGA and \emph{FAST} is $5 * 10^{-4}$ and $2.5 * 10^{-3}$, which reject the hypothesis that they are normally distributed. Similary to the above, proportional odds regression~\cite{mccullagh1980regression} is used and we introduce a variable ``group'' representing AGA and \emph{FAST} and take project size as a control variable. The results show that the \textit{p-value} of ``group'' is $1.35 * 10^{-6}$, indicating significant difference between AGA and \emph{FAST}, and the effect size (Cohen's $d$) is 1.28 (very large effect).} \rme{The Kruskal-Wallis rank sum test shows that the \textit{p-value} is 0.0009, which indicates statistically significant difference. To measure the magnitude of their time difference, we compute the effect size $\eta = 0.237$, indicating large effect.} \add{We also present the violin plot with included box plot in Figure~\ref{violin4}. Again, we find that on most projects, AGA has a large improvement than \emph{FAST}.} In~\cite{miranda2018fast}, the authors proposed \emph{FAST} to solve the scalability problem of TCP techniques with the decrease of effectiveness. Their approach is evaluated to be efficient when the project size grows up rapidly. However, our AGA approach is even more efficient than \emph{FAST}, and this means AGA may scale up better and is practical in real-world scenarios. \rme{Also, recall that when we compare AGA with \emph{FAST} on open-source projects, there is no significant difference between them, and we thus conjecture that this is due to the relatively small sizes of open-source projects.}\ins{Also, recall that when we compare AGA with \emph{FAST} on open-source projects, the \textit{p-value} is larger and the effect size is smaller than here, and we conjecture that this is due to the relatively small sizes of open-source projects.}

\ins{
Additionally, besides \emph{FAST}, which targets the TCP efficiency problem, we also compare AGA with other more general TCP techniques as we have done in Section~\ref{sec:comp}. Specifically, we run ART-D, GA-S, and GE on the 22 subjects. Considering the randomness of these techniques, each of them is run 10 times. The results are shown in Table~\ref{tab:case} (Appendix~\ref{appendd}). As we can see, these techniques are much slower than AGA, even GA. The average speedup ratio of AGA over ART-D, GA-S, and GE is 993.37X, 4230.53X, and 123.25X, respectively.
}

It is worth noting that \add{on one hand, \baidu~is sensitive to the positions of detected faults in history, thus these positions are not available to us. On the other hand,} they only provide coverage data after desensitization and we do not have access to the source code of these subjects (to create mutants) due to the confidential policy. As a result, we cannot compare the effectiveness of the three approaches in terms of APFD in industrial subjects.

\begin{tcolorbox}
	\textbf{Conclusion: } Our AGA approach achieves \new{44.27X} speedup ratio compared to GA. AGA even outperforms \emph{FAST} in terms of time efficiency (\new{4.58X}), and the difference is statistically significant. This indicates that AGA is practical in real-world scenarios. 
\end{tcolorbox}

\section{Discussion}
\label{sec:discussion}

\new{
\noindent{\bf Space comparison.}
From the space complexity analysis in Section~\ref{sec:app1}, AGA consumes at most twice more space than GA, which is acceptable in practice. Moreover, AGA does not require high performance servers, e.g., the time cost of AGA on the two largest open-source projects (i.e., commons-math \& camel-core) is only 153.42s and 187.82s (on a personal computer whose Intel Core-i5 with 8GB memory), almost the same as Table~\ref{opensource} \ins{(Appendix~\ref{appendc})}).
}

\add{
\noindent{\bf Impact of seeded faults and real faults.}
Previous work~\cite{just2014mutants,papadakis2018mutation,daran1996software} has explored the relationship between seeded faults and real faults and they may have different characteristics, which has potential influence on the evaluation on test case prioritization, fault localization, etc. In this paper, we evaluate our approach on both 55 open-source subjects with seeded faults and Defects4J dataset with real faults. The high performance of AGA on both of them can illustrate its superiority well.
}

\add{
\noindent{\bf Discussion on other TCP approaches.}
Researchers have put dedicated efforts in TCP and have proposed a large number of TCP techniques since then. Many approaches take other information rather than coverage information (e.g., test inputs, test outputs, mutants) as input, so they are in different dimensions. However, even taken all kinds of approaches into consideration, the GA approach remains one of the most effective strategies in terms of fault-detection rate~\cite{zhang2013bridging,li2007search,jiang2009adaptive,miranda2018fast}. So, we target GA in this paper and AGA can be better than other approaches.
}

\section{Related Work}
\label{related}

Test case prioritization attracts much attention since this problem was raised at the end of the 20th century, and the work on test case prioritization can be classified into prioritization algorithms~\cite{li2007search,jiang2009adaptive,fraser2007test,yoo2009clustering,saha2015information,ma2008test}, coverage criteria used in prioritization~\cite{rothermel1999test,elbaum2000prioritizing,elbaum2002test,do2004empirical,jones2003test,mei2012static,zhang2009prioritizing,korel2005test,mei2009test}, measurement used to estimate prioritization effectiveness~\cite{rothermel1999test,kapfhammer2007using,elbaum2001incorporating}, and empirical studies~\cite{shin2019empirical,elbaum2002test,rothermel2001prioritizing,rothermel1999test,do2008empirical,luo2016large,henard2016comparing,hao2016test,epitropakis2015empirical,hao2015optimal}. Moreover, a number of surveys on test case prioritization are also given in the literature~\cite{yoo2012regression,catal2013test,mohanty2011survey}. For example, Catal et al.~\cite{catal2013test} conducted a systematic study of TCP techniques in 2001-2011 including 120 papers published in that time period. Due to the space limit, we do not list all the prioritization work here, but introduce some very recently published work. Di et al.~\cite{di2018test} proposed Hypervolume-based Genetic Algorithm to prioritize test cases using multiple test coverage criteria. Azizi et al.~\cite{azizi2018graphite} proposed a graph-based framework to map the prioritization problem to a graph traversal algorithm. Chen et al.~\cite{chen2018test} gave an adaptive random sequences approach based on clustering techniques using black-box information. Different from them, our work targets the effective GA algorithm and attempts to solve its efficiency problem.

Moreover, some researchers noticed the efficiency problem of TCP and began to work on it. Henard et al.~\cite{henard2016comparing} said, ``if prioritization takes too long, then it eats into the time available to run the prioritized test suite.'' That is, for large software, it is necessary to take the scalability of TCP into consideration. Marijan et al.~\cite{marijan2013test} proposed ROCKET to prioritize test cases based on historical failure data, test execution time and domain-specific heuristics to improve the efficiency in the scenario of continuous integration. Knauss et al.~\cite{knauss2015supporting} proposed to analyze the correlation between test failures and source code changes to rapidly prioritize test cases. Elbaum et al.~\cite{elbaum2014techniques} introduced two techniques that use readily available test execution history data to determine what test cases are worth executing and execute them with higher priority. Recently, Miranda et al.~\cite{miranda2018fast} introduced the \emph{FAST} techniques to provide similarity-based test case prioritization techniques with scalable improvements. Our work is related to the above work because all of them target TCP efficiency problem. However, the above work either does not take advantage of the coverage information which results in lower effectiveness or addresses the efficiency problem alone without balancing or even sacrificing the effectiveness. That is, to our best knowledge, none of the existing work can improve the efficiency of GA while maintaining its widely-recognized effectiveness. Our work achieves this goal and AGA is particularly advantageous for large-scale industrial projects.

\section{Conclusions}
\label{conclusions}

In this paper, we make a deep analysis of the Greedy Additional algorithm (GA) for test case prioritization (TCP) problem and propose AGA to improve its efficiency while preserving effectiveness. On one hand, we find the redundant data accesses in GA and take the use of extra data structures to cut down them, which leads to an optimized time complexity from $\mathcal{O}(m^2n)$ to $\mathcal{O}(kmn)$ \add{given $n > m$}, where $m$ is the number of test cases, $n$ is the number of program elements, and $k$ is the iteration number. On the other hand, we notice the impacts of iteration numbers on the effectiveness and efficiency of GA and propose to reduce it to a relatively small value to improve efficiency while preserving effectiveness. Overall, we achieve an $\mathcal{O}(mn)$ algorithm for prioritization.

We performed comprehensive experiments on 55 open-source projects to show the effectiveness and efficiency of AGA. On one hand, AGA can achieve the same average effectiveness as the GA approach, whose performance is considered to be high, and at the same time, the efficiency of AGA is much higher than GA. Specifically, our AGA approach can achieve \new{5.95X/}27.72X speedup ratio over GA on average \new{on two input formats}. On the other hand, compared with \emph{FAST}, which was recently proposed to solve the TCP efficiency problem while sacrificing effectiveness to some extent, AGA achieves 0.1702 higher APFD values on average and surprisingly the average speedup ratio of AGA over \emph{FAST} is \new{4.29X}.

Additionally, we conducted an industrial case study on 22 industrial subjects, collected from \baidu, which is a famous Internet service provider with over 600M monthly active users. The experimental results show that the average speedup ratios of AGA over GA and \emph{FAST} are \new{44.27X/}61.43X and \new{4.58X} (with significant difference and \rme{large}\ins{very large} effect), respectively.

To the best of our knowledge, this is the first attempt to alleviating the efficiency problem of the Greedy Additional TCP approach while maintaining its effectiveness. It is worth noting that the efficiency of TCP algorithm is especially important when software becomes larger, that is to say, in real-world scenarios. Our empirical evidence indicates that AGA is particularly more advantageous for large-scale industrial projects.

\section*{Acknowledgments}

The authors would like to thank all the reviewers for their valuable comments and suggestions. This work was supported by the National Natural Science Foundation of China under Grant No. 61872008.

\bibliographystyle{unsrt}
\bibliography{sample-bibliography}

\begin{IEEEbiography}[{\includegraphics[width=1in,height=1.25in,clip,keepaspectratio]{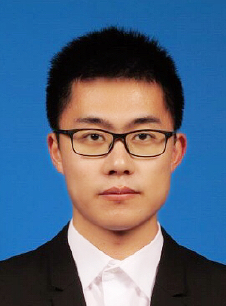}}]{Feng Li}
received his B.S. degree from Peking University in 2018. He is currently a Ph.D. candidate in School of Computer Science at Peking University. His research interests include software testing and analysis.
\end{IEEEbiography}	

\begin{IEEEbiography}[{\includegraphics[width=1in,height=1.25in,clip,keepaspectratio]{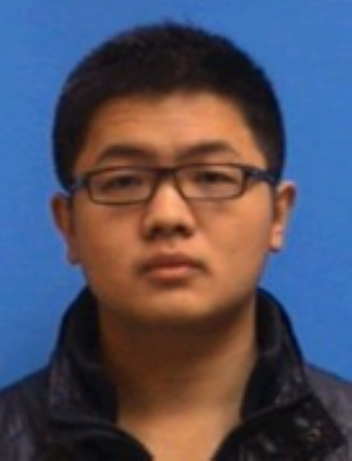}}]{Jianyi Zhou}
received his B.S. degree in 2014, and M.S. degree in 2017, both from Beihang University. He is currently a Ph.D. candidate in School of Computer Science at Peking University. His research interests include software testing and analysis.
\end{IEEEbiography}	

\begin{IEEEbiography}[{\includegraphics[width=1in,height=1.25in,clip,keepaspectratio]{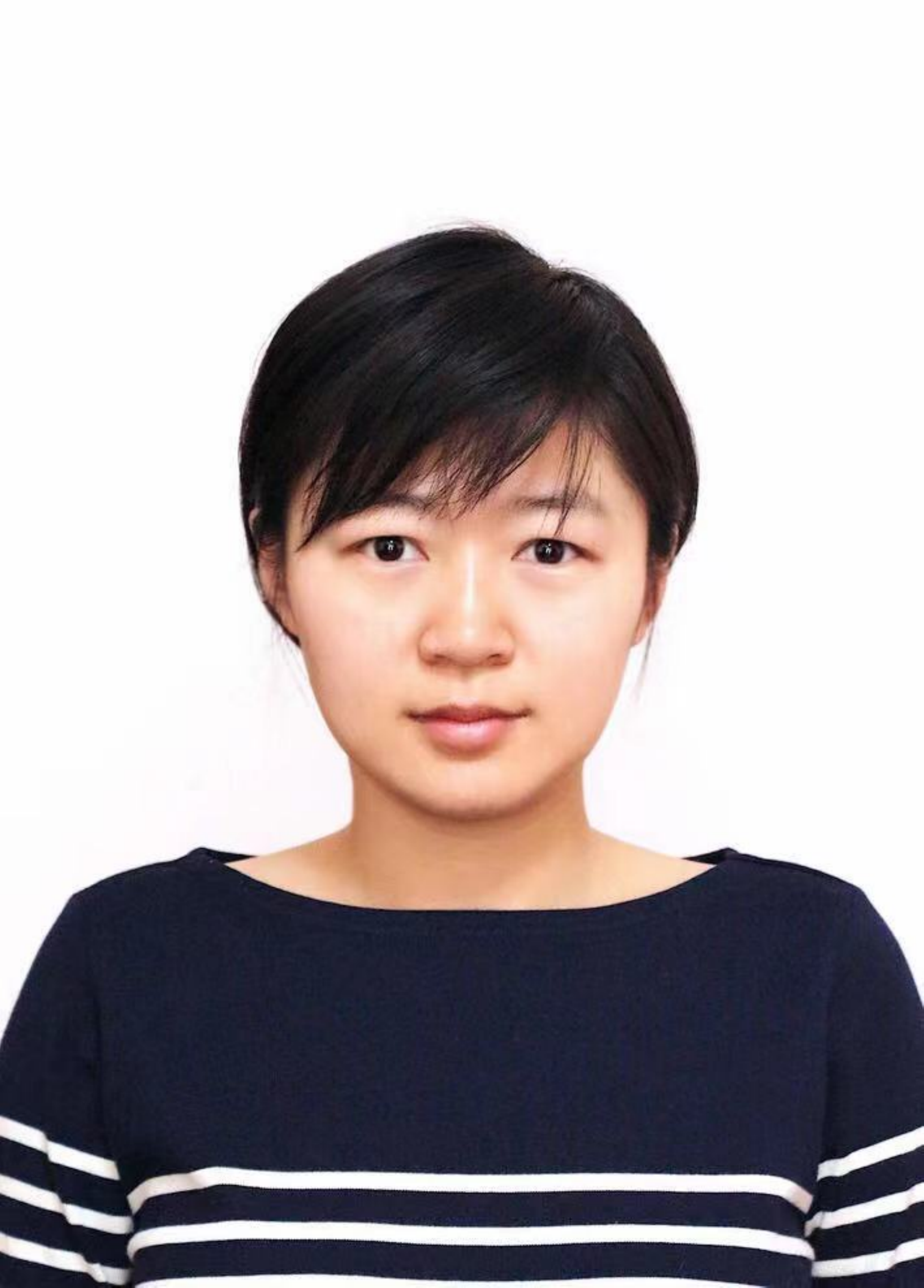}}]{Yinzhu Li}
received her M.S. degree in Computer Science and Technology in 2012 from Tianjin Normal University. She is now an employee at Baidu Online Network Technology (Beijing) Co., Ltd., mainly working on automation testing. Her research interest is intelligent testing, including test case selection, test case generation, and fault localization.
\end{IEEEbiography}	

\begin{IEEEbiography}[{\includegraphics[width=1in,height=1.25in,clip,keepaspectratio]{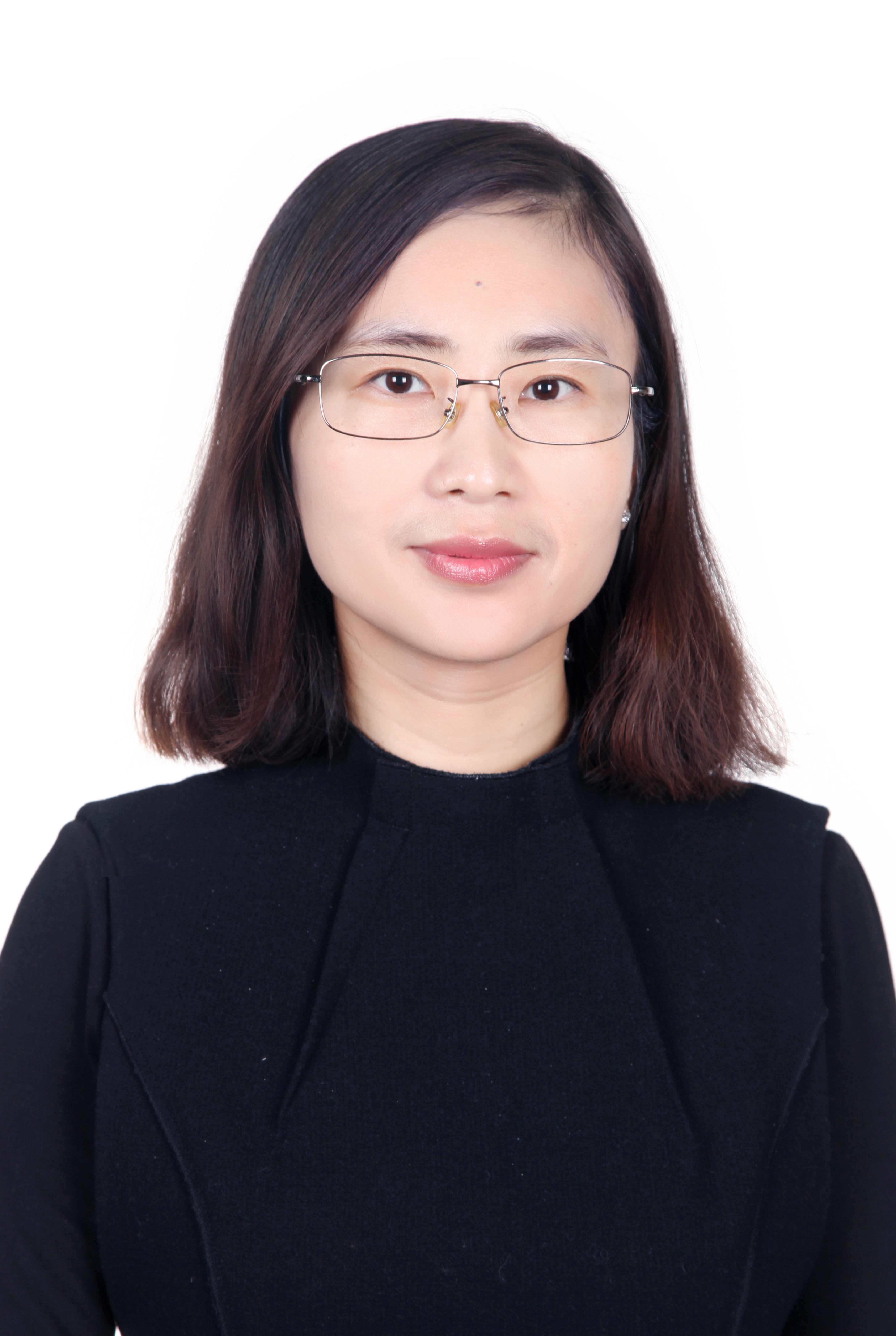}}]{Dan Hao}
is an associate professor at School of Computer Science, Peking University, P.R.China. She received her Ph.D. in Computer Science from Peking University in 2008, and the B.S. in Computer Science from the Harbin Institute of Technology in 2002.  She was a program co-chair of ASE 2021 and SANER 2022, a general co-chair of SPLC 2018, the program committees of many prestigious conferences (e.g., ICSE, FSE, ASE, and ISSTA). Her current research interests include software testing and debugging.
\end{IEEEbiography}	

\begin{IEEEbiography}[{\includegraphics[width=1in,height=1.25in,clip,keepaspectratio]{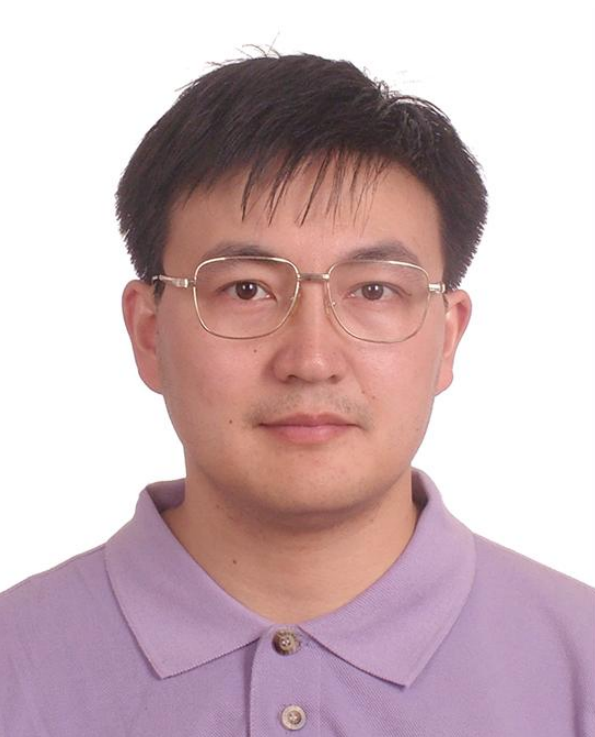}}]{Lu Zhang}
is a professor at School of Computer Science, Peking University, P.R. China. He received both Ph.D. and BSc in Computer Science from Peking University in 2000 and 1995 respectively. He was a postdoctoral researcher in Oxford Brookes University and University of Liverpool, UK. He served on the program committees of many prestigious conferences, such as FSE, OOPSLA, ISSTA, and ASE. He was a program co-chair of SCAM 2008 and a program co-chair of ICSME 2017. He has been on the editorial boards of Journal of Software Maintenance and Evolution: Research and Practice and Software Testing, Verification and Reliability. His current research interests include software testing and analysis, program comprehension, software maintenance and evolution, software reuse and component-based software development, and service computing.
\end{IEEEbiography}

\clearpage

\appendices

\section{charts of iteration number and time cost}
\label{appendb}

\add{To better analyze the relationship between iteration number and time cost, we put detailed results in Section~\ref{sec:rq2} here. We draw a line chart of iteration number and time cost for each project. Note that in order to see the trend, we only present the projects whose iteration number is no less than \rme{$5$}\ins{$20$} (\rme{$k \geq 5$}\ins{$k \geq 20$}). \ins{As we can see, all projects follow a similar trend. In some projects, the first several iterations cost more time than other iterations. It is reasonable because along with the decrease of the number of remaining test cases ($n$), prioritization also becomes faster.} The plots also support our claim that the iteration number contributes much to the time cost. As $k$ is the coefficient of time complexity, it largely determines the actual efficiency in practice, so, we think there is a large space to reduce time complexity.}

\begin{figure}[!h]
        \includegraphics[width=0.23\textwidth]{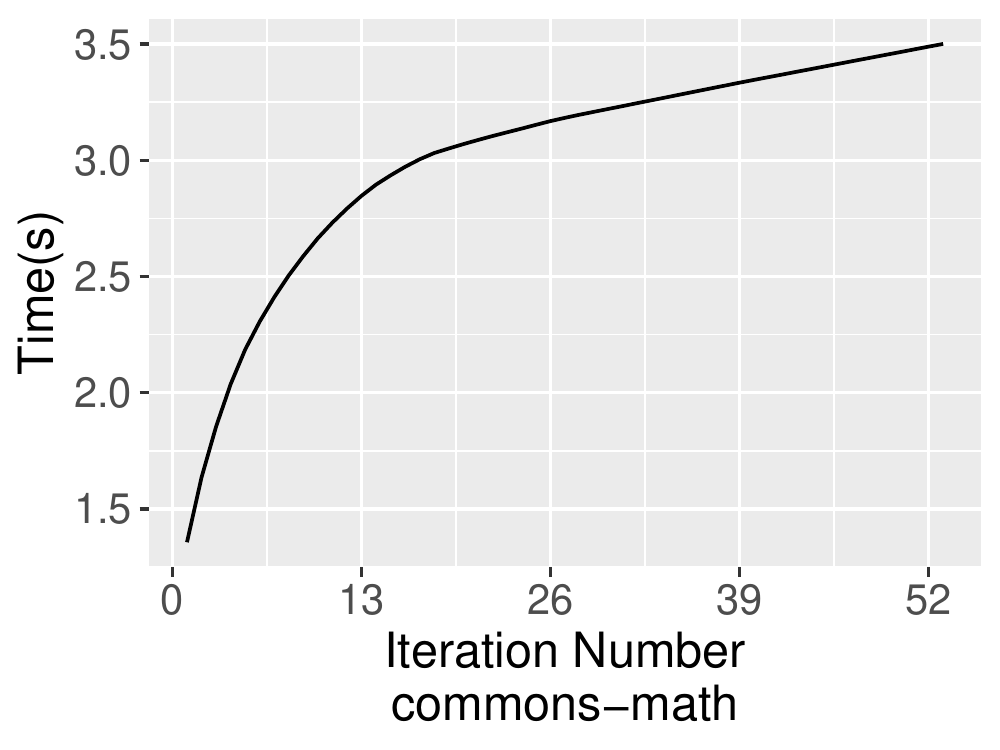}
        \includegraphics[width=0.23\textwidth]{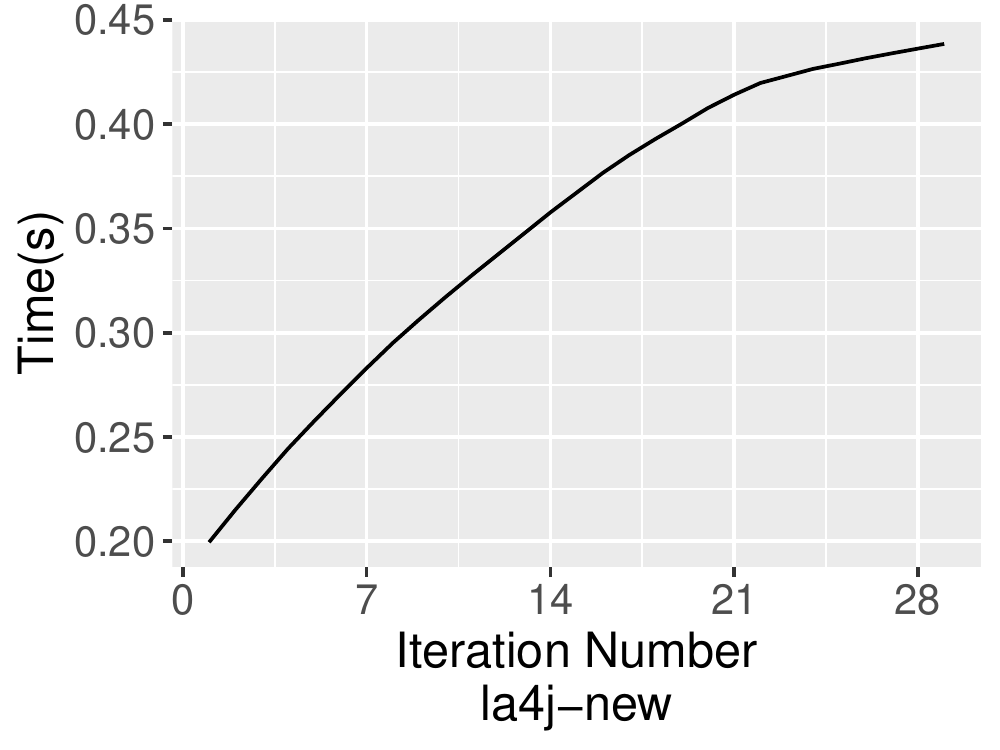}
\end{figure}

\begin{figure}[!h]
        \includegraphics[width=0.23\textwidth]{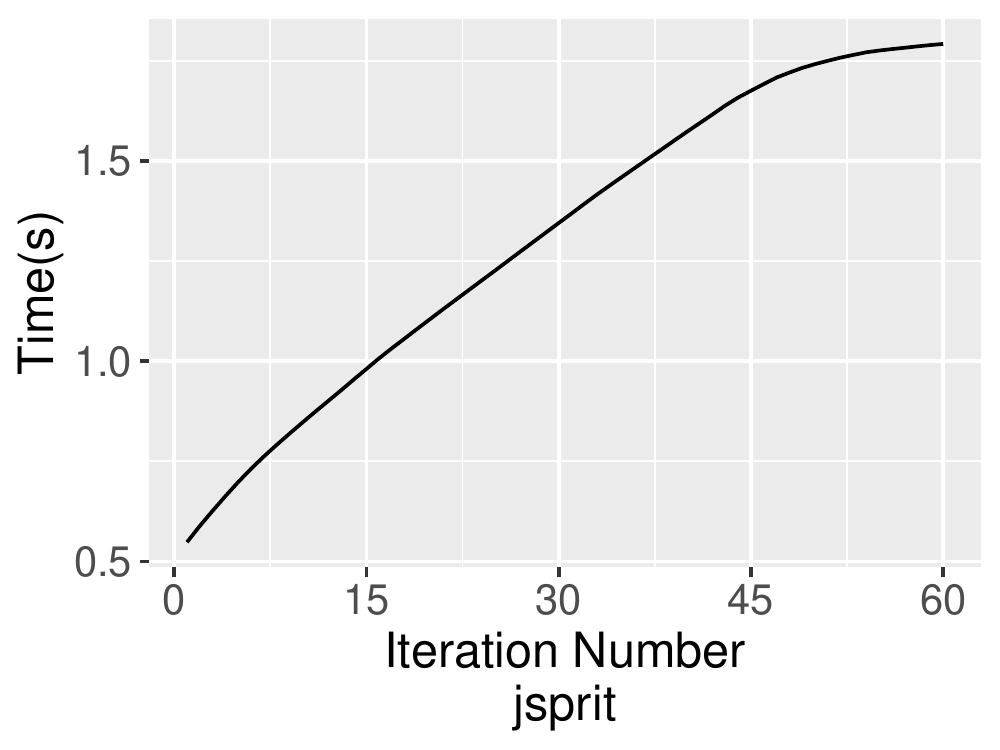}
        \includegraphics[width=0.23\textwidth]{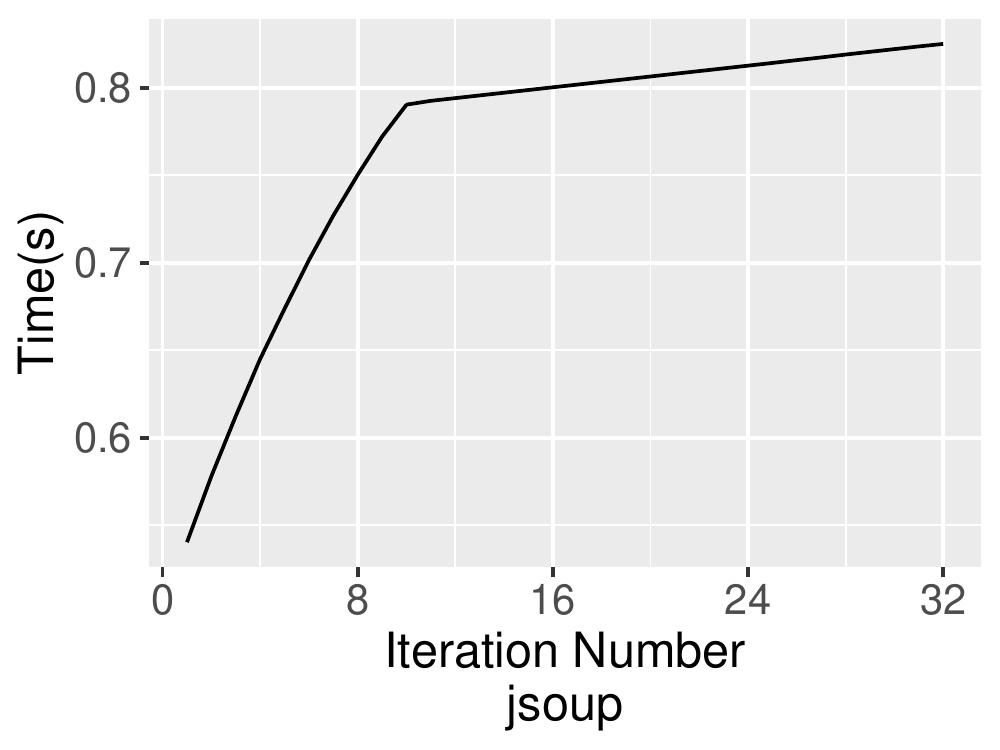}
\end{figure}
\begin{figure}[!h]
        \includegraphics[width=0.23\textwidth]{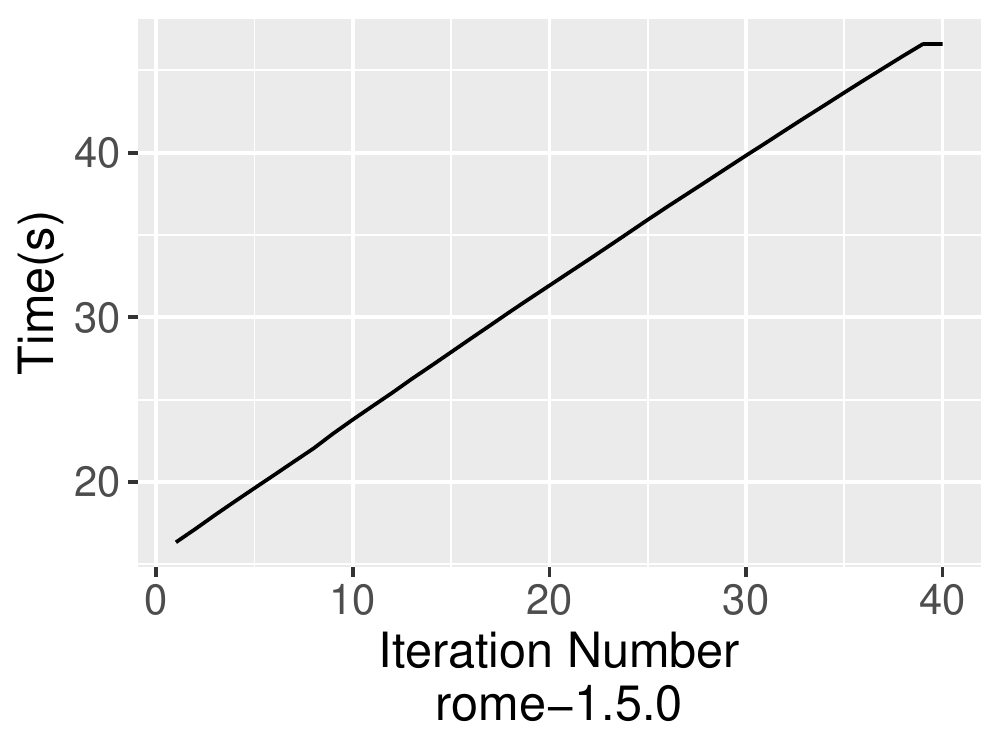}
        \includegraphics[width=0.23\textwidth]{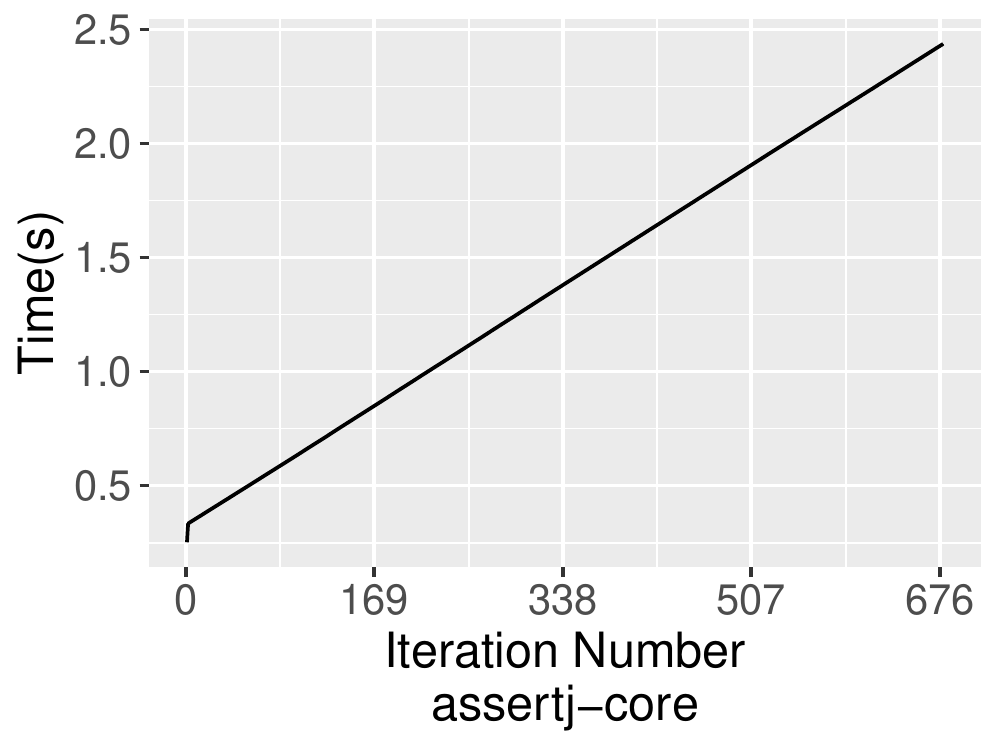}
\end{figure}

\begin{figure}[!h]
        \includegraphics[width=0.23\textwidth]{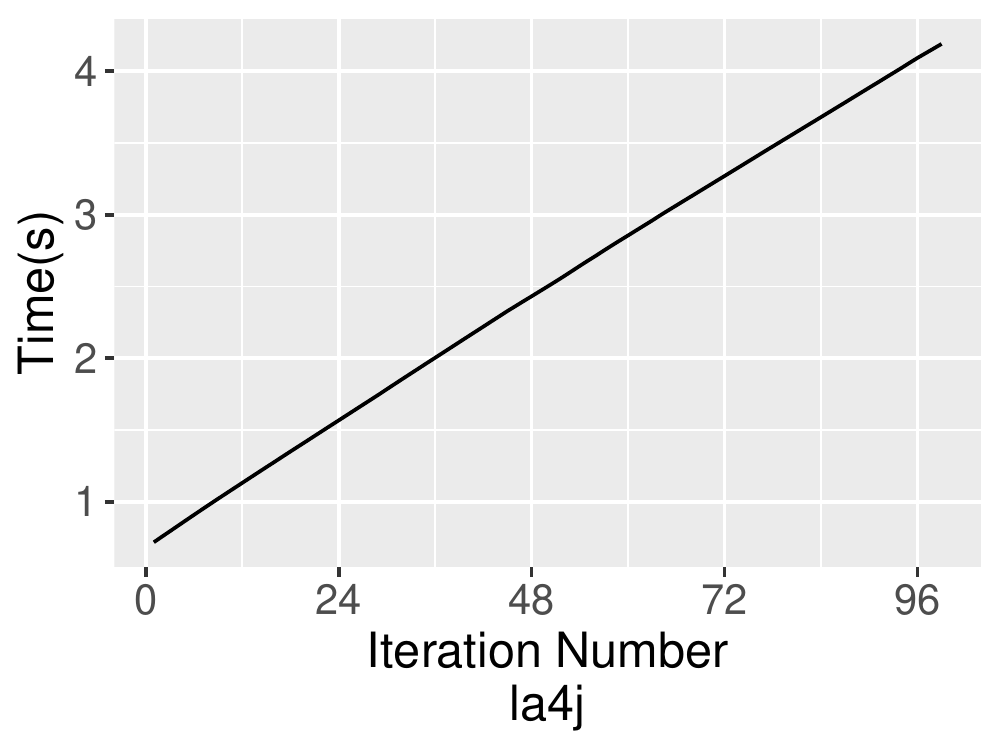}
        \includegraphics[width=0.23\textwidth]{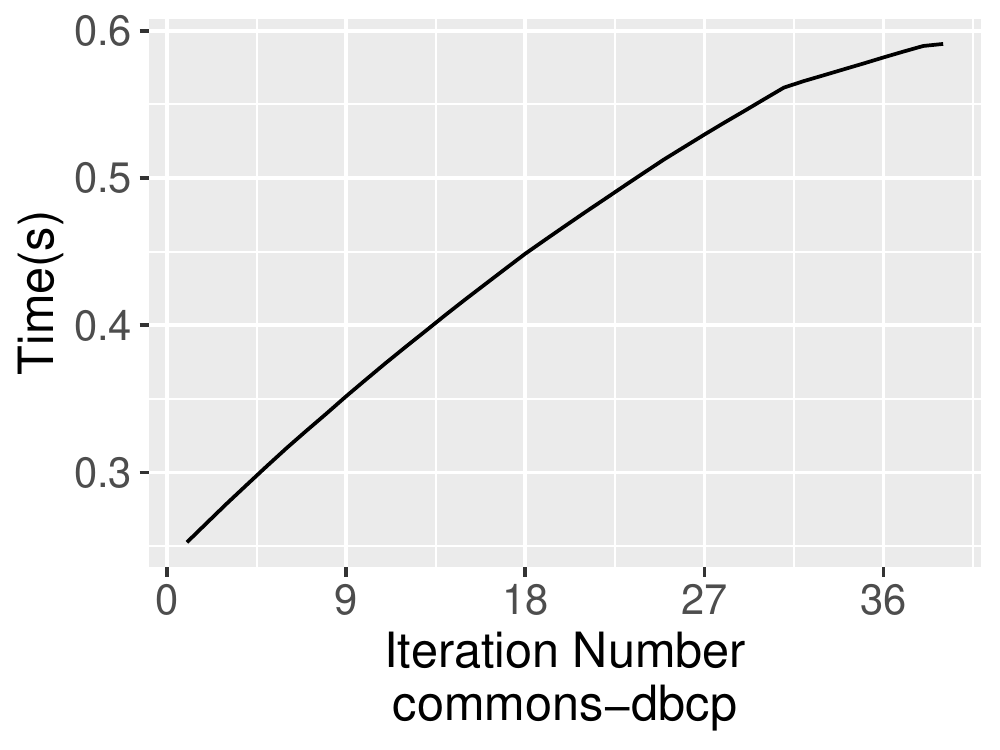}
\end{figure}

\begin{figure}[!h]
        \includegraphics[width=0.23\textwidth]{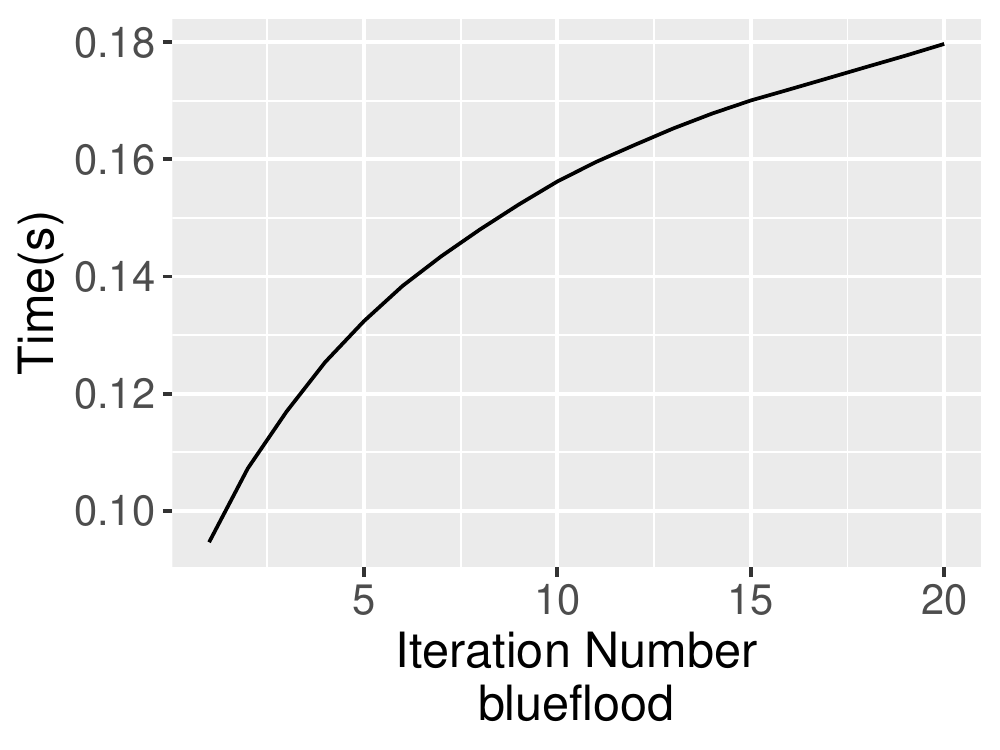}
        \includegraphics[width=0.23\textwidth]{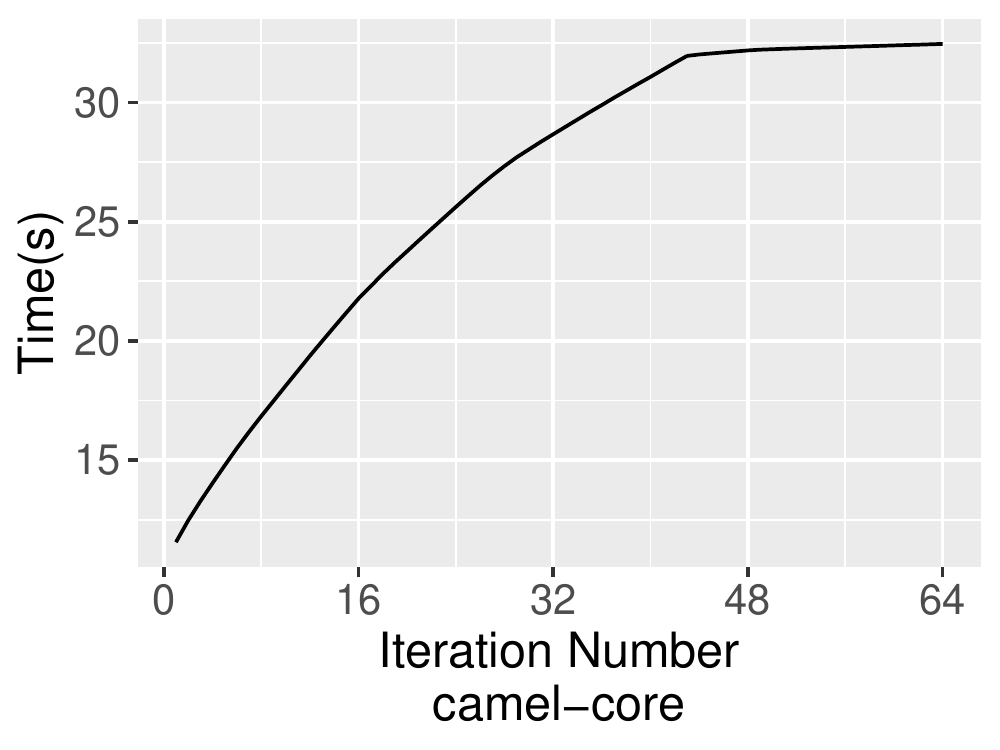}
\end{figure}

\begin{figure}[!h]
        \includegraphics[width=0.23\textwidth]{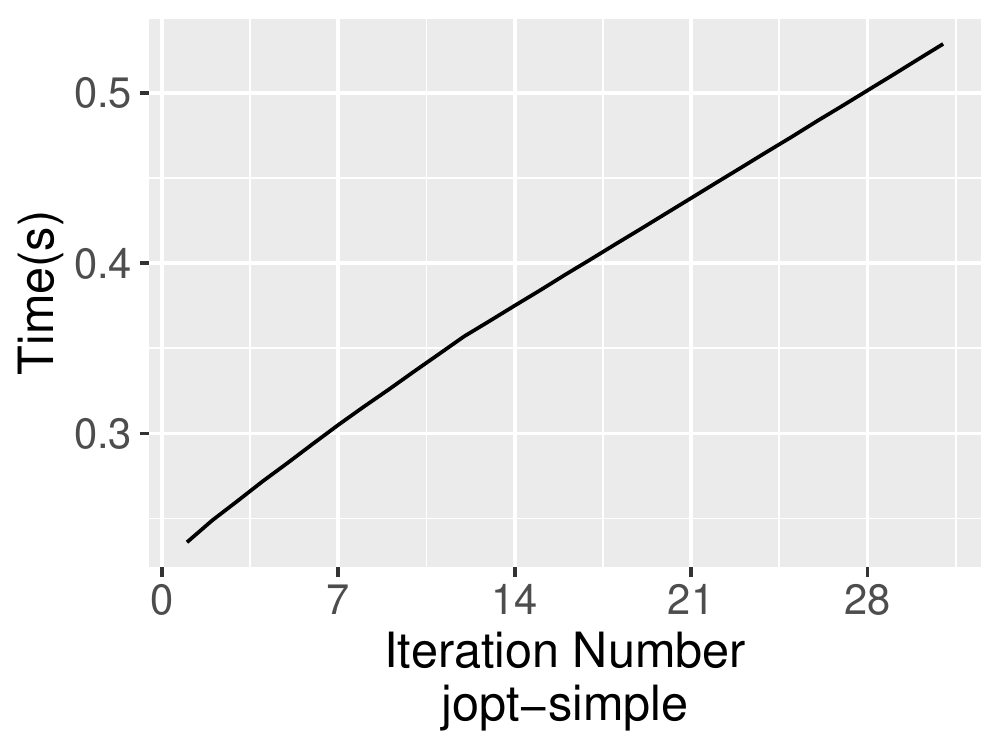}
        \includegraphics[width=0.23\textwidth]{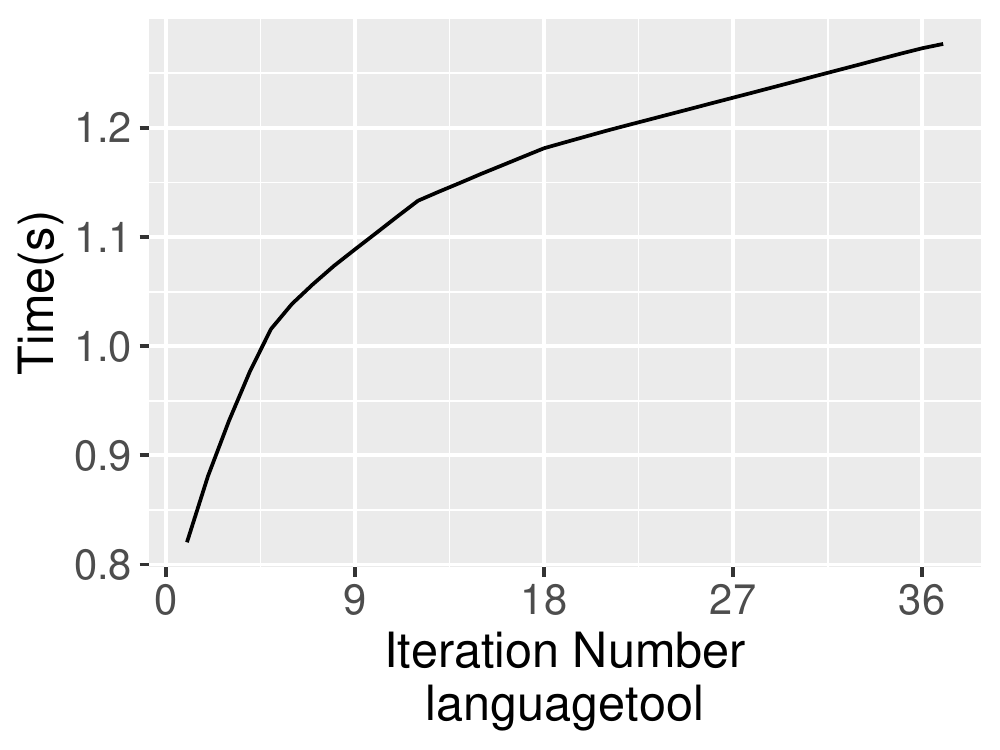}
\end{figure}

\begin{figure}[!h]
        \includegraphics[width=0.23\textwidth]{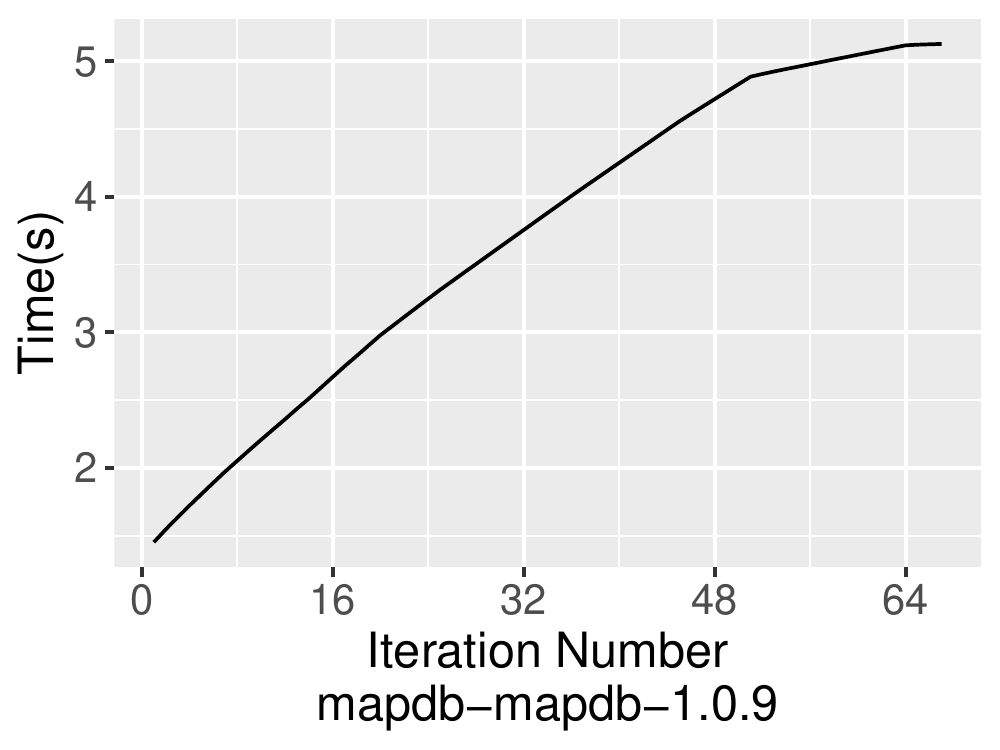}
\end{figure}

\section{\ins{basic information of open-source subjects}}
\label{appenda}

\ins{
Table~\ref{opensourceinfo} shows some basic information of our 55 open-source subjects. Specifically, for each subject, we present the source lines of code (SLOC), test lines of code (TLOC), number of test cases (\#Test cases), and number of mutants (\#Mutants), respectively. The projects are sorted in ascending order of source lines of code.
}

\begin{table*}[htbp]
    \centering
    \tiny
        \caption{Basic Information for Open-Source Subjects}
    \label{opensourceinfo}
    \resizebox{0.7\textwidth}{!}{
    \begin{tabular}{c|l|r|r|r|r}
        \toprule
            \textbf{ID} & \textbf{Subjects} & \textbf{SLOC} & \textbf{TLOC} & \textbf{\#Test Cases} & \textbf{\#Mutants} \\
        \midrule
$\mathcal{S}_{1}$& DiskLruCache& 780& 1,030& 61& 152\\
$\mathcal{S}_{2}$& gson-fire& 895& 726& 36&  520\\
$\mathcal{S}_{3}$& gson-fire-v2& 1,178& 952& 47& 202\\
$\mathcal{S}_{4}$& jumblr& 1,489& 1,243& 103&  167\\
$\mathcal{S}_{5}$& java-apns& 1,503& 1,724& 87& 412\\
$\mathcal{S}_{6}$& jasmine-maven-plugin& 1,671& 1,931& 102& 561\\
$\mathcal{S}_{7}$& java-uuid-generator& 1,790& 2,388& 45& 346\\
$\mathcal{S}_{8}$& gdx-artemis-master& 1,851& 1,492& 35& 961\\
$\mathcal{S}_{9}$& jopt-simple& 1,924& 5,903& 727& 1,677\\
$\mathcal{S}_{10}$& protoparser& 2,153& 3,227& 171&  864\\
$\mathcal{S}_{11}$& jackson-datatype-guava& 2,217& 1,035& 73& 845\\
$\mathcal{S}_{12}$& jackson-datatype-guava-v2& 2,366& 1,327& 80&  320\\
$\mathcal{S}_{13}$& JActor& 2,542& 4,418& 65& 56\\
$\mathcal{S}_{14}$& spring-retry& 2,765& 3,419& 185& 351\\
$\mathcal{S}_{15}$& scribe-java& 2,808& 2,536& 99& 563\\
$\mathcal{S}_{16}$& metrics-core& 2,835& 2,194& 150&  1,656\\
$\mathcal{S}_{17}$& javapoet& 2,986& 4,399& 332&  973\\
$\mathcal{S}_{18}$& low-gc-membuffers& 3,184& 9,782& 51& 780\\
$\mathcal{S}_{19}$& lambdaj-master& 3,634& 4,914& 265&  3,399\\
$\mathcal{S}_{20}$& LastCalc-0.1& 4,522& 581& 32& 2,499\\
$\mathcal{S}_{21}$& stream-lib& 4,835& 3,806& 141& 3,811\\
$\mathcal{S}_{22}$& webbit& 4,914& 8,463& 131& 349\\
$\mathcal{S}_{23}$& commons-pool& 5,206& 8,232& 272& 633\\
$\mathcal{S}_{24}$& redline-smalltalk-master& 5,648& 480& 43& 3,450\\
$\mathcal{S}_{25}$& la4j& 7,086& 4,050& 625&  5,023\\
$\mathcal{S}_{26}$& redline-smalltalk& 7,212& 2,414& 240& 833\\
$\mathcal{S}_{27}$& nv-websocket-client& 7,351& 657& 73& 277\\
$\mathcal{S}_{28}$& joss& 8,078& 6,035& 531& 1,289\\
$\mathcal{S}_{29}$& raml-java-parser-master& 8,696& 3,005& 192& 4,506\\
$\mathcal{S}_{30}$& raml-java-parser& 8,788& 5,061& 197& 1,288\\
$\mathcal{S}_{31}$& la4j-v2& 9,272& 4,035& 799& 3,141\\
$\mathcal{S}_{32}$& commons-io& 9,980& 19,189& 1,081& 7,773\\
$\mathcal{S}_{33}$& streamex& 10,427& 7,906& 450&  3,958\\
$\mathcal{S}_{34}$& jsoup& 10,507& 12,037& 666& 3,157\\
$\mathcal{S}_{35}$& commons-dbcp& 11,592& 8,752& 560&  2,601\\
$\mathcal{S}_{36}$& rome-1.5.0& 11,647& 2,705& 475& 4,929\\
$\mathcal{S}_{37}$& assertj-core& 13,361& 53,059& 2,470&  4,571\\
$\mathcal{S}_{38}$& vraptor-archive& 16,910& 16,213& 1,130& 7,245\\
$\mathcal{S}_{39}$& mapdb-mapdb-1.0.9& 17,589& 35,873& 1,776& 876\\
$\mathcal{S}_{40}$& RoaringBitmap& 17,807& 21,494& 1,148& 21,319\\
$\mathcal{S}_{41}$& blueflood& 19,517& 15,774& 961& 1,854\\
$\mathcal{S}_{42}$& lanterna& 20,682& 7,724& 34& 344\\
$\mathcal{S}_{43}$& jackson-core& 21,320& 10,924& 376& 6,215\\
$\mathcal{S}_{44}$& jsprit& 23,073& 18,373& 1,250& 12,350\\
$\mathcal{S}_{45}$& hivemall& 28,569& 3,975& 150&  6,557\\
$\mathcal{S}_{46}$& asterisk-java& 30,495& 4,263& 217&  3,226\\
$\mathcal{S}_{47}$& asterisk-java-v2& 31,074& 4,258& 217& 921\\
$\mathcal{S}_{48}$& restcountries& 31,324& 468& 40& 113\\
$\mathcal{S}_{49}$& chukwa& 32,654& 8,051& 131&  569\\
$\mathcal{S}_{50}$& ews-java-api& 45,313& 1,328& 90& 1,782\\
$\mathcal{S}_{51}$& languagetool& 47,589& 20,778& 719& 26,662\\
$\mathcal{S}_{52}$& OpenTripPlanner-otp-0.20.0& 64,718& 14,207& 379& 7,325\\
$\mathcal{S}_{53}$& hbase-1.2.2& 66,630& 17,385& 434& 1,781\\
$\mathcal{S}_{54}$& commons-math& 86,748& 90,798& 5,082& 84,476\\
$\mathcal{S}_{55}$& camel-core& 120,248& 134,036& 5,623&  13,005\\

\midrule
        \multicolumn{2}{c|}{\textbf{Total}} & 912,045 & 633,085 & 31,454 & 262,295 \\
        \bottomrule
    \end{tabular}
    }
\end{table*}

\section{\ins{results of open-source subjects}}
\label{appendc}

\ins{
Due to space limit, we show the complete results on open-source subjects in Table~\ref{opensource}. The subjects are sorted in ascending order of source lines of code (SLOC). The first three columns present the results for RQ1, the first five columns present the results for RQ2, the first nine columns present the results of RQ3, and the last four columns present the comparison results with \emph{FAST}. The detailed analysis can be found in Sections~\ref{results} and \ref{sec:comp}.
}

\begin{table*}[!htbp]
	\centering
	\caption{Results of Open-Source Subjects}
	\label{opensource}
	\resizebox{0.9\textwidth}{!}{
	\begin{threeparttable}
\begin{tabular}{c|rr|rr|cccr|ccrc}
\toprule
\midrule
& \multicolumn{2}{c|}{\textbf{\ins{RQ1}}} & \multicolumn{2}{c|}{\textbf{RQ2}}  & \multicolumn{4}{c|}{\textbf{RQ3}}  & \multicolumn{4}{c}{\textbf{Comparison with \emph{FAST}}}              \\
\multirow{-2}{*}{\textbf{Project}} & $\mathbf{Time}_{\mathbf{GA}}$ & \ins{$\mathbf{Time}_{\mathbf{I}}$} & $\mathbf{Time}_{\mathbf{GAF}}$ &  $\mathbf{Time}_{\mathbf{C}}$ & $\mathbf{APFD}_{\mathbf{GAF}}$ & $\mathbf{APFD}_{\mathbf{GA}}$ & $\mathbf{APFD}_{\mathbf{AGA}}$ & $\mathbf{Time}_{\mathbf{AGA}}$ & $\mathbf{APFD}_{\mathbf{FAST}}$ & $\mathbf{Win}_{\mathbf{APFD}}$ & $\mathbf{Time}_{\mathbf{FAST}}$ & $\mathbf{Win}_{\mathbf{Time}}$ \\
\midrule
$\mathcal{S}_{1}$        &  0.0197  & \ins{0.0197} &  0.0069  &  0.0157 \checkmark  &   0.8809             &   0.9070             &   0.9070             \checkmark  &  0.0157 \checkmark  &   0.8164        &   \checkmark        & 0.0717 &  \checkmark\\
$\mathcal{S}_{2}$        &  0.0072  & \ins{0.0072} &  0.0030  &  0.0058 \checkmark  &   0.8369             &   0.8380             &   0.8380             \checkmark  &  0.0058 \checkmark  &   0.7164       &   \checkmark         & 0.0348 &  \checkmark\\
$\mathcal{S}_{3}$        &  0.0074  & \ins{0.0074} &  0.0038  &  0.0222 \;\;\;  &   0.8868             &   0.8848             &   0.8848             \checkmark  &  0.0222 \;\;\;  &   0.6916       &   \checkmark         & 0.0228 &  \checkmark\\
$\mathcal{S}_{4}$         &  0.0140  & \ins{0.0140} &  0.0058  &  0.0789 \;\;\;  &   0.8505             &   0.8509             &   0.8509             \checkmark  &  0.0789 \;\;\;  &   0.7183        &   \checkmark        & 0.0140 & \;\;\;\\
$\mathcal{S}_{5}$        &  0.0314  & \ins{0.0314} &  0.0163  &  0.0089 \checkmark  &   0.8527             &   0.8527             &   0.8527             \checkmark  &  0.0089 \checkmark  &   0.7285        &   \checkmark        & 0.0542 &  \checkmark\\
$\mathcal{S}_{6}$        &  0.0115  & \ins{0.0115} &  0.0149  &  0.0046 \checkmark  &   0.8101             &   0.8101             &   0.8101             \checkmark  &  0.0046 \checkmark  &   0.6731        &   \checkmark        & 0.0315 &  \checkmark\\
$\mathcal{S}_{7}$        &  0.0045  & \ins{0.0045} &  0.0027  &  0.0076 \;\;\;  &   0.9045             &   0.9059             &   0.9059             \checkmark  &  0.0076 \;\;\;  &   0.7531        &   \checkmark        & 0.0163 &  \checkmark\\
$\mathcal{S}_{8}$        &  0.0202  & \ins{0.0202} &  0.0092  &  0.0180 \checkmark  &   0.8913             &   0.8898             &   0.8898             \checkmark  &  0.0180 \checkmark  &   0.6771        &   \checkmark        & 0.1102 &  \checkmark\\
$\mathcal{S}_{9}$        &  1.7455  & \ins{1.7105} &  1.7891  &  0.5288 \checkmark  &   0.9144             &   0.9144             &   0.9144             \checkmark  &  0.3363 \checkmark  &   0.8688        &   \checkmark        & 1.8445 &  \checkmark\\
$\mathcal{S}_{10}$        &  0.0871  & \ins{0.0871} &  0.0471  &  0.0247 \checkmark  &   0.9514             &   0.9518             &   0.9518             \checkmark  &  0.0247 \checkmark  &   0.7639       &   \checkmark         & 0.1326 &  \checkmark\\
$\mathcal{S}_{11}$        &  0.0975  & \ins{0.0975} &  0.0464  &  0.0553 \checkmark  &   0.8741             &   0.8766             &   0.8766             \checkmark  &  0.0553 \checkmark  &   0.6758        &   \checkmark        & 0.3315 &  \checkmark\\
$\mathcal{S}_{12}$        &  0.0243  & \ins{0.0243} &  0.0090  &  0.0108 \checkmark  &   0.8854             &   0.8864             &   0.8864             \checkmark  &  0.0106 \checkmark  &   0.6693        &   \checkmark        & 0.0452 &  \checkmark\\
$\mathcal{S}_{13}$        &  0.0383  & \ins{0.0383} &  0.0158  &  0.0385 \;\;\;  &   0.8500             &   0.8615             &   0.8615             \checkmark  &  0.0385 \;\;\;  &   0.7584       &   \checkmark         & 0.1346 &  \checkmark\\
$\mathcal{S}_{14}$        &  0.0876  & \ins{0.0870} &  0.0327  &  0.0407 \checkmark  &   0.9188             &   0.9188             &   0.9188             \checkmark  &  0.0385 \checkmark  &   0.6149       &   \checkmark         & 0.1315 &  \checkmark\\
$\mathcal{S}_{15}$        &  0.0221  & \ins{0.0221} &  0.0117  &  0.0055 \checkmark  &   0.8582             &   0.8582             &   0.8582             \checkmark  &  0.0055 \checkmark  &   0.7052        &   \checkmark        & 0.0381 &  \checkmark\\
$\mathcal{S}_{16}$        &  0.0723  & \ins{0.0723} &  0.0357  &  0.0152 \checkmark  &   0.8010             &   0.8031             &   0.8031             \checkmark  &  0.0152 \checkmark  &   0.6293       &   \checkmark         & 0.0793 &  \checkmark\\
$\mathcal{S}_{17}$        &  0.3706  & \ins{0.3693} &  0.2144  &  0.1357 \checkmark  &   0.9114             &   0.9183             &   0.9183             \checkmark  &  0.1317 \checkmark  &   0.7128       &   \checkmark         & 0.7061 &  \checkmark\\
$\mathcal{S}_{18}$        &  0.0323  & \ins{0.0323} &  0.0167  &  0.0226 \checkmark  &   0.9026             &   0.9028             &   0.9028             \checkmark  &  0.0226 \checkmark  &   0.7688        &   \checkmark        & 0.1115 &  \checkmark\\
$\mathcal{S}_{19}$        &  1.9204  & \ins{1.9201} &  1.5941  &  0.7017 \checkmark  &   0.9003             &   0.9033             &   0.9033             \checkmark  &  0.6995 \checkmark  &   0.6955       &   \checkmark         & 3.6103 &  \checkmark\\
$\mathcal{S}_{20}$        &  0.0655  & \ins{0.0655} &  0.0336  &  0.0469 \checkmark  &   0.8014             &   0.8013             &   0.8013             \checkmark  &  0.0469 \checkmark  &   0.6636       &   \checkmark         & 0.3444 &  \checkmark\\
$\mathcal{S}_{21}$        &  0.0942  & \ins{0.0942} &  0.0416  &  0.0248 \checkmark  &   0.8353             &   0.8328             &   0.8328             \checkmark  &  0.0248 \checkmark  &   0.6847        &   \checkmark        & 0.1313 &  \checkmark\\
$\mathcal{S}_{22}$        &  0.0413  & \ins{0.0413} &  0.0217  &  0.0215 \checkmark  &   0.8642             &   0.8642             &   0.8642             \checkmark  &  0.0215 \checkmark  &   0.7375       &   \checkmark         & 0.1019 &  \checkmark\\
$\mathcal{S}_{23}$         &  0.3202  & \ins{0.3202} &  0.2099  &  0.0962 \checkmark  &   0.8189             &   0.8198             &   0.8198             \checkmark  &  0.0962 \checkmark  &   0.6742        &   \checkmark        & 0.4346 &  \checkmark\\
$\mathcal{S}_{24}$         &  0.0239  & \ins{0.0217} &  0.0079  &  0.0198 \checkmark  &   0.9850             &   0.9858             &   0.9858             \checkmark  &  0.0187 \checkmark  &   0.9143        &   \checkmark        & 0.0919 &  \checkmark\\
$\mathcal{S}_{25}$        &  6.9852  & \ins{2.7593} &  1.0409  &  4.1901 \checkmark  &   0.7135             &   0.8450             &   0.8401             \;\;\;  &  1.0579 \checkmark  &   0.7500        &   \checkmark        & 5.7828 &  \checkmark\\
$\mathcal{S}_{26}$        &  0.0786  & \ins{0.0786} &  0.0397  &  0.0294 \checkmark  &   0.8322             &   0.8339             &   0.8339             \checkmark  &  0.0294 \checkmark  &   0.6079       &   \checkmark         & 0.0435 &  \checkmark\\
$\mathcal{S}_{27}$        &  0.0095  & \ins{0.0095} &  0.0041  &  0.0331 \;\;\;  &   0.9614             &   0.9614             &   0.9614             \checkmark  &  0.0303 \;\;\;  &   0.6707       &   \checkmark         & 0.0201 & \;\;\;\\
$\mathcal{S}_{28}$        &  0.5833  & \ins{0.5819} &  0.3976  &  0.1179 \checkmark  &   0.9153             &   0.9164             &   0.9164             \checkmark  &  0.1132 \checkmark  &   0.7159        &   \checkmark        & 0.5634 &  \checkmark\\
$\mathcal{S}_{29}$         &  8.5669  & \ins{8.5669} &  5.0644  &  2.3900 \checkmark  &   0.9482             &   0.9490             &   0.9490             \checkmark  &  2.3900 \checkmark  &   0.7234        &   \checkmark        & 14.0143 &  \checkmark\\
$\mathcal{S}_{30}$        &  0.4128  & \ins{0.4128} &  0.2461  &  0.1804 \checkmark  &   0.9620             &   0.9617             &   0.9617             \checkmark  &  0.1804 \checkmark  &   0.7599        &   \checkmark        & 0.9181 &  \checkmark\\
$\mathcal{S}_{31}$        &  2.0359  & \ins{1.9270} &  0.9671  &  0.4384 \checkmark  &   0.9511             &   0.9426             &   0.9426             \checkmark  &  0.3170 \checkmark  &   0.6473        &   \checkmark        & 1.5359 &  \checkmark\\
$\mathcal{S}_{32}$        &  1.0693  & \ins{1.0649} &  0.8031  &  0.1415 \checkmark  &   0.8889             &   0.8911             &   0.8911             \checkmark  &  0.1270 \checkmark  &   0.7007       &   \checkmark         & 0.3410 &  \checkmark\\
$\mathcal{S}_{33}$         &  0.6033  & \ins{0.6033} &  0.5780  &  0.0921 \checkmark  &   0.8659             &   0.8662             &   0.8662            \checkmark  &  0.0921 \checkmark  &   0.6105        &   \checkmark        & 0.4881 &  \checkmark\\
$\mathcal{S}_{34}$        &  5.7624  & \ins{5.6923} &  6.0388  &  0.8250 \checkmark  &   0.9277             &   0.9328             &   0.9328             \checkmark  &  0.7903 \checkmark  &   0.7515       &   \checkmark         & 4.2321 &  \checkmark\\
$\mathcal{S}_{35}$        &  1.5128  & \ins{1.3176} &  0.6824  &  0.5911 \checkmark  &   0.9163             &   0.9473             &   0.9467             \;\;\;  &  0.3631 \checkmark  &   0.7872        &   \checkmark        & 1.5940 &  \checkmark\\
$\mathcal{S}_{36}$        &  210.0549  & \ins{123.6547} &  32.9429  &  46.6052 \checkmark  &   0.8644             &   0.9418             &   0.9371             \;\;\;  &  23.7849 \checkmark  &   0.8092       &   \checkmark         & 129.8628 &  \checkmark\\
$\mathcal{S}_{37}$        &  5.2729  & \ins{3.2186} &  4.6264  &  2.4364 \checkmark  &   0.8508             &   0.8507             &   0.8507             \checkmark  &  0.3582 \checkmark  &   0.6925        &   \checkmark        & 1.0458 &  \checkmark\\
$\mathcal{S}_{38}$        &  3.6650  & \ins{3.6650} &  3.9989  &  0.2794 \checkmark  &   0.8657             &   0.8657             &   0.8657             \checkmark  &  0.2794 \checkmark  &   0.7608        &   \checkmark        & 1.2374 &  \checkmark\\
$\mathcal{S}_{39}$        &  42.3862  & \ins{35.0646} &  17.5059  &  5.1276 \checkmark  &   0.8679             &   0.9545             &   0.9545             \checkmark  &  2.2078 \checkmark  &   0.8279        &   \checkmark        & 11.6840 &  \checkmark\\
$\mathcal{S}_{40}$        &  4.0109  & \ins{4.0064} &  3.0029  &  0.6129 \checkmark  &   0.9198             &   0.9244             &   0.9244             \checkmark  &  0.5942 \checkmark  &   0.6993        &   \checkmark        & 1.6539 &  \checkmark\\
$\mathcal{S}_{41}$        &  0.9968  & \ins{0.9890} &  0.7100  &  0.1797 \checkmark  &   0.9040             &   0.9106             &   0.9106             \checkmark  &  0.1562 \checkmark  &   0.7181        &   \checkmark        & 0.3880 &  \checkmark\\
$\mathcal{S}_{42}$        &  0.0034  & \ins{0.0034} &  0.0035  &  0.0287 \;\;\;  &   0.8574             &   0.8569             &   0.8569              \checkmark  &  0.0287 \;\;\;  &   0.6954        &   \checkmark        & 0.0267 & \;\;\;\\
$\mathcal{S}_{43}$        &  1.4103  & \ins{1.4103} &  1.4142  &  0.2558 \checkmark  &   0.8913             &   0.8924             &   0.8924             \checkmark  &  0.2558 \checkmark  &   0.6681        &   \checkmark        & 1.5863 &  \checkmark\\
$\mathcal{S}_{44}$        &  7.2241  & \ins{5.9892} &  4.4672  &  1.7918 \checkmark  &   0.9159             &   0.9261             &   0.9240             \;\;\;  &  0.8465 \checkmark  &   0.7696       &   \checkmark         & 4.2414 &  \checkmark\\
$\mathcal{S}_{45}$         &  0.1297  & \ins{0.1297} &  0.1095  &  0.0597 \checkmark  &   0.8466             &   0.8464             &   0.8464             \checkmark  &  0.0597 \checkmark  &   0.6643         &   \checkmark       & 0.2205 &  \checkmark\\
$\mathcal{S}_{46}$        &  0.2842  & \ins{0.2842} &  0.2118  &  0.0858 \checkmark  &   0.8648             &   0.8656             &   0.8656             \checkmark  &  0.0858 \checkmark  &   0.6642       &   \checkmark         & 0.5231 &  \checkmark\\
$\mathcal{S}_{47}$        &  0.1310  & \ins{0.1310} &  0.0754  &  0.0629 \checkmark  &   0.8751             &   0.8750             &   0.8750             \checkmark  &  0.0629 \checkmark  &   0.7217        &   \checkmark        & 0.1188 &  \checkmark\\
$\mathcal{S}_{48}$        &  0.0025  & \ins{0.0025} &  0.0016  &  0.0025 \;\;\;  &   0.7939             &   0.7939             &   0.7939             \checkmark  &  0.0025 \;\;\;  &   0.6446       &   \checkmark         & 0.0064 &  \checkmark\\
$\mathcal{S}_{49}$         &  0.1545  & \ins{0.1545} &  0.0867  &  0.1120 \checkmark  &   0.7997             &   0.8009             &   0.8009             \checkmark  &  0.1120 \checkmark  &   0.6620        &   \checkmark        & 0.2295 &  \checkmark\\
$\mathcal{S}_{50}$        &  0.0071  & \ins{0.0070} &  0.0029  &  0.0049 \checkmark  &   0.8476             &   0.8517             &   0.8517             \checkmark  &  0.0047 \checkmark  &   0.7722       &   \checkmark         & 0.0069 &  \checkmark\\
$\mathcal{S}_{51}$         &  8.5532  & \ins{8.4018} &  8.6768  &  1.2770 \checkmark  &   0.8571             &   0.8671             &   0.8671             \checkmark  &  1.1036 \checkmark  &   0.6025        &   \checkmark        & 5.8874 &  \checkmark\\
$\mathcal{S}_{52}$        &  1.5933  & \ins{1.5887} &  1.5922  &  0.4241 \checkmark  &   0.9530             &   0.9542             &   0.9542             \checkmark  &  0.4052 \checkmark  &   0.7708       &   \checkmark         & 1.8314 &  \checkmark\\
$\mathcal{S}_{53}$        &  1.0478  & \ins{1.0459} &  0.9088  &  0.2427 \checkmark  &   0.8673             &   0.8710             &   0.8710             \checkmark  &  0.2298 \checkmark  &   0.6630        &   \checkmark        & 0.8469 &  \checkmark\\
$\mathcal{S}_{54}$        &  100.2525  & \ins{98.7254} &  78.0955  &  3.5015 \checkmark  &   0.9089             &   0.9089             &   0.9089             \checkmark  &  2.6648 \checkmark  &   0.7524       &   \checkmark         & 7.8017 &  \checkmark\\
$\mathcal{S}_{55}$         &  1,288.1519  & \ins{1,236.9016} &  734.9618  &  32.4581 \checkmark  &   0.9516             &   0.9544             &   0.9544             \checkmark  &  18.1040 \checkmark  &   0.8269        &   \checkmark        & 88.3705 &  \checkmark\\

\midrule
\textbf{Total} & & & & 48 & 0.8870 & 0.8870 & 51 & 48 & & 55 & & 52 \\
\midrule
\bottomrule
\end{tabular}
\end{threeparttable}
}
\end{table*}

\section{\ins{results of industrial subjects}}
\label{appendd}

\ins{
Due to space limit, we present the complete results on industrial subjects in Table~\ref{tab:case}. For each subject, we present its SLOC, \#Test cases, and the time cost of GA, AGA, \emph{FAST}, ART-D, GA-S, and GE, respectively. The detailed analysis can be found in Section~\ref{sec:case}.
}

\begin{table*}[t]
\setlength{\tabcolsep}{5.4pt}
	\centering
	\caption{Results of Industrial Subjects}
	\label{tab:case}
	\resizebox{.7\textwidth}{!}{
	\begin{threeparttable}
\begin{tabular}{c|rr|rrrrrr}
\toprule
& \multicolumn{2}{c|}{\textbf{Basic Information}}  & \multicolumn{6}{c}{\textbf{Time cost (s)}}         \\
\multirow{-2}{*}{\textbf{Subject}\tnote{*}} & \textbf{SLOC}\tnote{**} & \textbf{\#Test Cases} & \textbf{GA} & \textbf{AGA} & \textbf{\emph{FAST}} & \ins{\textbf{ART-D}} & \ins{\textbf{GA-S}} & \ins{\textbf{GE}}  \\
\midrule

$\mathcal{I}_1$        &  $>$500K  &  4,246  & 29,278.9102 & 359.9679 \checkmark & 1,860.1473  & \ins{543,106.2852} & \ins{2,680,036.2615} & \ins{54,969.0830} \\
$\mathcal{I}_2$        &  $>$200K  &  2,546  & 3,018.6473 & 89.9239 \checkmark & 398.8814   & \ins{32,938.6045} & \ins{315,888.7090} & \ins{13,887.3160} \\
$\mathcal{I}_3$        &  $>$200K  &  2,566  & 3,228.2772 & 86.0066 \checkmark & 417.8356   & \ins{30,458.8672} & \ins{304,555.6710} & \ins{14,124.1435} \\
$\mathcal{I}_4$        &  $>$200K  &  2,550  & 2,833.4841 & 80.5940 \checkmark & 383.9494  & \ins{24,944.4404} & \ins{265,881.1139} & \ins{19,345.1543} \\
$\mathcal{I}_5$        &  $>$200K  &  2,556  & 3,289.5958 & 94.0641 \checkmark & 428.5125 & \ins{31,799.7539} & \ins{366,902.7648} & \ins{8,424.3798} \\
$\mathcal{I}_6$        &  $>$500K  &  4,123  & 22,118.0296 & 329.4710 \checkmark & 1,439.6848 & \ins{402,039.4240} & \ins{1,766,206.5003} & \ins{49,274.3782} \\
$\mathcal{I}_7$        &  $>$500K  &  4,139  & 21,963.5968 & 336.3432 \checkmark & 1,600.3634 & \ins{411,725.1937} & \ins{2,390,410.2541} & \ins{54,897.2351} \\
$\mathcal{I}_8$        &  $>$200K  &  2,529  & 4,250.2729 & 89.2680 \checkmark & 446.4625 & \ins{36,610.5757} & \ins{461,096.5509} & \ins{3,857.2345} \\
$\mathcal{I}_9$        &  $>$500K  &  4,134  & 22,057.8564 & 335.8682 \checkmark & 1,450.5679   & \ins{28,328.2207} & \ins{2,091,910.4123} & \ins{37,817.4141} \\
$\mathcal{I}_{10}$       &  $>$200K  &  2,542  & 3,238.5423 & 96.6740 \checkmark & 418.7254   & \ins{769,960.0223} & \ins{265,087.9653} & \ins{7,134.1514} \\
$\mathcal{I}_{11}$       &  $>$500K  &  4,133  & 23,749.9149 & 348.1437 \checkmark & 1,531.0934 & \ins{398,946.3216} & \ins{2,537,854.1564} & \ins{39,417.0345} \\
$\mathcal{I}_{12}$       &  $>$500K  &  4,137  & 22,194.6776 & 342.6023 \checkmark & 1,466.4241 & \ins{398,254.6365} & \ins{2,016,031.3451} & \ins{38,741.9410} \\
$\mathcal{I}_{13}$       &  $>$500K  &  4,128  & 22,545.8684 & 362.5389 \checkmark & 1,470.3869 & \ins{446,056.7049} & \ins{2,018,768.3295} & \ins{49,287.1451} \\
$\mathcal{I}_{14}$       &  $>$200K  &  2,234  & 571.9417 & 22.2583 \checkmark & 85.0108 & \ins{4,999.5140} & \ins{37,081.3254} & \ins{487.0905} \\
$\mathcal{I}_{15}$       &  $>$500K  &  2,201  & 6,517.1065 & 190.7537 \checkmark & 926.5795 & \ins{71,541.1456} & \ins{513,769.5738} & \ins{19,481.4108} \\
$\mathcal{I}_{16}$       &  $>$20K  &  202  & 7.4382 & 3.5816 \checkmark & 9.7204 & \ins{87.4167} & \ins{601.2848} & \ins{42.7104} \\
$\mathcal{I}_{17}$       &  $>$200K  &  2,216  & 599.1948 & 16.0822 \checkmark & 85.3307 & \ins{12,411.9608} & \ins{32,268.7012} & \ins{7,015.4581} \\
$\mathcal{I}_{18}$       &  $>$20K  &  299  & 11.6980 & 2.2721 \checkmark & 10.5942 & \ins{83.9378} & \ins{988.3095} & \ins{38.6094} \\
$\mathcal{I}_{19}$       &  $>$500K  &  3,993  & 21,482.4772 & 335.6216 \checkmark & 1,750.2093 & \ins{444,997.4857} & \ins{2,295,089.0447} & \ins{64,510.4519} \\
$\mathcal{I}_{20}$       &  $>$200K  &  2,206  & 586.5093 & 18.7069 \checkmark & 87.0280 & \ins{6,905.6778} & \ins{75,574.2453} & \ins{1,048.8951} \\
$\mathcal{I}_{21}$       &  $>$20K  &  281  & 8.0470 & 1.8397 \checkmark & 9.1955 & \ins{34.1523} & \ins{610.4776} & \ins{19.9627} \\
$\mathcal{I}_{22}$       &  $>$500K  &  4,034  & 24,446.3671 & 335.9041 \checkmark & 1,778.7107 & \ins{466,512.4680} & \ins{2,636,222.8890} & \ins{52,941.8715} \\

\midrule
\textbf{Total} & $>$6,860K & 61,995 & & 22 & \\
\bottomrule
\end{tabular}
	\begin{tablenotes}
		\footnotesize
		\item[*] We hide project names for \new{the confidential policy}.
		\item[**] We report rough scale of SLOC due to the confidential policy.
	\end{tablenotes}
\end{threeparttable}
}
\end{table*}

\end{document}